\documentclass[fleqn]{article}
\usepackage[utf8]{inputenc}
\usepackage[english]{babel}
\usepackage{hyperref}

\hypersetup{
  colorlinks=true,
  linkcolor=blue,
  citecolor=black,
  urlcolor=black,
  breaklinks=true,
  pdfencoding=auto
}
\usepackage{mathtools}  
\usepackage{diffcoeff}
\usepackage{authblk}
\usepackage{csquotes}
\usepackage{color}
\usepackage{xcolor}
\usepackage{subcaption}
\usepackage{booktabs}
\usepackage{amsmath}
\usepackage{amssymb} 
\usepackage{graphicx}
\usepackage{mathtools}
\usepackage{float}
\usepackage{natbib}
\setcitestyle{authoryear} 
\usepackage{graphicx} 
\graphicspath{ {./images/} }
\usepackage[rightcaption]{sidecap}
\usepackage{wrapfig}
\usepackage{oplotsymbl}
\usepackage{tikz}
\usepackage{wasysym}
\usepackage{subcaption}
\DeclareRobustCommand\dashed{\tikz[baseline=-0.4ex]\draw[thick,dashed] (0,0)--(0.54,0);}
\DeclareRobustCommand\full{\tikz[baseline=-0.4ex]\draw[thick] (0,0)--(0.54,0);}
\newcommand{\blue}[1]{{\color{black}#1}}
\newcommand{\green}[1]{{\color{black}#1}}

\title{The effect of contact angle hysteresis on a droplet in a viscoelastic two-phase system}
\author[$,a$]{Kazem Bazesefidpar\thanks{Corresponding author : kazemba@kth.se}}
\author[a]{Outi Tammisola}
\affil[a]{\small \textit{SeRC (Swedish e-Science Research Centre) and FLOW, Dept. of Engineering Mechanics, KTH Royal Institute of Technology, SE-10044 Stockholm, Sweden}}
\date{} 

\begin{document}

\maketitle

\section*{abstract}
We investigate the dynamic behaviour of a two-dimensional (2D) droplet adhering to a wall in Poiseuille flow at low Reynolds numbers, in a system where either the droplet is viscoelastic (V/N) or the surrounding medium (N/V). The fluid viscoelasticity has been modeled by the Giesekus constitutive equation, and the Cahn–Hilliard Phase-Field method is used to capture the interface between the two phases. The contact angle hysteresis is represented by an advancing contact angle $\theta_{A}$ and a receding contact angle $\theta_{R}$. The results reveal that the deformation of the viscoelastic drop over time is changed due to the presence of polymeric molecules, and it can be categorized in two stages prior to depinning of the contact lines. In the first stage, the viscoelastic droplet speeds up and deforms faster, while in the second stage, the Newtonian counterpart accelerates and its deformation outpaces the viscoelastic droplet. The deformation of viscoelastic drop is retarded significantly in the second stage with increasing Deborah number $De$. In the V/N case, the viscous bending is enhanced on the receding side for small $De$, but it is weakened by further increase in $De$, and this non-monotonic behavior brings about an increase in the receding contact line velocity at small $De$ and a decrease at large $De$. \green{On the advancing side, the viscous bending is decreased monotonically for $Ca<0.25$, and hence the advancing contact line velocity is decreased with increasing $De$; however, the viscous bending presents a non-monotonic behavior for $Ca=0.25$ at the advancing side}. The non-monotonic behavior on the receding side is attributed to the emergence of outward pulling stresses in the vicinity of the receding contact line and the inception of strain-hardening at higher $De$, while the reduction in the viscous bending at the advancing side is the result of just strain-hardening due to the presence of dominant extensional flow on the advancing side. Finally, in the N/V system, the viscoelasticity of the medium suppresses the droplet deformation on both receding and advancing sides, and this effect is more pronounced with increasing De; the weakening effect of viscous bending is enhanced significantly at the advancing side by increasing the Giesekus mobility parameter. These results give a thorough understanding of viscoelastic effect on both drop deformation and depinning at both contact lines over a surface with contact angle hysteresis.

\vspace{0.5cm}
\textbf{Keywords:}
Wetting, viscoelasticity, contact angle hysteresis, superhydrophobic surface, Diffuse-interface method
\section{Introduction}\label{sec1}
The wetting of a solid substrate by a polymeric drop is an important process in many applications such as inkjet printing, coating, and self-cleaning surfaces. Among these applications, superhydrophobic surfaces have recently drawn the attention of many researchers due to their application in drag reduction and self-cleaning \citep{Cheng2005,Neinhuis1997}. These surfaces are inspired by natural surfaces, such as gecko skin \citep{Watson2015} and lotus leaves that give rise to the so-called lotus effect \citep{Shirtcliffe2010}. Superhydrophobic surfaces are usually characterized (on a macroscopic length scale) by the advancing contact angle $\theta_{A}$ and the receding contact angle $\theta_{R}$, and the difference between these two contact angles called Contact Angle Hysteresis (CAH). The properties of a surface such as microscopic roughness and inhomogeneities may lead to CAH \citep{Joanny1990}, and the movement of a droplet on the surface can be significantly affected by CAH. Furthermore, the non-Newtonian properties of a droplet affect its motion on a solid surface. For example, the elasticity of a droplet moderately speeds up the contact-line motion on a hydrophilic flat and chemically homogeneous surface at late stages of wetting and for a high-viscosity liquid \citep{Yue2012,Wang2015,Wang2017}, but a recent study \citep{Yada2023} revealed that increasing the polymer concentration does not modify the contact-line friction for aqueous polymer (shear-thinning) solutions on the smooth surface in the initial wetting regime. Thus, the elasticity of a drop has a negligible effect on the early stages of the wetting process, and the contact line motion at early times can be controlled by changing the viscosity of the solvent or adding roughness/inhomogeneities to the surface.

In contrast to the smooth hydrophilic and hydrophobic surfaces, the elasticity of a droplet plays an important role on the superhydrophobic surfaces. Experiment results suggest that the superhydrophobic surface may lose its self-cleaning property when they interact with highly elastic fluids \citep{Xuve2018}. The motion of a Boger fluid drop, which shows viscoelasticity without shear thinning, on the superhydrophobic surface is strikingly different from a Newtonian drop. The velocity of the viscoelastic drop is notably reduced in comparison with the Newtonian drop, and complex branch-like patterns are left on the tilted superhydrophobic surface. Previous studies also showed that the polyacrylamide drops at higher polymer concentrations were noticeably more elongated when sliding down a hydrophilic surface with small CAH \citep{Varagnolo2017}. The wetting area of a viscoelastic drop in a Newtonian surrounding fluid has been studied by numerical simulations using the Oldroyd-B model \citep{Wang2023}. They found that the wetting area decreased when the elasticity of the fluid increased, and the same was observed when the polymeric viscosity ratio decreased. Also droplet motion on lubricant surfaces has been analysed. A lubricant surface is constructed by trapping a suitable low surface tension lubricating liquid inside a texture \citep{Lafuma2011}, and these surfaces hardly pin drops. An oscillatory motion has been reported for the viscoelastic drop having both shear-thinning and elasticity on lubricant surfaces \citep{Sartori2022}.

Our aim is to model the effect of CAH on viscoelastic fluids on a continuum level, neglecting the microscopic details of the roughness, using a model that can handle a dynamic contact angle, but still allows receding and advancing contact angles to have different values. The most popular approach has been developed by \citet{Spelt2005} for a level-set method. An intermediate contact angle is computed, and the contact line pins if it is within the hysteresis window $[\theta_{R},\theta_{A}]$, otherwise it moves. This approach has then been extended to the phase-field method \citep{Ding2008}, volume-of-fluid method \citep{Dupont2010}, lattice Boltzmann method \citep{Ba2013}, and front-tracking method \citep{Shin2018}. It should be noted that the above methods need ghost cells outside the boundary to impose the contact angle condition and a slip model to circumvent the well-known singularity in the contact line \citep{Batchelor2003}; however, the Cahn–Hilliard equation automatically regularizes the singularity at the contact line \citep{Jacqmin2000}. A CAH model capable of capturing pinning, advancing and receding automatically without needing any explicit knowledge of contact line velocity was proposed by \citet{Yue2020}, and it has been further extended to level-set method \citep{Zhang2020}. Recently, an angle-dependent line friction model for the Cahn–Hilliard equation has been proposed by \citet{Amberg2022}, and the numerical results of the proposed model were in a very good agreement with the experimental data.

In the present work, we investigate the effect of a fluid's viscoelastic properties on the deformation and movement of a droplet with contact line hysteresis in a 2D Poiseuille flow at low Reynolds number ($Re$) by direct numerical simulation. The deformation and yield condition (depinning of contact lines) of both 2D and three-dimensional (3D) Newtonian drops were the subject of many previous studies at low and high $Re$, which are summarized below.

The yield condition of a 2D droplet in a shear flow and stokes regime was examined for a broad range of capillary numbers (Ca), viscosity ratios ($\lambda_{\mu}$), and CAH ($\Delta\theta=\theta_{A}-\theta_{R}$) \citep{Dimitrakopoulos1997}. It is found that increasing $\Delta\theta$ results in an increase in the critical capillary number ($Ca_{cr}$), above which the droplet starts to move. Their study also demonstrated that a droplet with a higher viscosity (higher $\lambda_{\mu}$) on a surface with a specific CAH starts to move at a smaller $Ca_{cr}$ than a droplet with a lower viscosity.

The effect of $Ca$, $\lambda_{\mu}$, slip velocity, and surfactants on the 2D droplet deformation and yield condition in Stokes flow regime was studied by \citet{Schleizer1999} in both shear-driven and pressure-driven flows. They demonstrated that surfactants reduce the deformation of a drop due to the generation of marangoni stresses along the droplet surface; in addition, a decrease in the deformation of the drop when the surface is hydrophilic and an increase in the deformation when the surface is hydrophobic have been observed in the presence of a slip velocity without considering CAH.

Later, 2D droplet deformation in shear flow with pinned and moving contact lines was studied at moderate $Re$ using a level-set method by \citet{Spelt2006}. The author showed that the critical Weber number ($We_{cr}$) approaches a constant value when $Re$ increases. The droplet becomes unstable for $We>We_{cr}$, and the transient behaviour is very different at different Reynolds numbers. It was demonstrated that the slip at the wall increases the $We_{cr}$, and the slip length has a significant effect on the contact line speed. It should be mentioned that the behavior of the droplet with slip velocity was reported to be dependent on the slip length.

Afterward, the critical conditions for the motion of a 3D droplet at both low and moderate $Re$ in the presence of CAH were studied \citep{Ding2008}. The inertial effects led to a complicated structure around the deformed drop with a strong 3D effect for the moderate Re, whereas their results were in agreement with previous 2D studies at low Re.

In the present study, we conduct direct numerical simulations to study the effect of fluid viscoelasticity on the deformation and depinning of a 2D drop either in V/N or N/V systems over a surface in the presence of CAH at low $Re$. In particular, we focus on the physical effects of polymers on the deformation and depinning processes. We apply the Cahn–Hilliard equation with a dynamic contact angle boundary condition for capturing the interface between the two phases, and the Giesekus constitutive equation to model the viscoelasticity of the drop. First, the qualitative effect of the elasticity is investigated over a wide range of the $Ca$ at a low constant $De=1$. We compare the time evolution \green{of both dynamic contact angle and velocity} of the receding and advancing contact lines for a viscoelastic droplet (V/N) and a Newtonian droplet in a Newtonian medium over surfaces with a varying but significant degree of contact line hysteresis ($\Delta\theta=[100^{\circ},130^{\circ},140^{\circ}]$). Next, the effect of $De$ is studied in the V/N system. \blue{Finally, the effect of $De$ and $\alpha$ has been investigated on the drop's deformation and the depping of contact lines in the N/V system}.
\section{Equations}\label{sec2}
\subsection{Governing equations}
Consider an incompressible two-phase system composed of a Newtonian fluid with viscosity $\mu_{n}=\mu_{s}$ and a viscoelastic (Giesekus) fluid with solvent viscosity $\mu_{s}$, and polymeric viscosity $\mu_{p}$. A phase-field variable $\phi$ varies smoothly from phase one ($\phi=1$) to another phase ($\phi=-1$) with $\phi=0$ denoting the fluid/fluid interface. This two-phase system is modeled by the following coupled system of equations \citep{Yue2004,Abels2012}:
\\
the linear momentum and continuity equations
\begin{eqnarray}
\rho(\frac{\partial{\mathbf{u}}}{\partial{t}}+({\mathbf{u}}\cdot\nabla){\mathbf{u}})+
{\mathbf{J}}\cdot\nabla{\mathbf{u}}&=& \nonumber -\nabla{p}+\nabla\cdot\mu(\nabla{{\mathbf{u}}}+
\nabla{{\mathbf{u}}^T})+G\nabla\phi\\
+\nabla\cdot({\frac{(1\pm\phi)}{2}\boldsymbol\tau_{p}}),
\label{NS1}
\end{eqnarray}
\begin{eqnarray}
\nabla\cdot{\mathbf{u}}=0,
\label{NS2}
\end{eqnarray}
the Giesekus constitutive model for the viscoelastic fluid
\begin{eqnarray}
\boldsymbol\tau_{p}+\lambda_H(\frac{\partial{\boldsymbol\tau_{p}}}{\partial{t}}+
{\mathbf{u}}\cdot\nabla{\mathbf{\boldsymbol\tau_{p}}}-\boldsymbol\tau_{p}\nabla{{\mathbf{u}}}-
\nabla{{\mathbf{u}}^T}\boldsymbol\tau_{p})+\frac{\alpha\lambda_H}{\mu_p}(\boldsymbol\tau_{p}\cdot\boldsymbol\tau_{p})&=& \nonumber \\
\mu_p(\nabla{{\mathbf{u}}}+\nabla{{\mathbf{u}}^T}), \quad \quad \quad \quad \quad \quad \quad \quad \quad \quad \quad \quad \quad \quad \quad \quad \quad \quad \quad
\label{NS3}
\end{eqnarray}
and the Cahn-Hilliard equation for the movement of the interface
\begin{eqnarray}
\frac{\partial{\phi}}{\partial{t}}+\nabla\cdot({{\mathbf{u}}\phi})= M\nabla^{2} G,
\label{NS4}
\end{eqnarray}
\begin{eqnarray}
G=-\lambda\nabla^2{\phi}+h(\phi).
\label{NS5}
\end{eqnarray}
In the above equations, $\mathbf{u}(\mathbf{x},t)$ is the velocity vector, $p(\mathbf{x},t)$ is the pressure, and $\boldsymbol\tau_p$ is the extra stress due to the polymers. In the Giesekus equation, $\lambda_H$ is the polymer relaxation time, and $\alpha$ is the Giesekus mobility parameter. In the Cahn-Hilliard equation, $G$ is the chemical potential, $M$ is the mobility parameter, $\lambda$ is the mixing energy density which is related to the surface tension in the sharp-interface limit \citep{Yue2004} as
\begin{eqnarray}
\sigma=\frac{2\sqrt{2}}{3}\frac{\lambda}{\eta}.
\label{NS6}
\end{eqnarray} 
The function $h(\phi)$ in Eq. (\ref{NS5}) is defined as 
\begin{eqnarray}
h(\phi)=F^{\prime}(\phi)=\frac{\lambda}{\eta^2}\phi(\phi^2-1),
\label{NS7}
\end{eqnarray}
with $F(\phi)=\frac{\lambda}{4\eta^2}(\phi^2-1)^{2}$ being the double-well potential, and where $\eta$ is the capillary width of the interface. The density $\rho$ and the dynamic viscosity $\mu$ of the mixture are a linear function of the phase-field variable as:
\begin{eqnarray}
\rho = \frac{(1+\phi)}{2}{\rho_1}+\frac{(1-\phi)}{2}{\rho_2},
\label{NS8}
\end{eqnarray}
\begin{eqnarray}
\mu=\frac{(1+\phi)}{2}{\mu_{s1}}+\frac{(1-\phi)}{2}{\mu_{s2}}.
\label{NS9}
\end{eqnarray}
The total viscosity of the non-Newtonian phase is $\mu=\mu_{s}+\mu_{p}$. The flux of density due to the diffusion of components $\mathbf{J}$ \citep{Abels2012} in Eq. (\ref{NS1}) is given by
\begin{eqnarray}
\mathbf{J}=-\frac{(\rho_1-\rho_2)}{2}M\mathbf{\nabla}{G}.
\label{NS10}
\end{eqnarray}

The equations (\ref{NS1}), (\ref{NS3}), and (\ref{NS4}) should be supplemented with appropriate boundary and initial conditions. The following boundary conditions are imposed:
\begin{eqnarray}
\boldsymbol{u}=\mathbf{0},
\label{NS11}
\end{eqnarray}
\begin{eqnarray}
{\mathbf{n}}\cdot\nabla{G}=0,
\label{NS12}
\end{eqnarray}
\begin{eqnarray}
-\eta\mu_{f}(\frac{\partial{\phi}}{\partial{t}}+\mathbf{u}\cdot\nabla\phi)=\frac{3}{2\sqrt{2}}\sigma\eta \mathbf{n}\cdot\nabla{\phi}+f^{\prime}_{w}(\phi) \hspace{1cm} on \hspace{0.5cm} \partial\Omega_{w},
\label{NS13}
\end{eqnarray}
\begin{eqnarray}
\mathbf{n}\cdot\nabla{\phi}=0 \hspace{1cm} on \hspace{0.5cm} \partial\Omega \backslash \partial\Omega_{w}
\label{NS14}
\end{eqnarray}
Equations (\ref{NS11}) and (\ref{NS12}), respectively, impose the no-slip and zero diffusive flux at the wall, where $\mathbf{n}$ is the outward pointing normal vector to the boundary. Equations (\ref{NS13}) and (\ref{NS14}) correspond to the wall energy relaxation \citep{Jacqmin2000,Carlson2009} and neutral wettability condition, respectively. $\partial\Omega \backslash \partial\Omega_{w}$ denotes the boundaries other than the solid walls intersected with contact line, and $\mu_{f}$ is the contact line friction. The surface chemical potential $L({\phi,\nabla\phi})$ is defined as
\begin{eqnarray}
L({\phi,\nabla\phi})=\lambda\mathbf{n}\cdot\nabla{\phi}+f^{\prime}_{w}(\phi),
\label{NS15}
\end{eqnarray}
where $f_{w}=-\sigma \frac{\phi (3-\phi^{2})}{4} cos\theta$ is the wall energy.
\section{Numerical method}\label{sec3}
The governing equations are discretized by a second-order finite difference method on a dual-resolution grid. The numerical method used in the present study has been described in detail and validated in \citep{Bazesefidpar2022}, except for the Cahn-Hilliard equation which is described in the following section.
\subsection{Cahn-Hilliard equation}
Different approaches have been proposed to deal with the non-linear term in the Cahn-Hilliard equation (Eq. \ref{NS4}) embedded in the chemical potential $G$ \citep{Gao2014,Shen2010,Guill2013,Shen2018}, but scalar auxiliary variable (SAV) proposed by \citep{Shen2018} has the advantages that the resulting discretized equations are linear, unconditionally energy stable, and relatively straight-forward to implement. Challenges still remain when using this method, and the accuracy of SAV scheme declines if too large time steps are chosen so that the results are generally no longer accurate. One strategy to overcome this issue is to use adaptive time stepping strategy \citep{Shen2019} or the relaxed-SAV (RSAV) method \citep{Jiang2022,Shen2022}, and the RSAV method proposed by \citet{Jiang2022} has been used in the present study. The Equations (\ref{NS4}-\ref{NS5}) with the boundary conditions (\ref{NS12}-\ref{NS14}) can be reformulated and discretized in time by using RSAV-BDF2 scheme \citep{Jiang2022} as
\begin{eqnarray}
\frac{\gamma_{0}\phi^{n+1}-\hat{\phi}}{\Delta t}+\nabla\cdot({{\mathbf{u^{*,n+1}}}\phi^{*,n+1}})= M\nabla^{2} G^{n+1},
\label{NS17}
\end{eqnarray}
\begin{eqnarray}
G^{n+1}= -\lambda\nabla^2{\phi^{n+1}}+\frac{\tilde{r}^{n+1}}{\sqrt{E(\phi^{*,n+1})}}h(\phi^{*,n+1}),
\label{NS18}
\end{eqnarray}
\begin{eqnarray}
\frac{\gamma_{0}\tilde{r}^{n+1}-\hat{r}}{\Delta t}= \int_{\Omega}\frac{h(\phi^{*,n+1})}{2\sqrt{E(\phi^{*,n+1})}} \frac{\gamma_{0}\phi^{n+1}-\hat{\phi}}{\Delta t},
\label{NS19}
\end{eqnarray}
\begin{eqnarray}
{\mathbf{n}}\cdot\nabla{G}^{n+1}=0,
\label{NS20}
\end{eqnarray}
\begin{eqnarray}
\eta\mu_{f}(\frac{\gamma_{0}\phi^{n+1}-\hat{\phi}}{\Delta t})=-L^{n+1}_{A,R} \hspace{1cm} on \hspace{0.5cm} \partial\Omega_{w},
\label{NS21}
\end{eqnarray}
\begin{eqnarray}
\mathbf{n}\cdot\nabla{\phi^{n+1}}=0 \hspace{1cm} on \hspace{0.5cm} \partial\Omega \backslash \partial\Omega_{w},
\label{NS22}
\end{eqnarray}
\begin{eqnarray}
r^{n+1}= \zeta_{0}\tilde{r}^{n+1}+(1-\zeta_{0})\sqrt{E(\phi^{n+1})},
\label{NS23}
\end{eqnarray}
where $E(t)=\int_{\Omega}F(\phi)$, $g^{*,n+1}=2g^{n}-g^{n-1}$ represents the second-order explicit approximation of $g^{n+1}$, $\hat{g}=2g^{n}-\frac{1}{2}g^{n-1}$, and $\gamma_{0} = \frac{3}{2}$. The $L^{n+1}_{A,R}$ on the right-hand side of Eq. \ref{NS21} is defined as \citep{Yue2020}
\begin{equation}
    L^{n+1}_{A,R}=
	\begin{cases}
	L_{A}({\phi^{*,n+1},\nabla\phi^{n+1})} \qquad if \qquad 0>L^{*,n+1}_{A}>L^{*,n+1}_{R},\\
	L_{R}({\phi^{*,n+1},\nabla\phi^{n+1})} \qquad if \qquad L^{*,n+1}_{A}>L^{*,n+1}_{R}>0,\\
    0 \qquad otherwise,
	\end{cases}  
	\label{NS16}
\end{equation}
where the explicit second order approximation $\phi^{*,n+1}$ of the phase field function has been used in computing $L_{A}$ and $L_{R}$ with A and R corresponding to advancing and receding contact angles respectively. If the contact angle is between the hysteresis window $[\theta_{R},\theta_{A}]$, the contact line is pinned; otherwise it moves. \blue{The speed of the receding contact line (when it is not pinned) depends on the $\theta_{d,R}$, and the advancing contact line speed on the $\theta_{d,A}$, where $\theta_{d}$ is the dynamic contact angle defined on the wall ($y=0$)}. The above system of equations (\ref{NS17}-\ref{NS22}) can be transformed into two coupled equations \citep{Yang2019} by introducing new field functions as
\begin{eqnarray}
\phi^{n+1}=\phi_{1}^{n+1}+z^{n+1}\phi_{2}^{n+1},
\label{NS24}
\end{eqnarray}
\begin{eqnarray}
G^{n+1}=G_{1}^{n+1}+z^{n+1}G_{2}^{n+1},
\label{NS25}
\end{eqnarray}
resulting in the following final set of equations. 

The first coupled system of equations for $\phi_{1}$ and $G_{1}$ is
\begin{eqnarray}
\nabla^{2} G_{1}^{n+1}-\frac{\gamma_{0}}{M\Delta t}\phi_{1}^{n+1}= \frac{-\hat{\phi}}{M\Delta t}+\frac{1}{M}\nabla\cdot({{\mathbf{u^{*,n+1}}}\phi^{*,n+1}}),
\label{NS26}
\end{eqnarray}
\begin{eqnarray}
\nabla^{2} \phi_{1}^{n+1}+\frac{1}{\lambda}G_{1}^{n+1}= \frac{1}{\lambda\gamma_{0}}(\hat{r}-\frac{1}{2}\int_{\Omega}\frac{h(\phi^{*,n+1})}{\sqrt{E(\phi^{*,n+1})}}\hat{\phi})\frac{h(\phi^{*,n+1})}{\sqrt{E(\phi^{*,n+1})}},
\label{NS27}
\end{eqnarray}
with the following boundary conditions
\begin{equation}
\begin{cases}
	\mathbf{n}\cdot\nabla{\phi_{1}}^{n+1}+\frac{\gamma_{0}\eta\mu_{f}}{\lambda\Delta t}\phi_{1}^{n+1}=\frac{\eta\mu_{f}}{\lambda\Delta t}\hat{\phi}-\\
    \hspace{3cm} \frac{1}{\lambda}f^{\prime}_{w}(\phi^{*,n+1})_{A,R}\qquad if \qquad L^{*,n+1}_{A}L^{*,n+1}_{R}>0,\\
    \phi_{1}^{n+1} = \frac{1}{\gamma_{0}}\hat{\phi} \qquad otherwise \qquad on \qquad \partial\Omega_{w}.
\end{cases}  
\label{NS28}
\end{equation}

\begin{eqnarray}
{\mathbf{n}}\cdot\nabla{G}_{1}^{n+1}=0 \qquad on \qquad \partial\Omega_{w},
\label{NS29}
\end{eqnarray}
and the second coupled system of equations for $\phi_{2}$ and $G_{2}$ is
\begin{eqnarray}
\nabla^{2} G_{2}^{n+1}-\frac{\gamma_{0}}{M\Delta t}\phi_{2}^{n+1}=0,
\label{NS30}
\end{eqnarray}
\begin{eqnarray}
\nabla^{2} \phi_{2}^{n+1}+\frac{1}{\lambda}G_{2}^{n+1}= \frac{1}{2\lambda}\frac{h(\phi^{*,n+1})}{\sqrt{E(\phi^{*,n+1})}},
\label{NS31}
\end{eqnarray}
with the following boundary conditions
\begin{equation}
\begin{cases}
	\mathbf{n}\cdot\nabla{\phi_{2}}^{n+1}+\frac{\gamma_{0}\eta\mu_{f}}{\lambda\Delta t}\phi_{2}^{n+1}=0\qquad if \qquad L^{*,n+1}_{A}L^{*,n+1}_{R}>0,\\
    \phi_{2}^{n+1} = 0 \qquad otherwise \qquad on \qquad \partial\Omega_{w},
\end{cases}  
\label{NS32}
\end{equation}

\begin{eqnarray}
{\mathbf{n}}\cdot\nabla{G}_{2}^{n+1}=0 \qquad on \qquad \partial\Omega_{w},
\label{NS33}
\end{eqnarray}
First, the coupled equations (\ref{NS26}-\ref{NS27}) with the boundary conditions (\ref{NS28}-\ref{NS29}) are solved, and then the coupled equations (\ref{NS30}-\ref{NS31}) with the boundary conditions (\ref{NS32}-\ref{NS33}) are solved. The resultant coupled equations are non-symmetric linear system so that GMRES iterative solver with geometric multigrid preconditioner with the Message-Passing Interface (MPI) for distributed-memory parallelization provided by PETSc library \citep{petsc2019} has been used to solve them. Then, $\phi^{n+1}$ and $G^{n+1}$ is computed by using Eq. (\ref{NS24}-\ref{NS25}). At the end, the relaxation technique Eq. (\ref{NS23}) is used to improve the accuracy and consistency of the SAV scheme. The interested readers are referred to \citep{Jiang2022} for the definition of $\zeta_{0}$.  
\begin{figure}[tbp]
	\centering
	\includegraphics[width=0.7\textwidth]{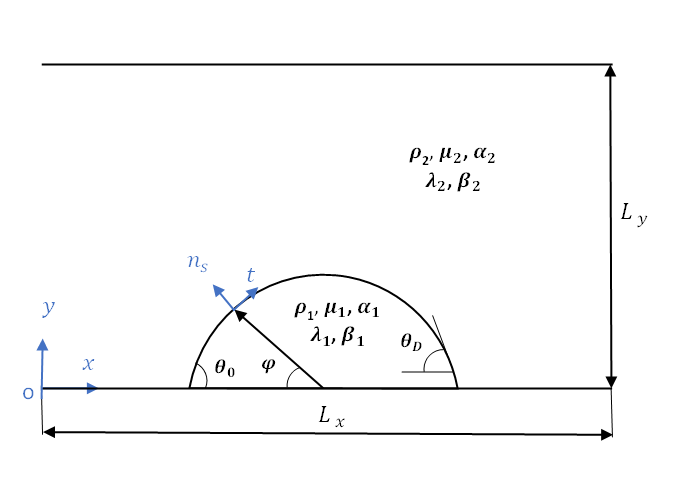}
	\caption{The Initial configuration for the droplet in the 2D with $\theta_{0}$. $\theta_{D}$ has been extracted at the height $y=2\hspace{0.05cm}Cn$ above the wall.}
	\label{fig:fig1}
\end{figure}
\begin{figure}[ht!]
    \centering
    \subfloat[]{\includegraphics[width=0.5\textwidth]{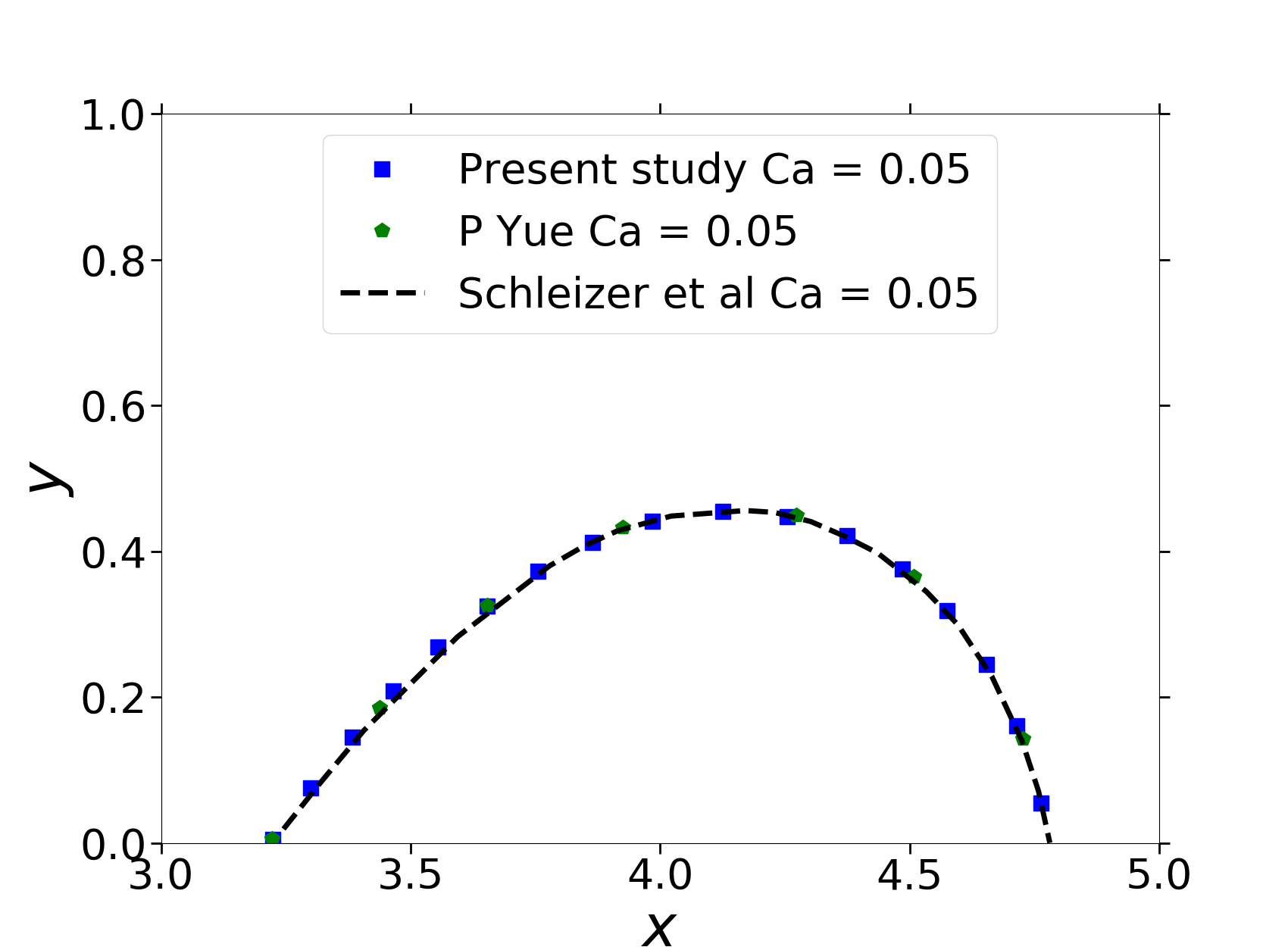}}
    \subfloat[]{\includegraphics[width=0.5\textwidth]{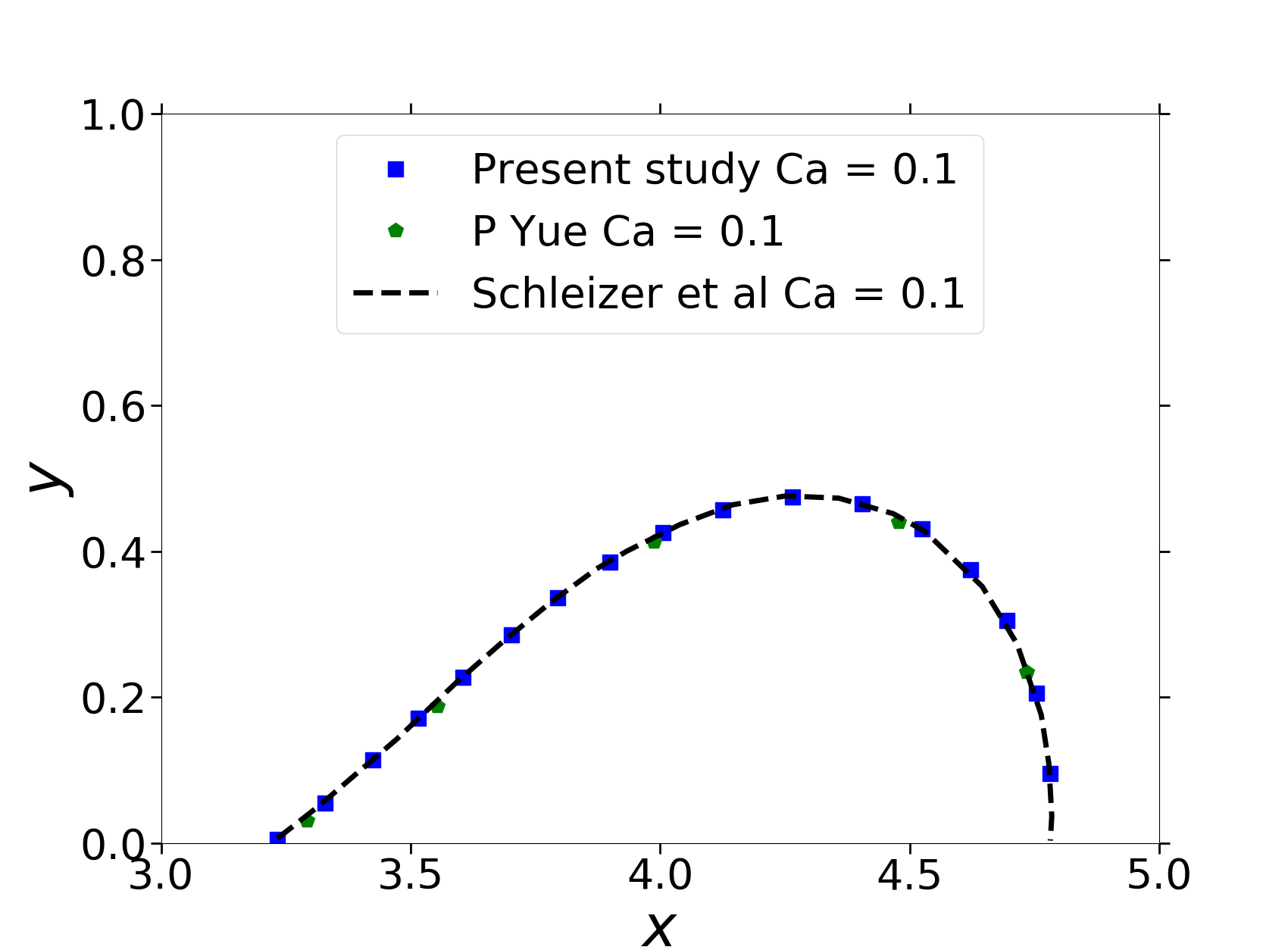}}\\
    \subfloat[]{\includegraphics[width=0.5\textwidth]{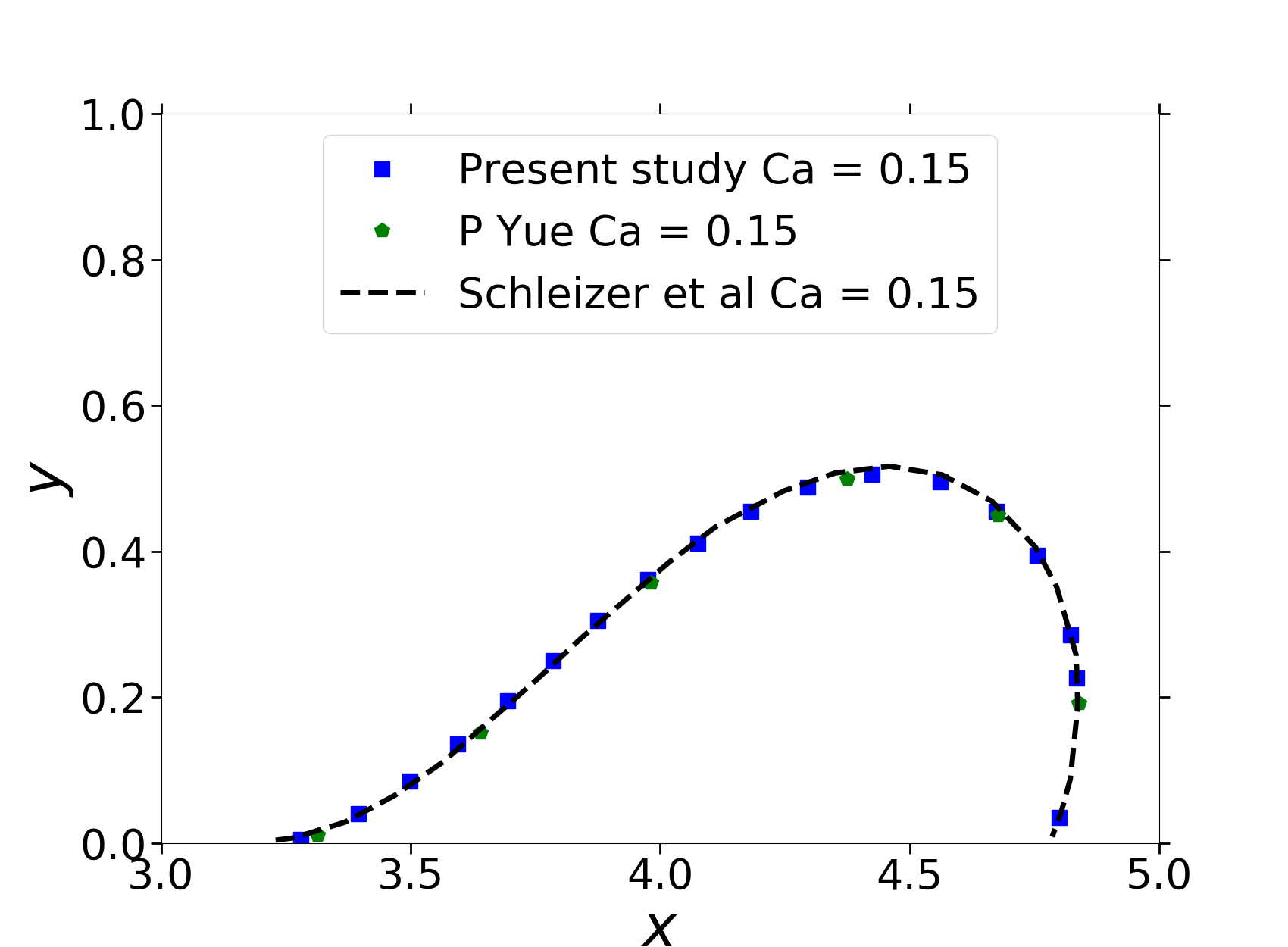}}
	\caption{The steady-state deformed droplet profile $\phi=0$ at $Re=0.01$, $\theta_{R}=10^{\circ}$, and $\theta_{A}=170^{\circ}$ (a) $Ca=0.05$ (b) $Ca=0.1$ (c) $Ca=0.15$. The simulation data are compared with the results in \citep{Schleizer1999} and \citep{Yue2020}}
	\label{fig:fig2}
\end{figure}
\section{Validation}\label{sec4}
\subsection{Droplet deformation with pinned contact lines}\label{sec4_1}
The implementation of the Cahn-Hilliard equation with CAH has been validated against a test case from the literature \citep{Schleizer1999,Yue2020}: a 2D Newtonian droplet in the Newtonian medium (N/N) with a large hysteresis window $[\theta_{R},\theta_{A}]$. The computational domain is a 2D rectangle,  $\Omega=[0,8H]\times[0,2H]$, with the droplet placed on the bottom wall, see Fig. \ref{fig:fig1}. The no-slip and no-penetration conditions are imposed on the both walls, and a fully developed parabolic velocity $u=6\overline{U}((\frac{y}{L_{y}})-(\frac{y}{L_{y}})^{2})$ is imposed at the inlet and outlet of the channel. The droplet initially is semicircle with contact angle $\theta_{0}=60^{\circ}$, radius $R=0.9023H$, and height $h=0.4511H$. This test case is defined by the length scale $L_{ref}=H$, which is the half channel width; $u_{ref}=(\frac{3\overline{U}}{H})h$, which is the velocity scale. This results in the following dimensionless numbers \citep{Schleizer1999,Dupont2010,Liu2015,Yue2020}. From the physical parameters of the problem, we obtain the Reynolds number, $Re=\frac{\rho_{1} u_{ref} H}{\mu_{1}}$, the ratio between inertial and viscous forces; the Capillary number, $Ca=\frac{\mu_{1} u_{ref}}{\sigma}$, the ratio between viscous and surface tension forces, density ratio, $\lambda_{\rho}=\frac{\rho_{2}}{\rho_{1}}$, the ratio between ambient density to droplet density, and the viscosity ratio, $\lambda_{\mu}=\frac{\mu_{2}}{\mu_{1}}$, which is the ratio between ambient viscosity to droplet viscosity. Furthermore, from the phase field model, we obtain the Peclet number $Pe = \frac{2\sqrt{2}}{3}\frac{u_{ref}H\eta}{M\sigma}$, the ratio between the advection and diffusion; the Cahn number, $Cn=\frac{\eta}{H}$, which represents the ratio between the interface width and the characteristic length scale; $\Lambda=\frac{\mu_{f}Cn}{\mu_{1}}$, which characterizes the wall energy relaxation rate. 

Following \citet{Yue2020}, we consider the following values of the relevant dimensionless numbers:  
\[Re=0.01, \quad \lambda_{\mu}=1, \quad \lambda_{\rho}=1, \quad Cn=0.01,\]
\[Pe=\frac{2\sqrt{2}}{3}\frac{Ca}{Cn}, \quad \frac{\mu_{f}}{\mu_{1}}=\frac{0.1}{Cn}, \quad \theta_{R}=10^{\circ}, \quad \theta_{A}=170^{\circ}.\]

The comparison of the droplet interface $\phi=0$ with the results by \citet{Schleizer1999,Yue2020} at different Capillary numbers when they reach steady state represents a good agreement except at $Ca=0.15$ at the receding contact angle, see Fig. \ref{fig:fig2}; the same agreement and error at $Ca=0.15$ is indeed reported by \citep{Yue2020}, and it is due to a very small receding contact angle.
\section{Results}\label{sec5}
We scrutinize the effect of the viscoelasticity on both deformation and depinning of a drop over a surface with different degrees of the hysteresis window, $\Delta\theta=140^{\circ}$ with the choice of $\theta_{R}=30^{\circ}$ and $\theta_{A}=170^{\circ}$, $\Delta\theta=130^{\circ}$ with the choice of $\theta_{R}=30^{\circ}$ and $\theta_{A}=160^{\circ}$, and $\Delta\theta=100^{\circ}$ with the choice of $\theta_{R}=40^{\circ}$ and $\theta_{A}=140^{\circ}$. These hysteresis windows are larger than encountered in physical experiments, and are chosen to prolong the pinning of the contact lines, following previous numerical works of \citep{Dimitrakopoulos1997,Schleizer1999,Dimitrakopoulos2007,Ding2010,Dupont2010}. A rectangular domain, $\Omega=[0,2H]\times[0,8H]$, is used for the V/N system with the droplet placed on the middle of the bottom wall, and a larger domain, $\Omega=[0,2H]\times[0,16H]$, for the N/V system in order to allow a fully developed polymeric stresses in the inlet and outlet boundaries. It should be noted that the velocity inside the droplet has been set to zero in the beginning of the simulation, but a parabolic velocity profile is imposed elsewhere. 

In addition to the Newtonian dimensionless numbers introduced in section \ref{sec4_1}, the following dimensionless numbers have been used to study the effect of the viscoelasticity : the dimensionless cross-sectional area $A=({4A_{d}}/{L^{2}_{y}})$ representing the ratio between the droplet area ($A_{d}$) and the square of channel half height, the Deborah number $De=(\lambda_{H}u_{ref}/H)$ representing the ratio between elastic forces to the viscous forces, and the polymeric viscosity ratio $\beta=(\mu_{s}/\mu)$ which is the ratio between the solvent viscosity and the total viscosity. The dimensionless cross-sectional area is kept constant at $A=0.5$ in the following sections (section \ref{sec5_1} and section \ref{sec5_2}), and the semi-circular droplet's initial radius $r_{0}=\sqrt{A_{d}H/(\theta_{0}-\sin{\theta_{0}}\cos{\theta_{0}})}$ and the droplet's initial height $h=r_{0}(1-\cos{\theta_{0}})$ are computed for a specific initial contact angle ($\theta_{0}=90^{\circ}$). Furthermore, the following dimensionless numbers are kept constant in all simulations: $Re=0.5$ (so that the inertial effects are negligible), $\lambda_{\mu}=1$, $\lambda_{\rho}=1$, and $\beta=0.1$. It should be noted that $\lambda_{\mu}$ is the ratio between total viscosities of the two fluids; Since polymeric viscosity increases with polymer concentration, the viscosity of the Newtonian fluid also needs to increase if the polymer concentration changes (for example by adding more viscous fluids).

Regarding numerical parameters related to the phase field model, we choose an affordable but small capillary length ($Cn=\eta/H=0.00375$), and the Peclet number is set to $Pe=(2\sqrt{2}/3)(Ca/Cn)$ according to the guidelines \citep{Yue2010,Yue2020} in order to reach the so-called sharp-interface limit; the contact line friction is set to $\mu_{f}=(\Lambda/Cn)\mu_{1}$ \citep{Carlson2011,Yue2020} with $\Lambda=0.1$. The contact line is pinned when $\frac{\partial{\phi}}{\partial{t}}|^{t^{n+1}}=0$, so the droplet's deformation and the contact line's pinning/depinning are independent of the choice of $\mu_{f}$ prior to depinning.

To analyse the results further, we have extracted different quantities in the simulation, e.g., the locations of the receding contact line $X_{cl,R}$, and the advancing contact line $X_{cl,A}$, and the values of the dynamic contact angle at the receding side $\theta_{D,R}$, and at the advancing side $\theta_{D,A}$. Here, the dynamic contact angle has been computed using the geometric expression $\cos{\theta_{D}}=(\mathbf{n}\cdot\nabla{\phi})/(|\nabla{\phi}|)$ at the interface ($\phi=0$) at the height $y=2\hspace{0.05cm}Cn$ above the wall \citep{Wang2015}. \green{The velocity of the contact line ($U_{cl}=\frac{\sigma U^{*}_{cl}}{\mu_{f}}$) is computed by using second-order central difference with extracted $X_{cl}$ in the simulation, and the average contact line velocity ($\overline{U}_{cl}$) is computed by taking average over a nominal interval $[0,t_{end}]$.}

In the following sections, we present results for non-Newtonian cases, where either the droplet is laden with polymers and therefore viscoelastic (V/N), or the doplet is Newtonian but the ambient fluid is viscoelastic (N/V). In both cases, polymers affect the droplet deformation and dynamics in two ways: the polymeric stresses affect the interfacial deformation, and they also induce changes in the flow field which can affect the droplet dynamics \citep{Ramaswamy1999,Yue2005b,Aggarwal2007}. The stress tensor excluding the term due to mixing energy can be written as \citep{Yue2005}
\begin{eqnarray}
\boldsymbol\tau=-p\boldsymbol{I}+\mu(\nabla{{\mathbf{u}}}+\nabla{{\mathbf{u}}^T})+\frac{(1\pm\phi)}{2}\boldsymbol\tau_{p}.
\label{NS34}
\end{eqnarray}
Components of the stress tensor normal (${\mathbf{n}}_{s}$) and tangential (${\mathbf{t}}$) to the interface ($\phi=0$), see Fig. \ref{fig:fig1}, for the N/N, V/N, and N/V systems are shown along the interface in order to perceive the dominant forces there.

Furthermore, our numerical experiments showed that the polymeric stresses inside the droplet were growing without bound for $De\ge5$ when using Oldroyd-B model ($\alpha=0$). It is known that results from Oldroyd-B model may not be accurate \green{in} extensionally dominated flows, and it will be shown later that there are such regions in these flows. Therefore, we employ the Giesekus model by setting $\alpha=0.05$ for the V/N and $\alpha=[0.05,0.1]$ for N/V systems in order to prevent this issue while producing sufficiently large polymeric stresses to be able to observe viscoelastic effects.
\begin{figure}[ht]
	\centering
    \subfloat[]{\includegraphics[width=0.45\textwidth]{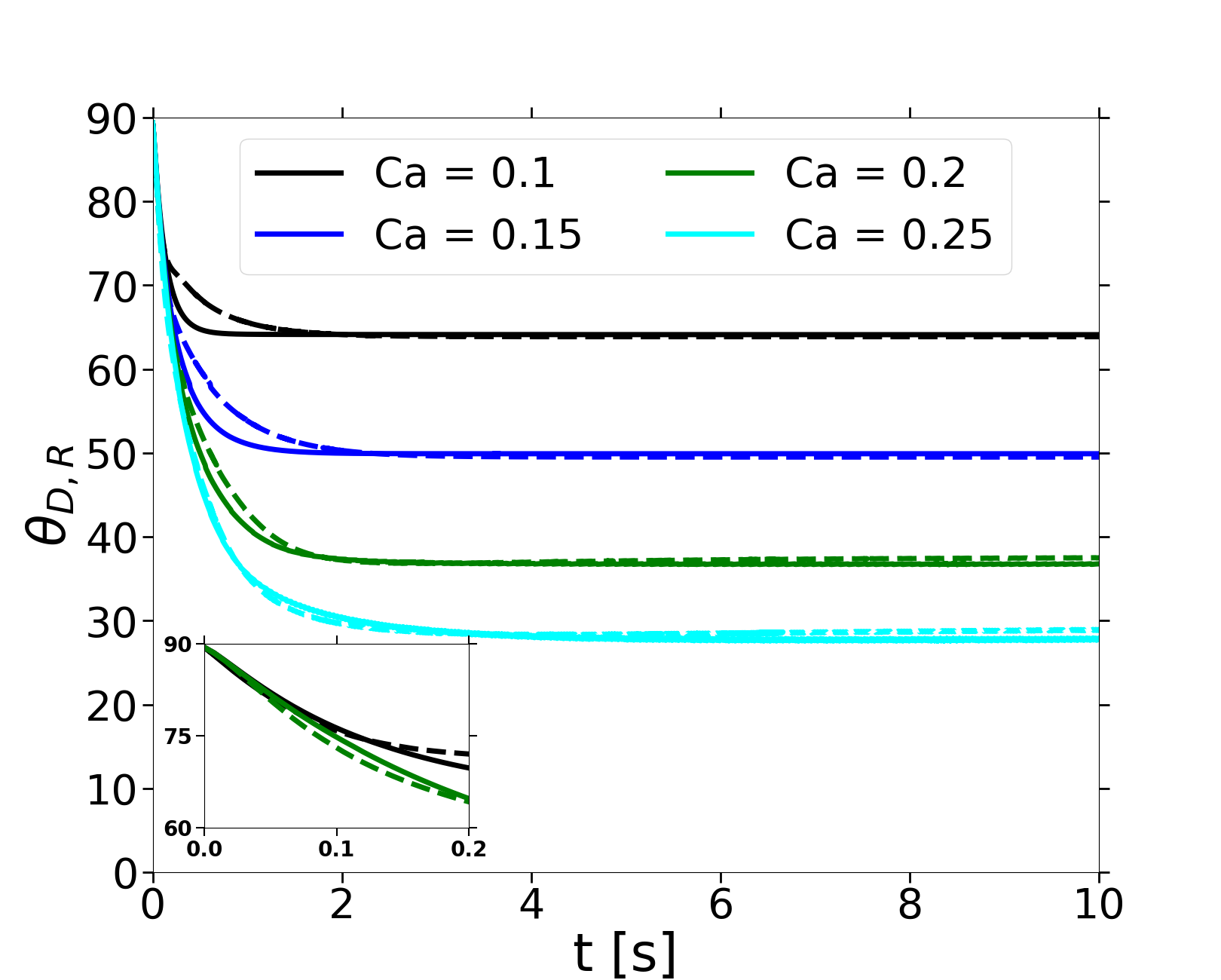}}
    \subfloat[]{\includegraphics[width=0.45\textwidth]{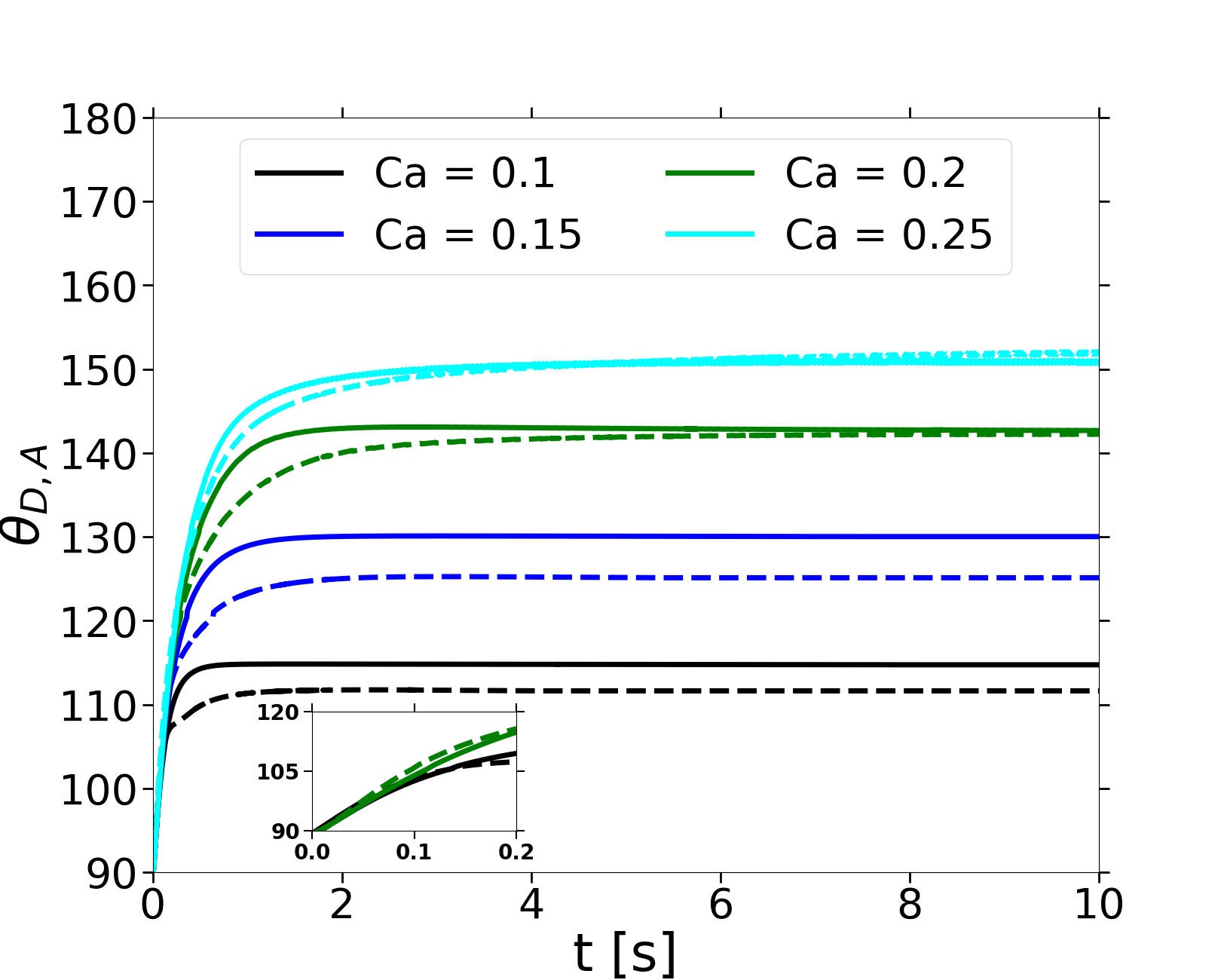}}\\
	\subfloat[]{\includegraphics[width=0.45\textwidth]{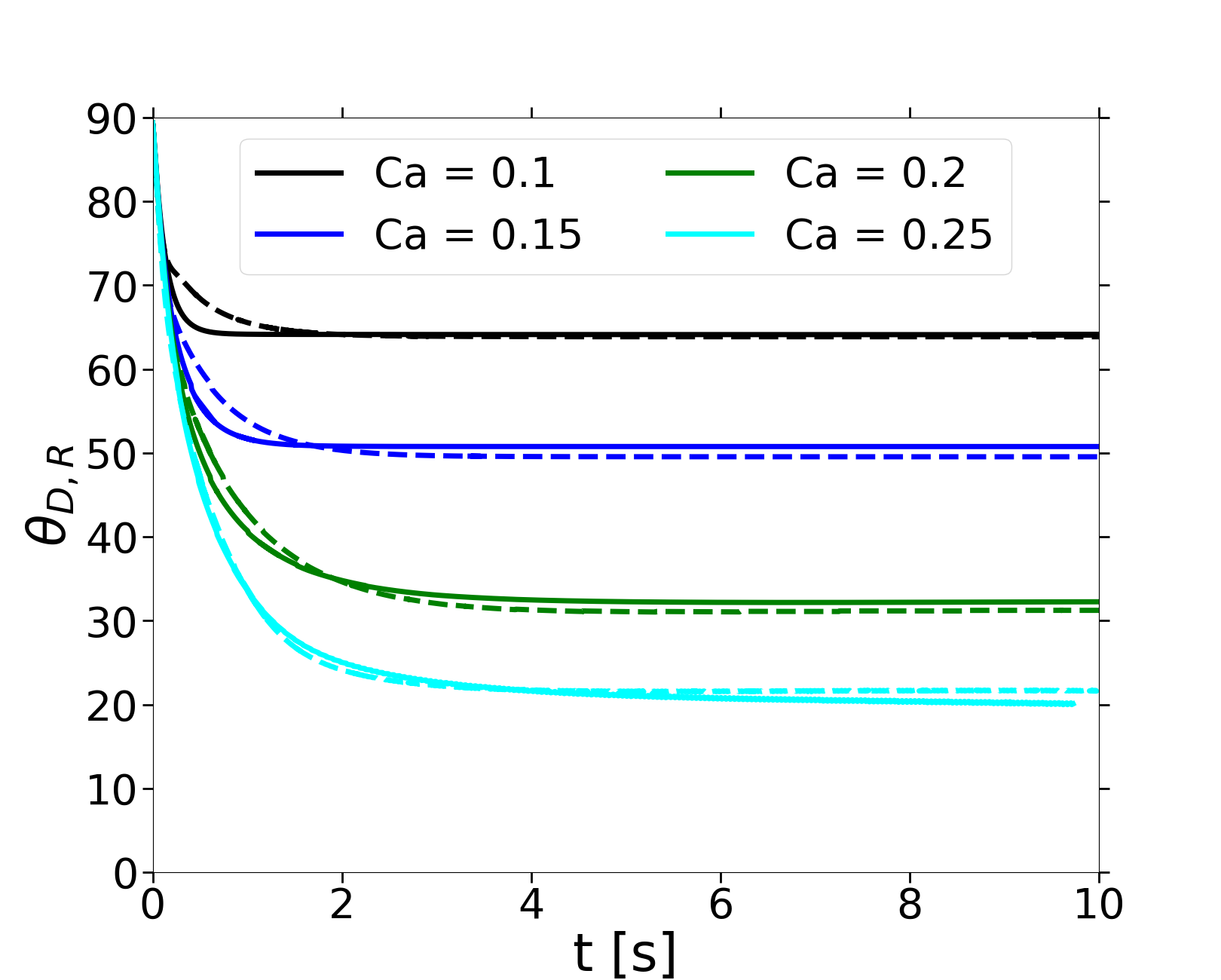}}
    \subfloat[]{\includegraphics[width=0.45\textwidth]{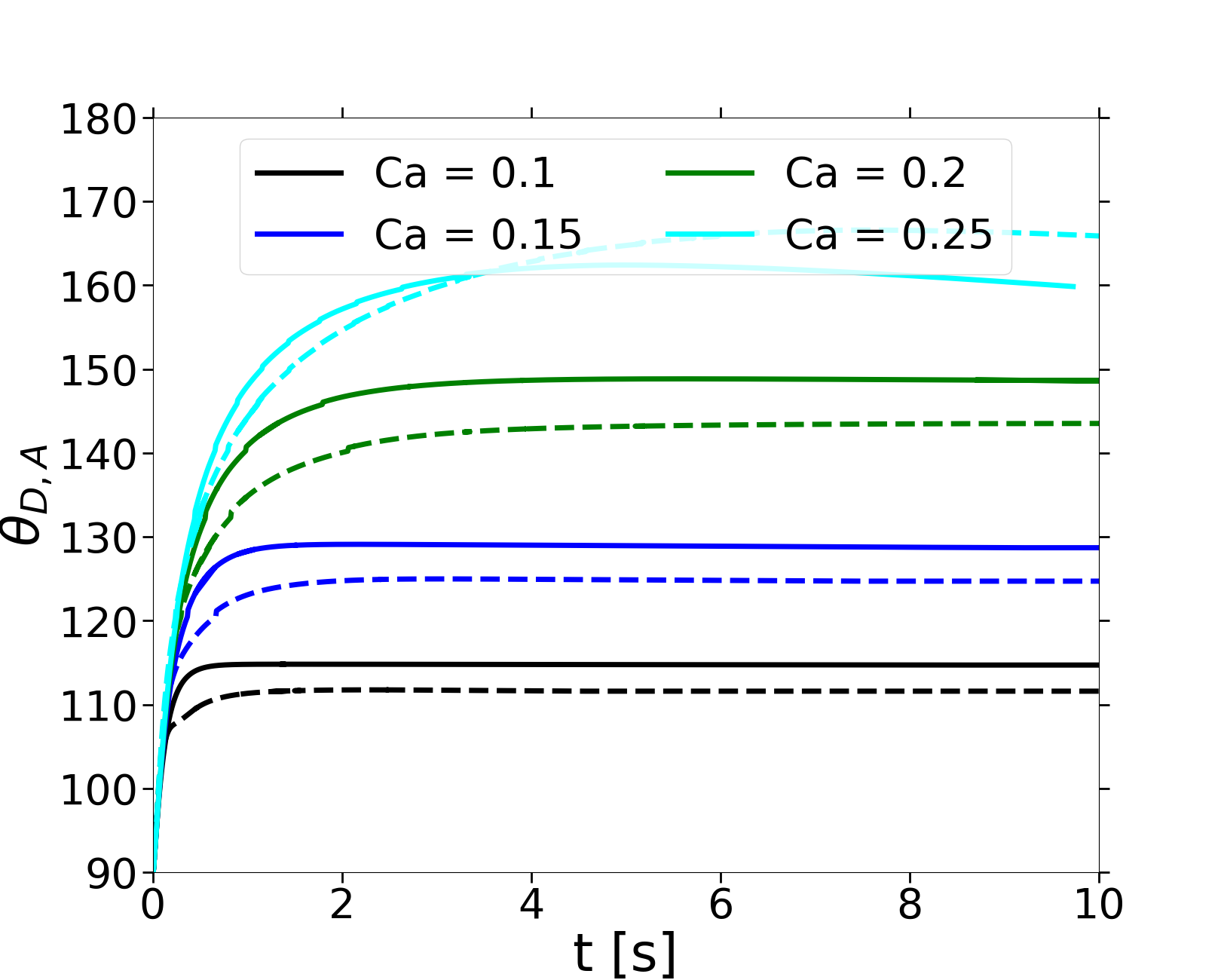}}\\
	\caption{The time evolution of $\theta_{D}$ for the Newtonian and Giesekus drops with $De=1$. N (\full) and VE (\dashed) refer to Newtonian and Giesekus drops respectively. (a) $\theta_{D,R}$ over the surface with $\Delta\theta=100^{\circ}$ (b) $\theta_{D,A}$ over the surface with $\Delta\theta=100^{\circ}$ (c) $\theta_{D,R}$ over the surface with $\Delta\theta=140^{\circ}$ (d) $\theta_{D,A}$ over the surface with $\Delta\theta=140^{\circ}$.}
	\label{fig:fig4}
\end{figure}
\begin{figure}[ht]
	\centering
    \subfloat[]{\includegraphics[width=0.36\textwidth]{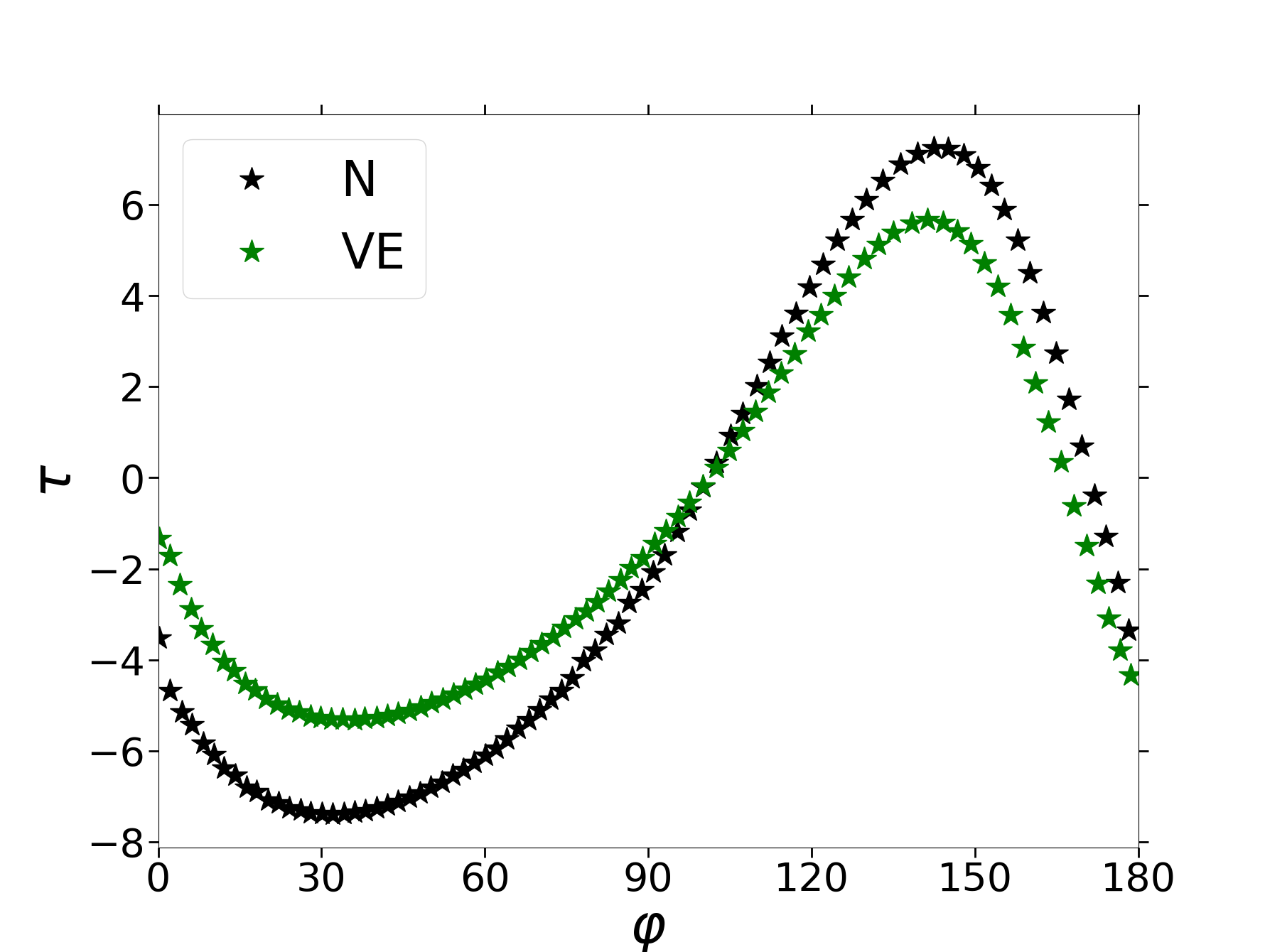}}
	\subfloat[]{\includegraphics[width=0.36\textwidth]{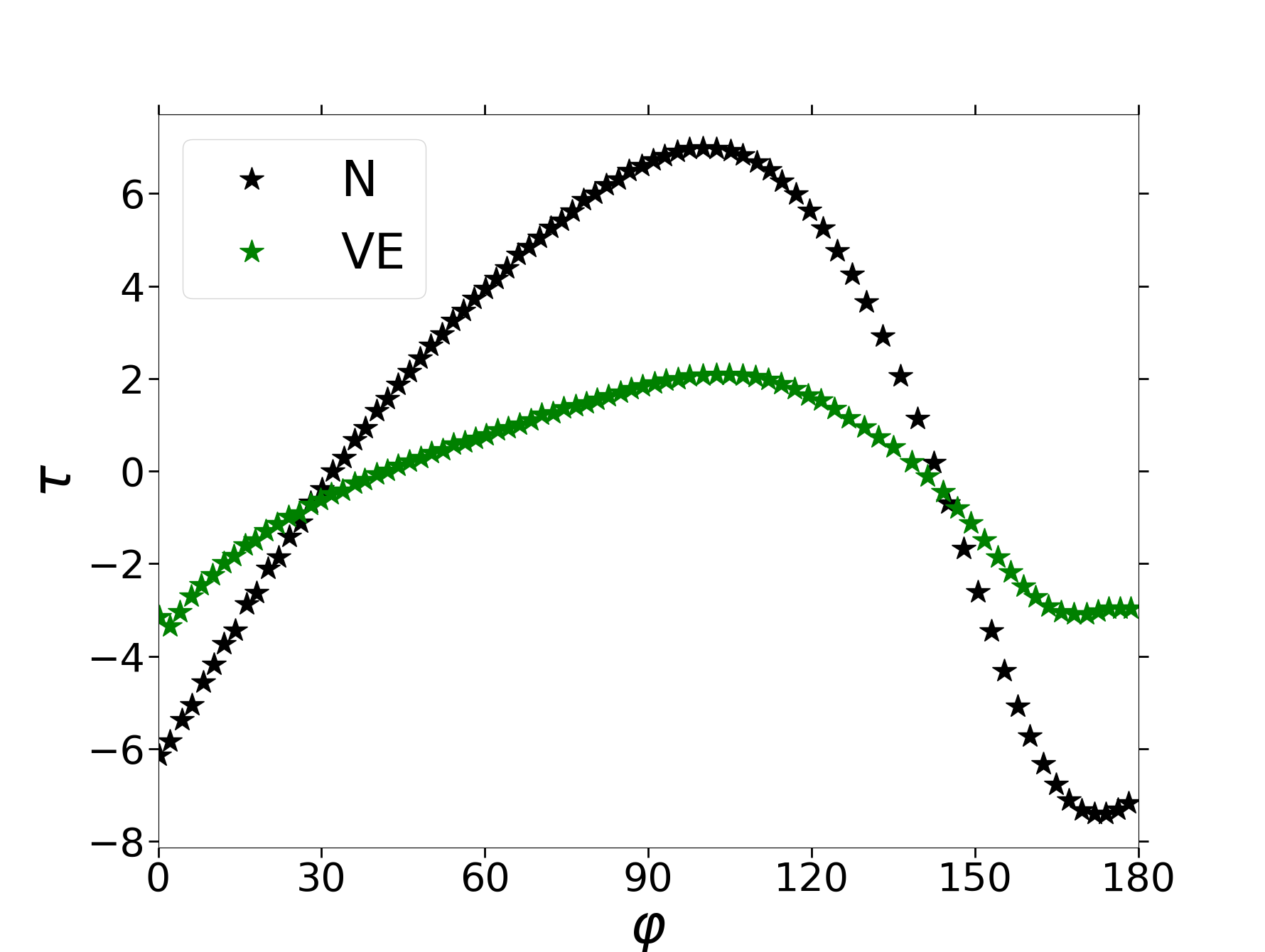}}\\
    \subfloat[]{\includegraphics[width=0.36\textwidth]{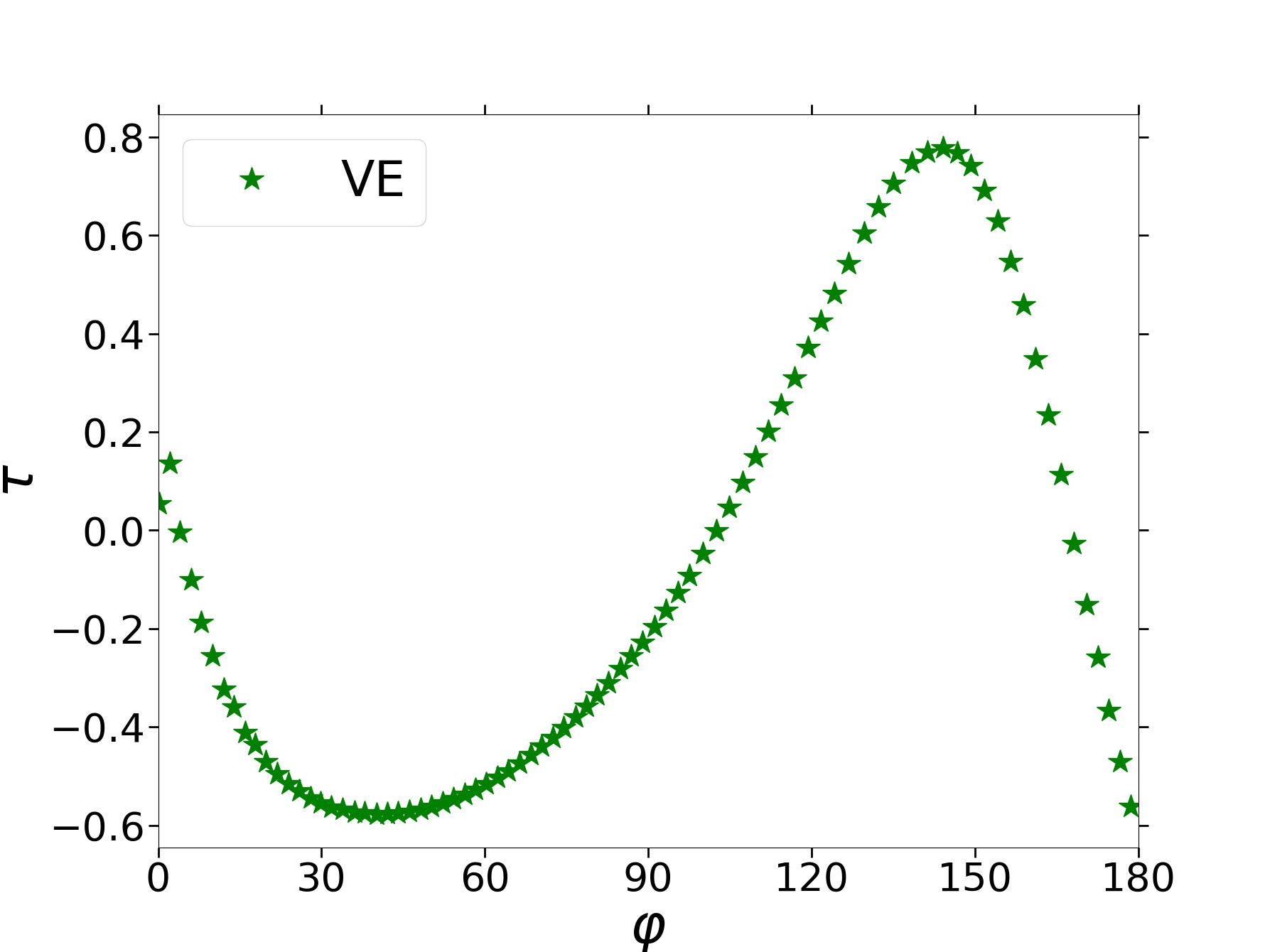}}
	\subfloat[]{\includegraphics[width=0.36\textwidth]{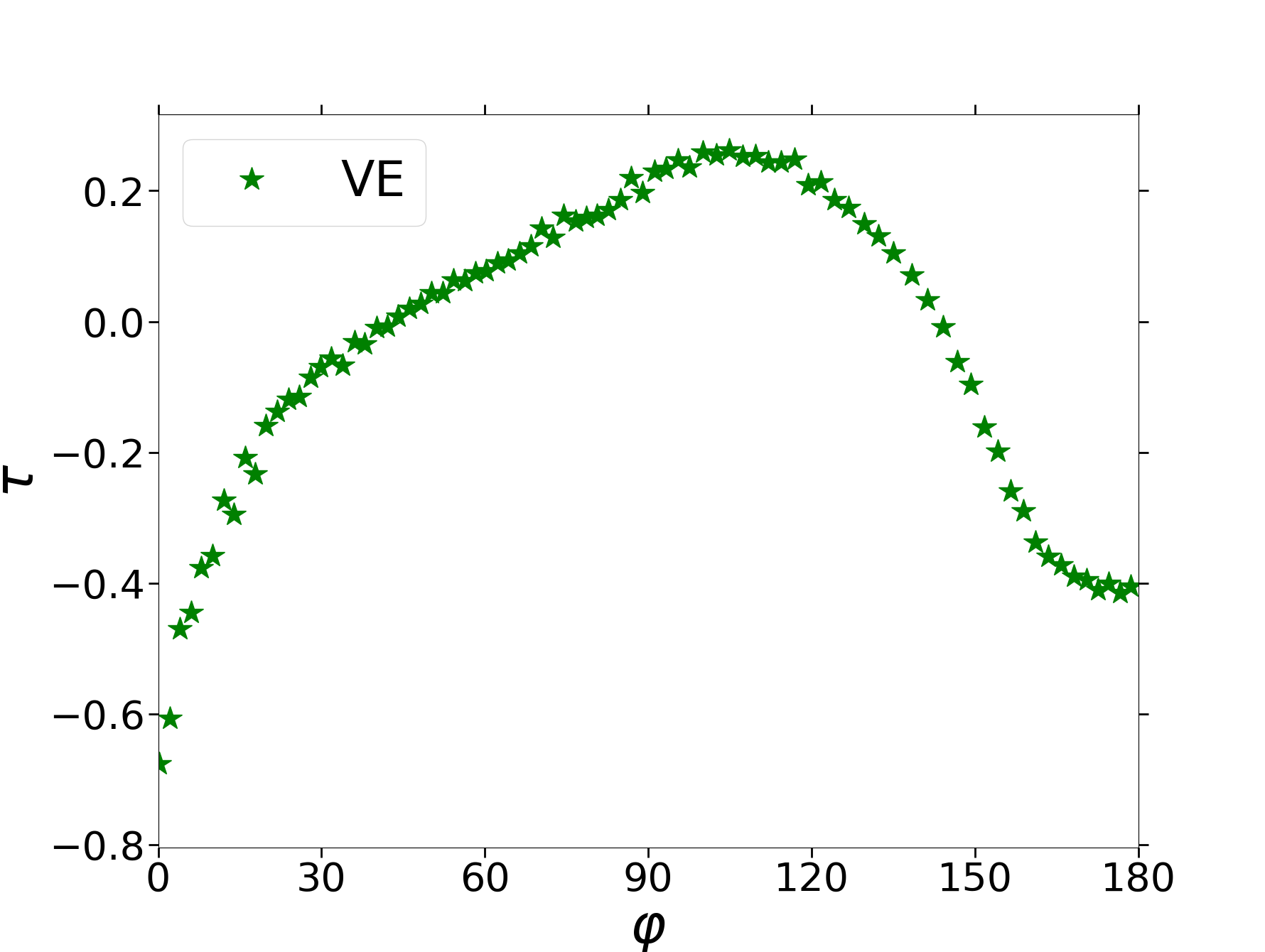}}\\
    \subfloat[]{\includegraphics[width=0.36\textwidth]{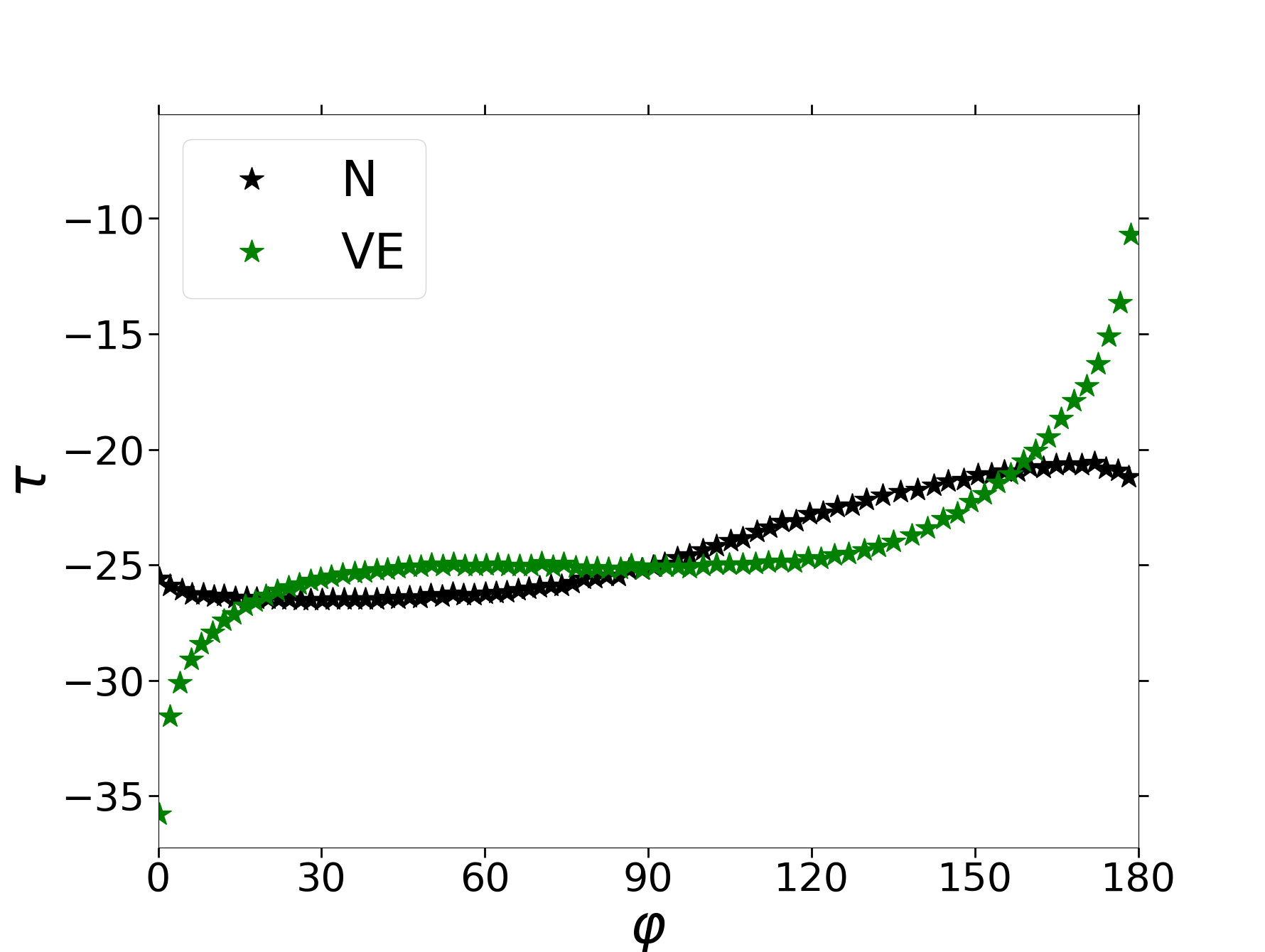}}
	\caption{Different components of the stresses along $\phi=0$ at time $t=0.1\hspace{0.05cm}$s for drops with $Ca=0.2$ and $De=1$ over the surface with $\Delta\theta=140^{\circ}$. (a) $\tau^{n}_{vi}$ (b) $\tau^{t}_{vi}$ (c) $\tau^{n}_{p}$ (d) $\tau^{t}_{p}$ (e) $\tau^{n}_{pre}$.}
	\label{fig:fig5}
\end{figure}

\begin{figure}[ht]
	\centering
    \subfloat[]{\includegraphics[width=0.45\textwidth]{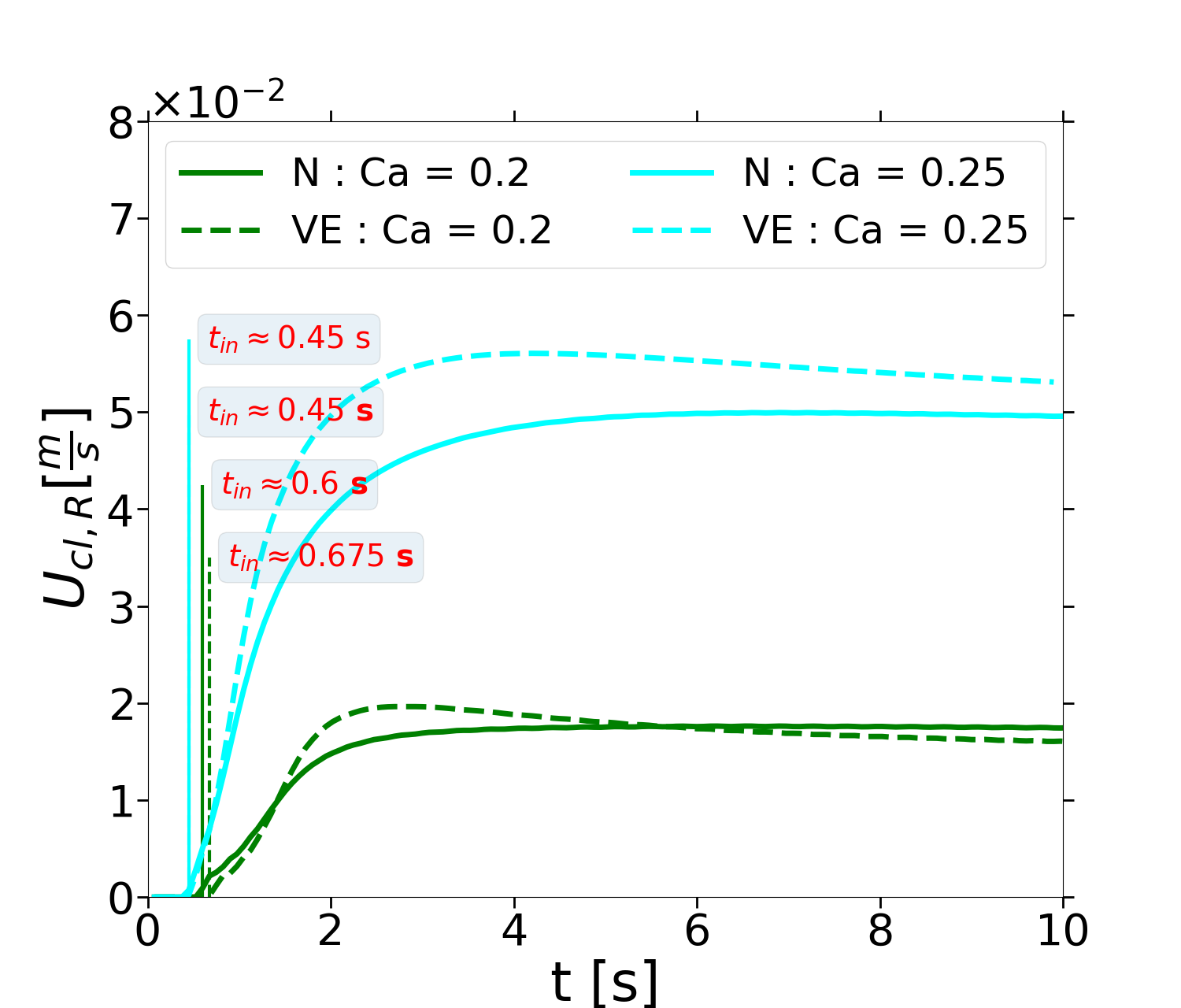}}
    \subfloat[]{\includegraphics[width=0.45\textwidth]{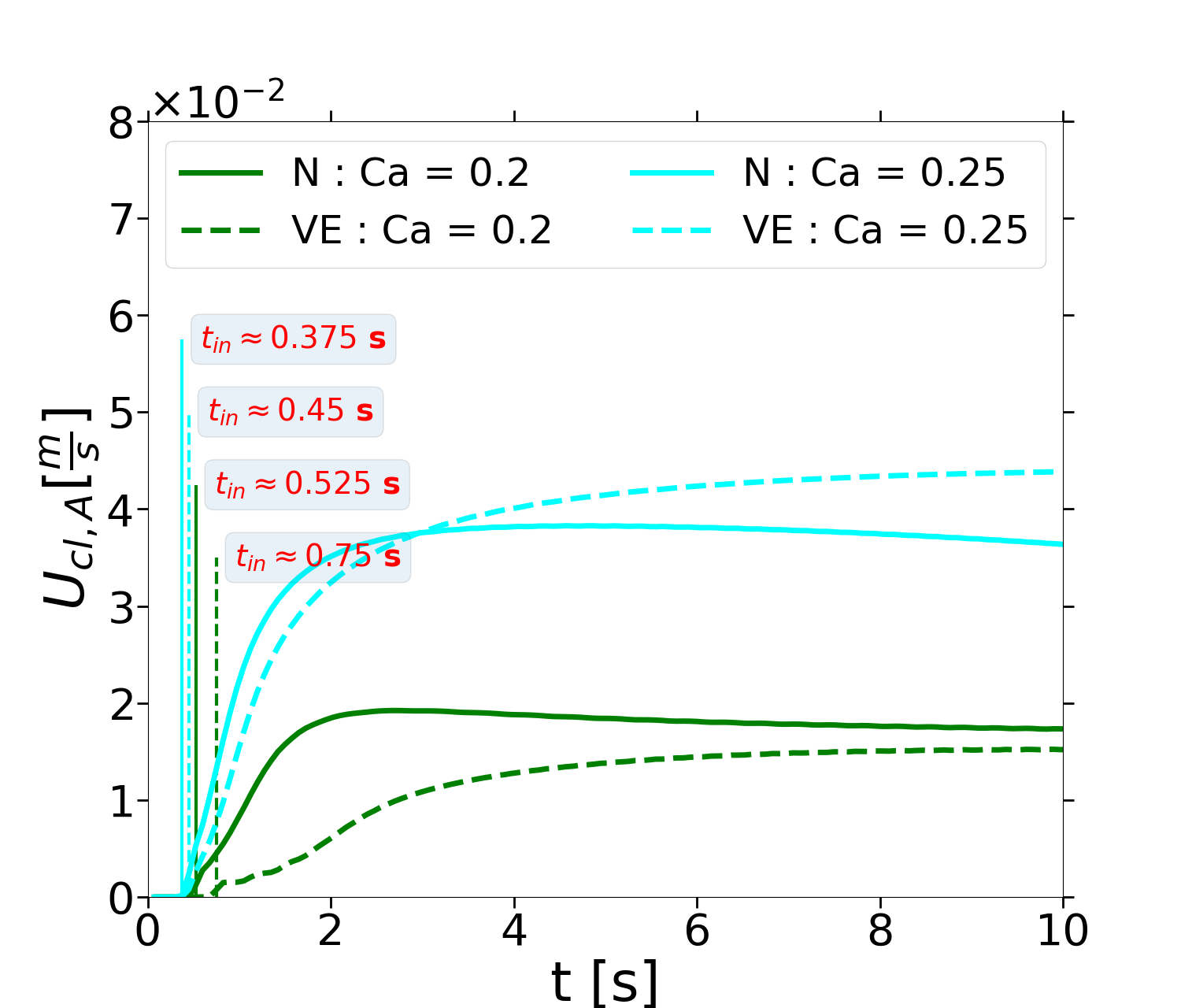}}\\
    \subfloat[]{\includegraphics[width=0.45\textwidth]{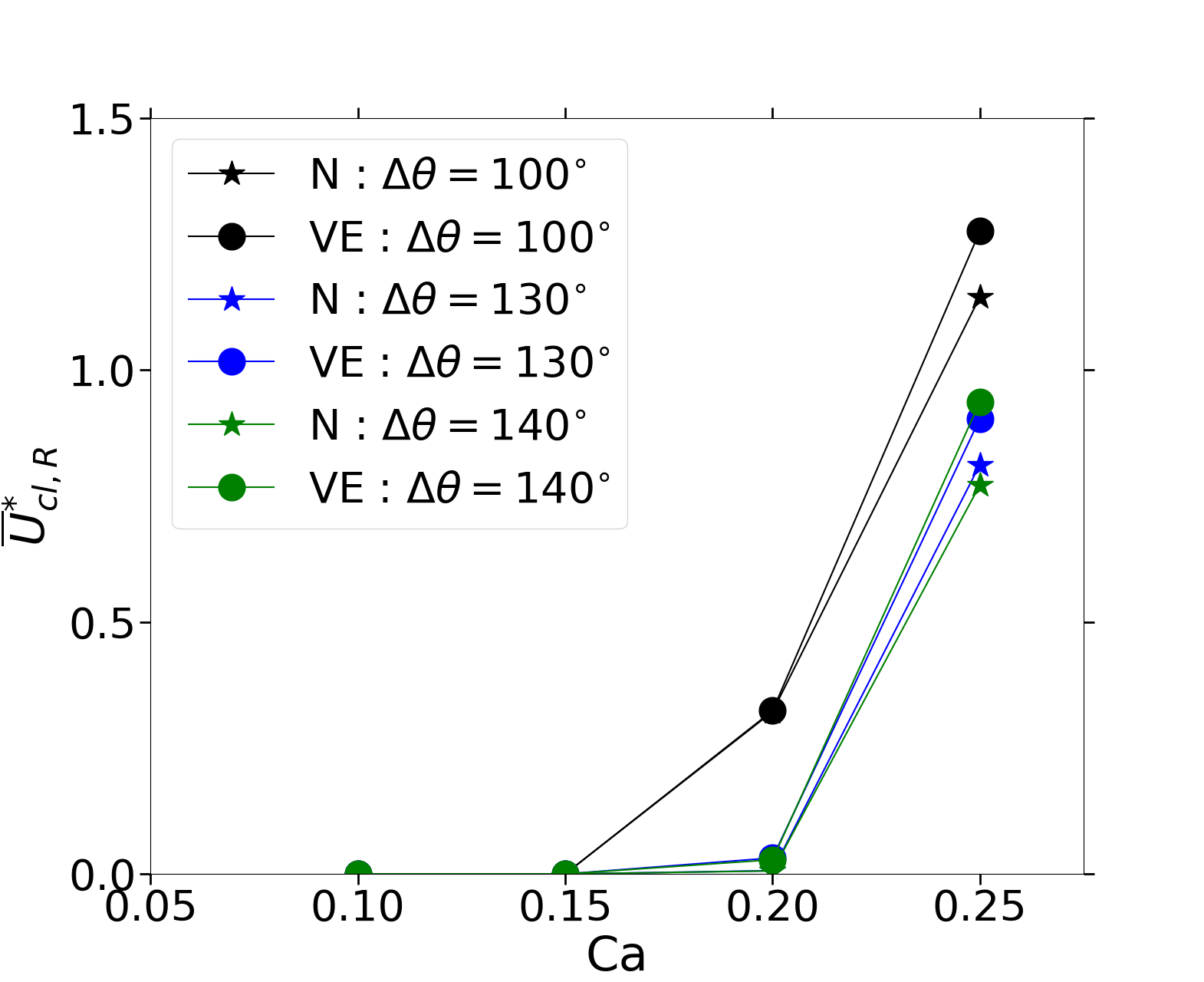}}
    \subfloat[]{\includegraphics[width=0.45\textwidth]{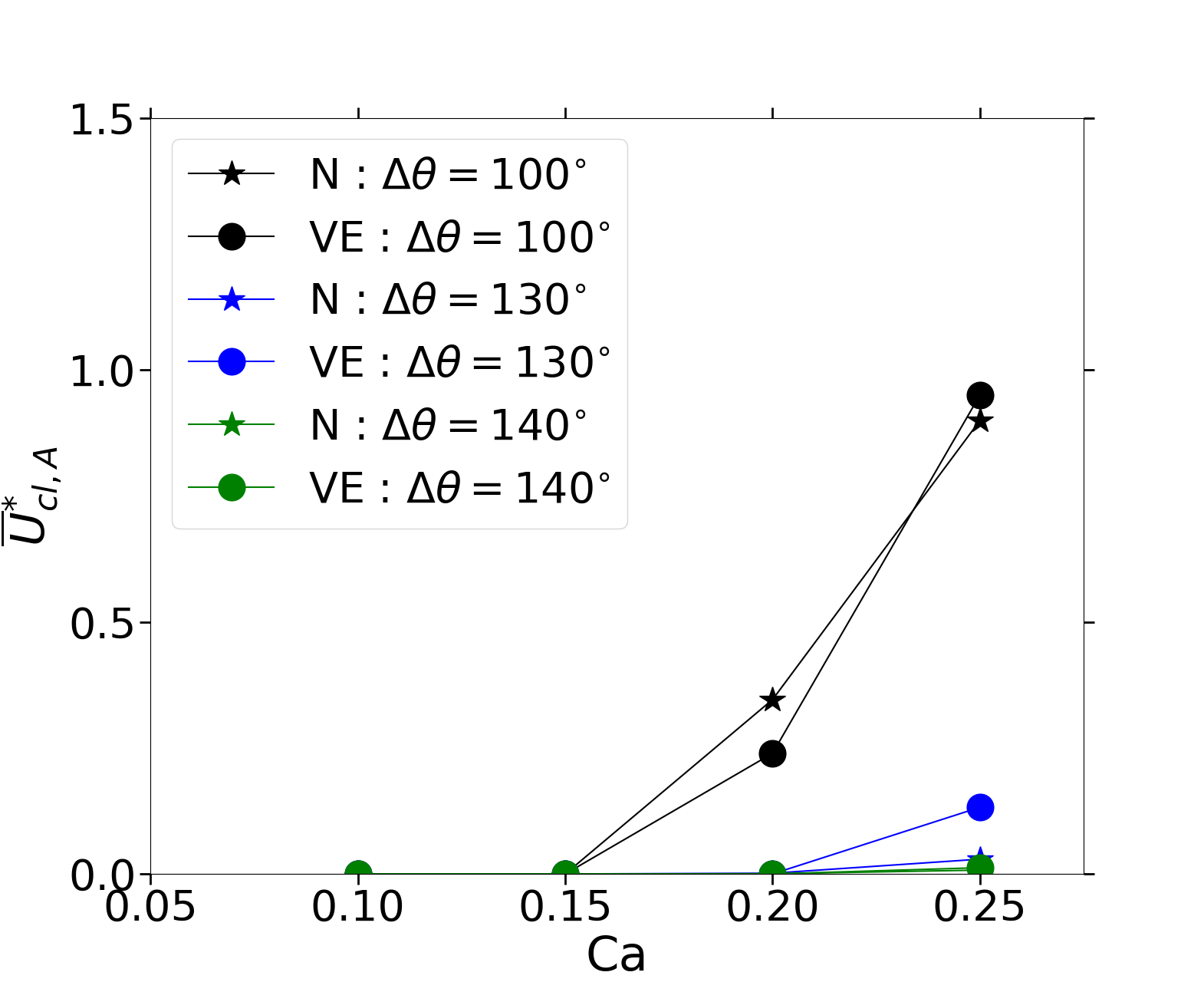}}
	\caption{\green{The time evolution of $U_{cl,R}$ ($X^{t=0}_{cl,R}=3.436$) and $U_{cl,A}$ ($X^{t=0}_{cl,A}=4.564$) for the Newtonian and Giesekus drops with $De=1$. N and VE refer to Newtonian and Giesekus drops respectively. (a) $U_{cl,R}$ for the surface with $\Delta\theta=100^{\circ}$ (b) $U_{cl,A}$ for the surface with $\Delta\theta=100^{\circ}$ (c) $\overline{U}^{*}_{cl,R}$ vs $Ca$ (d) $\overline{U}^{*}_{cl,A}$ vs $Ca$}.}
	\label{fig:fig6}
\end{figure}

\subsection{Elastic effects in the V/N system}\label{sec5_1}
In this section, we investigate the case where a viscoelastic droplet is surrounded by a Newtonian fluid (V/N system), while the capillary number, contact line hysteresis window and Deborah number are varied.  
\subsubsection{Effect of the capillary number}\label{sec5_1_1}
In this section, we investigate the effect of the capillary number on the deformation of the viscoelastic droplet, at a fixed Deborah number ($De=1$) and at different values of receding and advancing contact angles ($\theta_{R}$ and $\theta_{A}$). Figure \ref{fig:fig4} presents the time evolution of the instantaneous contact angle $\theta_{D}$ at both the receding and advancing sides over two different surfaces with $\Delta\theta=100^{\circ}$ and $\Delta\theta=140^{\circ}$, and we observe that viscoelasticity of the droplet affects them differently. Figures \ref{fig:fig4}(a) and \ref{fig:fig4}(c) show that the elasticity of the droplet causes $\theta_{D,R}$ to attain a smaller steady-state value in comparison to the Newtonian counterpart \green{for small $Ca$ when the contact line is pinned, and $\theta_{D,R}$ gains negligibly larger value when contact line is moving for ($De\le1$), see Fig. \ref{fig:fig6}(a)}. The advancing contact angle $\theta_{D,A}$ reaches a smaller equilibrium value for $Ca\le0.2$ and a larger value for $Ca>0.2$, see Figs. \ref{fig:fig4}(b) and  \ref{fig:fig4}(c); \green{It can be observed that $\theta^{VE}_{D,A}$ is approaching $\theta^{N}_{D,A}$ and surpassing it as the advancing contact line moves with considerable velocity which depends on both $Ca$ and $\Delta\theta$, see Fig. \ref{fig:fig6}(b).} It is worth mentioning that for larger values of $De$, the behaviour of $\theta_{D,R}$ is different, as will be discussed in section \ref{sec5_1_2}. 

The transient behavior of the $\theta_{D}$ reveals that the difference between dynamic contact angles ($\theta^{VE}_{D}-\theta^{N}_{D}$) for a specific $Ca$ is not monotonic in time, and this trend is more pronounced for the ($\theta_{D,R}$) in the early times, see Figs. \ref{fig:fig4}(a) and \ref{fig:fig4}(c) for $t\le2\hspace{0.05cm}$s. The reason is that the Newtonian and viscoelastic droplets deform at different rates. The viscoelastic drop speeds up and deforms faster than the Newtonian counterpart at early times, and this behavior has been observed in all of the cases in our simulation regardless of the $Ca$ and $\Delta\theta$. Since the steady contact angle value is lower for viscoelastic droplets, this causes the contact angle curves in fig. \ref{fig:fig4} to intersect.

To find the reason for the faster deformation of viscoelastic droplets at early times, we show different components of the stresses in Eq. (\ref{NS34}) normal and tangential to the interface ($\phi=0$) of the Newtonian and viscoelastic droplets at $Ca=0.2$ over the surface with $\Delta\theta=140^{\circ}$ in figure \ref{fig:fig5}. We show the instantaneous stresses along the interface at $t=0.1\hspace{0.05cm}$s, where the angle $\varphi$ along the interface is measured clockwise from the negative $x$-direction. Let us first consider the action of viscous stress in the Newtonian case (black markers in figure Fig. \ref{fig:fig5} (a)). The normal component of the viscous stress $\tau^{n}_{vi}$ at $t=0.1\hspace{0.05cm}$s pushes the droplet inward at the back side for $\varphi\lesssim100^{\circ}$ and outward in the front side $\varphi\approx[100^{\circ},172^{\circ}]$, both of which are in favor of the droplet deformation. Near the advancing contact line however, the viscous normal stress pushes the droplet inward at $\varphi\gtrsim172^{\circ}$, which resists the increase of $\theta_{D,A}$ near the wall. Let us now consider the viscoelastic case, green markers in figure \ref{fig:fig5} a. Viscous normal stress has the same effects as in the Newtonian case, but weaker. Also, the tangential component of viscous stress  (Fig. \ref{fig:fig5} (b)) which is related to the acceleration of the fluid elements near the interface is weaker in the viscoelastic flow. Polymeric stresses act similarly to viscous stresses, but have a significantly smaller amplitude (Figs. \ref{fig:fig5} (c,d)).

Thus, neither the viscous stress nor the polymeric stress is influential in the faster deformation of the viscoelastic droplet at the early times. The reason is instead found from the pressure distribution at the interface (Fig. \ref{fig:fig5} (e)). The pressure pulls the droplet inward for all values of the $\varphi$, and its magnitude is larger in comparison to other components due to the presence of the surface tension in the interface (Young–Laplace equation). The magnitude of the pressure is larger for $\varphi\lesssim20^{\circ}$ and smaller for $\varphi\gtrsim158^{\circ}$ for the viscoelastic droplet in comparison to the Newtonian counterpart; thus, the pressure is responsible for the faster decrease of the dynamic contact angle in the receding side, and faster increase on the advancing side, both of them increasing the deformation of the viscoelastic droplet at $t=0.1\hspace{0.05cm}$s. The change in the pressure is the consequence of the modification of the overall flow field by polymeric stresses.

Next, we will look at the effect of contact line hysteresis window $\Delta\theta=\theta_A-\theta_R$ on pinning and velocity of the contact lines. Firstly, we observe that as hysteresis increases, both Newtonian and viscoelastic droplet are pinned at higher capillary numbers. \green{Secondly, for droplets that are not pinned, the steady state contact line speed is higher for viscoelastic droplets when $Ca$ is larger ($Ca\ge0.25$)}. Figs. \ref{fig:fig6}(a,b) display the time evolution of the \green{contact line velocity} for different $Ca$ over surfaces with \green{$\Delta\theta=100^{\circ}$, and Figs. \ref{fig:fig6}(c,d) present the variation of averaged dimensionless contact line velocity ($\overline{U}^{*}_{cl}$) vs. $Ca$ over surfaces with different $\Delta\theta$. To characterise the depinning process, the time for the inception of motion ($U_{cl}(t_{in})\approx10^{-4}\hspace{0.05cm}\frac{m}{s}$) in both contact lines have been marked on their corresponding velocity curve for the hysteresis window $\Delta\theta=100^{\circ}$, Figs. \ref{fig:fig6}(a,b). Both of the contact lines are pinned when $Ca<0.2$. When $Ca\ge0.2$, the advancing contact line starts to move first (when the $\theta_{D,A}|_{y=0}\le\theta_{A}$ at $t\le t_{in}$) for the Newtonian drop; the receding contact line of the Giesekus drop moves first in this cases. At this moderate $De$, the both advancing and receding contact lines of the viscoelastic droplet move faster over all surfaces (different $\Delta\theta$) with $Ca=0.25$, but the advancing contact line of the Giesekus drop moves faster when $Ca=0.25$ and slower when $Ca=0.2$ in comparison with its Newtonian counterpart, see Figs. \ref{fig:fig6}(c,d).} 

\green{The contact line speed can be approximated by the difference between the dynamic contact angle $\theta$ and the equilibrium contact angle $\theta_e$ ($\theta_{A}$ and $\theta_{R}$ in the present of CAH) for minor variation of $\theta$ from $\theta_e$ \citep{Yue2011,Amberg2022,Yada2023}: $\frac{\mu_f}{\sigma} U_{cl} \propto \theta-\theta_e$}, where the contact line friction $\mu_f$ is a constant determined by surface and fluid properties. Hence, the contact line speed is enhanced if the droplet experiences more bending causing the dynamic contact angle to differ from the equilibrium contact angle. At the receding contact line, more bending implies a decrease in $\theta_{D,R}$.The elasticity ($De=1$) of the droplet enhances the viscous bending at the receding contact line ($\theta^{VE}_{D,R}\le\theta^{N}_{D,R}$) after its depinning over a large span of time, see Figs. \ref{fig:fig4}(a,c), for the simulations in this section and therefore $\theta^{VE}_{D,R}\le\theta^{N}_{D,R}$ implies that $U^{Ve}_{cl}\ge U^{N}_{cl}$. However, the viscous bending at the advancing contact line is initially weakened by the elasticity of the droplet ($\theta^{VE}_{D,A}\le\theta^{N}_{D,A}$). Thus, the elasticity of the droplet causes $U^{Ve}_{cl}\leq U^{N}_{cl}$ when $\theta^{VE}_{D,A}<\theta^{N}_{D,A}$ and $U^{Ve}_{cl}\geq U^{N}_{cl}$ when $\theta^{VE}_{D,A}>\theta^{N}_{D,A}$.

In this subsection, we have studied the effects of moderate viscoelasticity on the deformation of the droplet and movement of both contact lines by comparing a Newtonian droplet and a viscoelastic droplet with $De=1$.  Some of the conclusions will change when viscoelasticity is further increased in the next subsection. 

\begin{figure}[htp]
	\centering
    \subfloat[]{\includegraphics[width=0.45\textwidth]{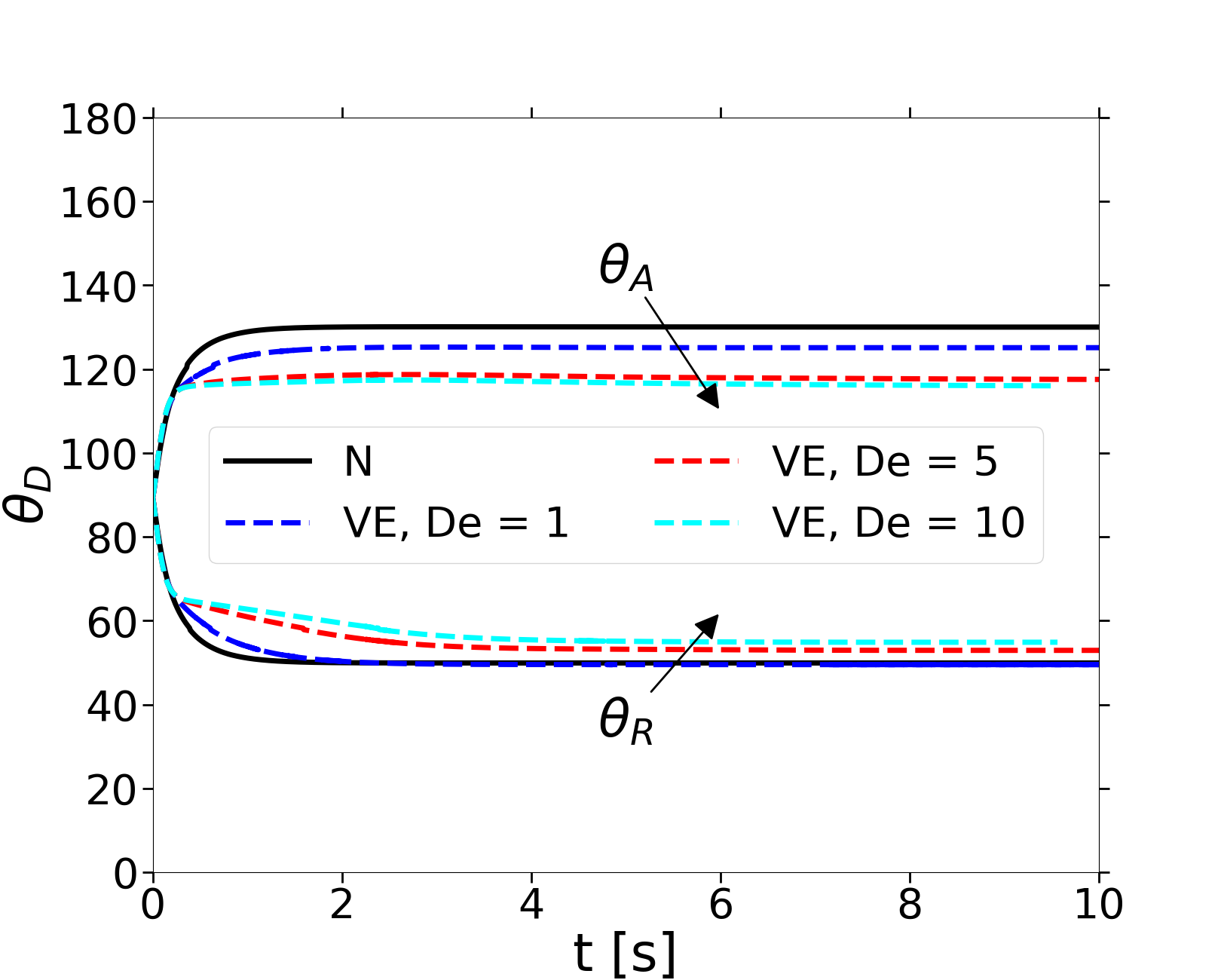}}
    \subfloat[]{\includegraphics[width=0.45\textwidth]{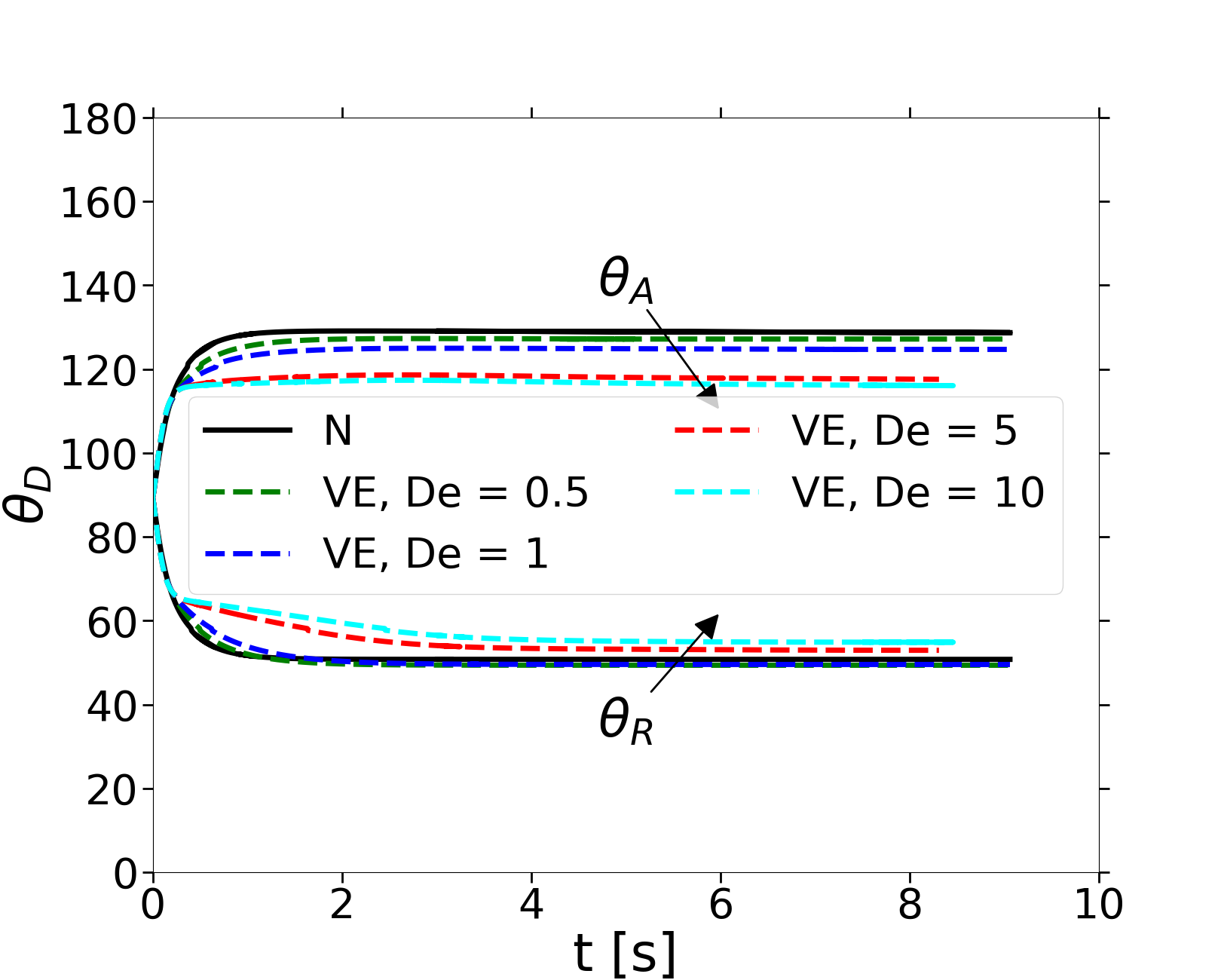}}\\
	\subfloat[]{\includegraphics[width=0.45\textwidth]{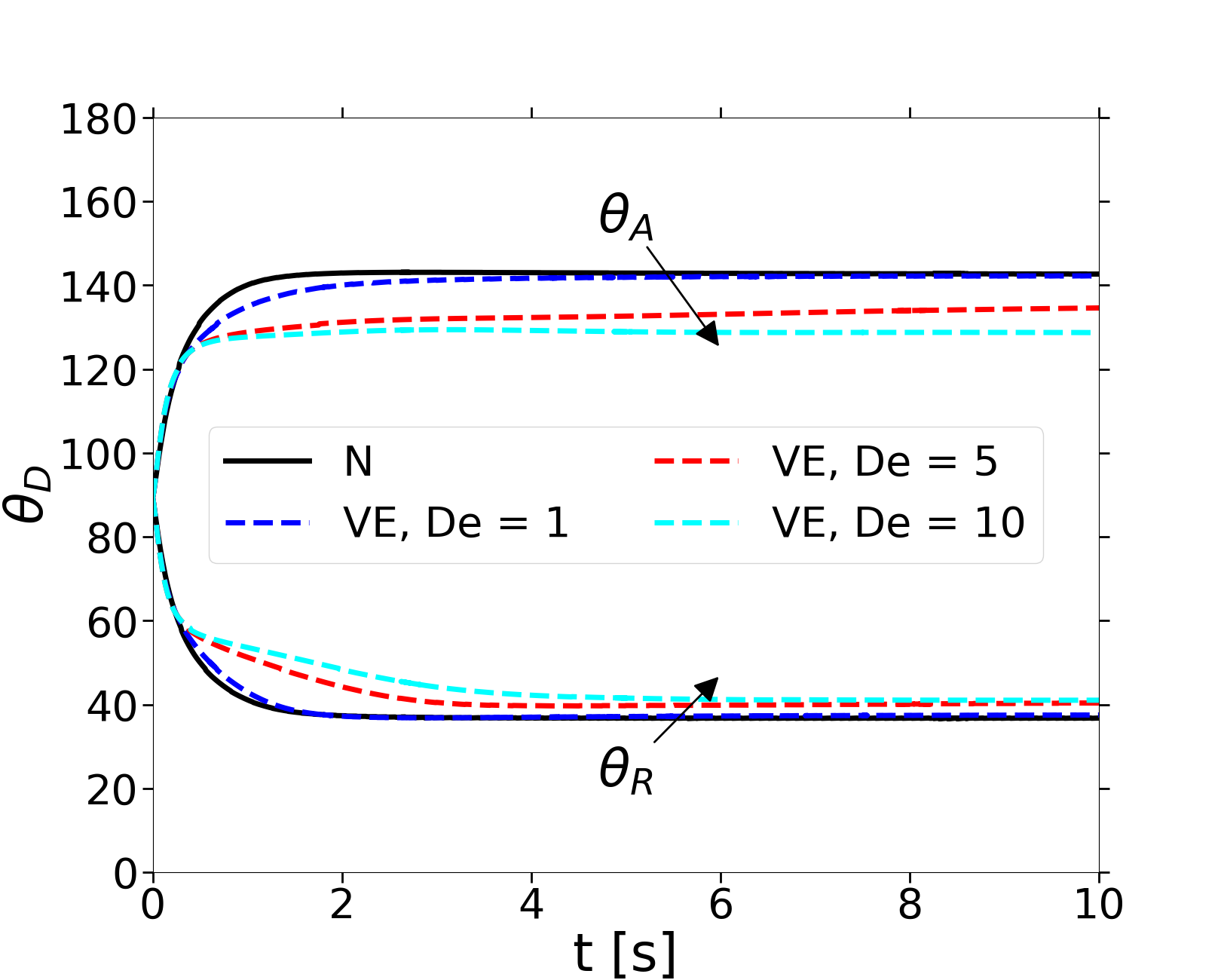}}
    \subfloat[]{\includegraphics[width=0.45\textwidth]{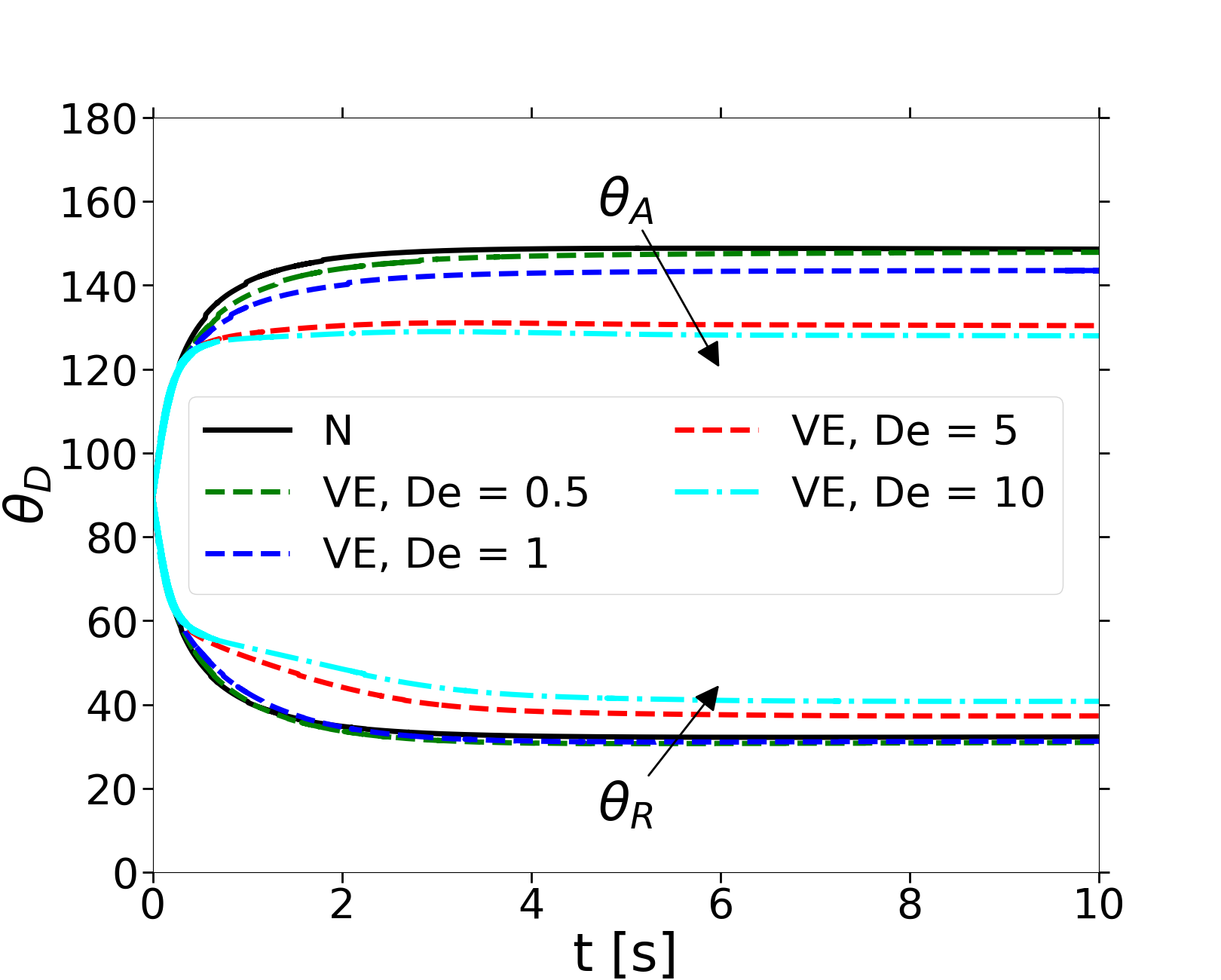}}\\
    \subfloat[]{\includegraphics[width=0.45\textwidth]{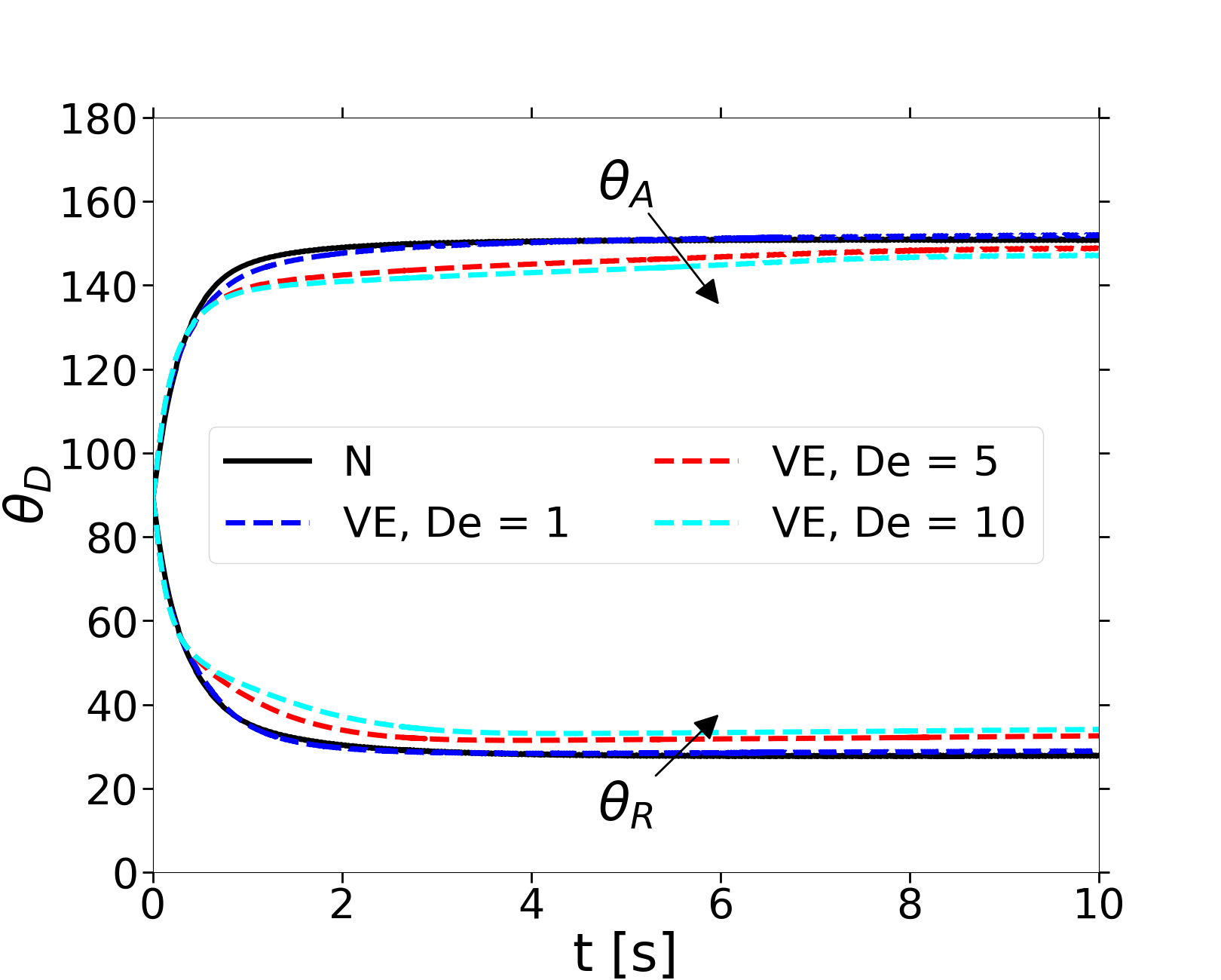}}
    \subfloat[]{\includegraphics[width=0.45\textwidth]{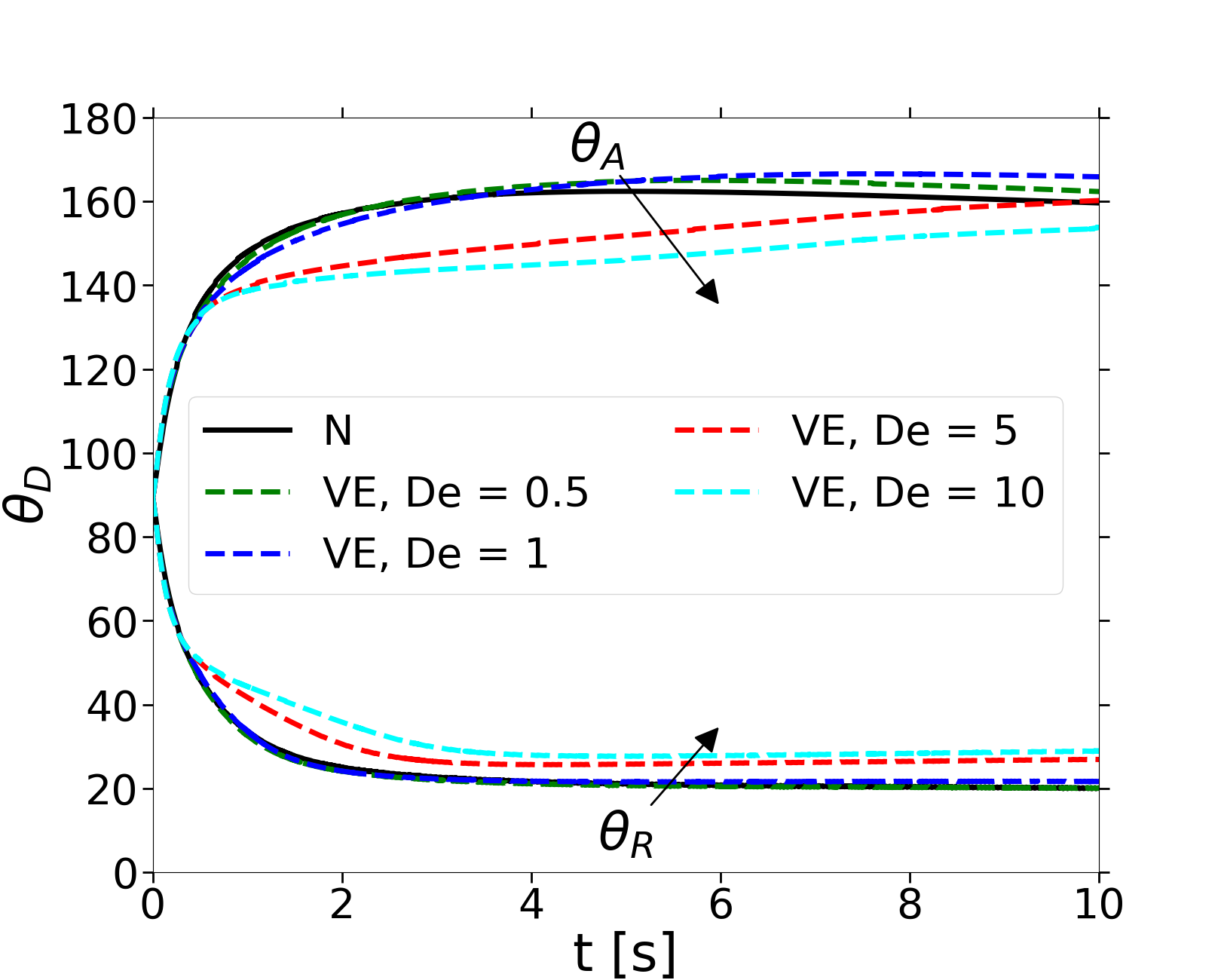}}
	\caption{The effect of $De$ on the time evolution of $\theta_{D}$ at constant capillary numbers in the V/N system. N and VE refer to Newtonian and Giesekus drops respectively. (a) $Ca=0.15$ and $\Delta\theta=100^{\circ}$ (b) $Ca=0.15$ and $\Delta\theta=140^{\circ}$ (c) $Ca=0.2$ and $\Delta\theta=100^{\circ}$ (d) $Ca=0.2$ and $\Delta\theta=140^{\circ}$ (e) $Ca=0.25$ and $\Delta\theta=100^{\circ}$ (f) $Ca=0.25$ and $\Delta\theta=140^{\circ}$.}
	\label{fig:fig7}
\end{figure}
\subsubsection{Effect of increasing Deborah number}\label{sec5_1_2}
\begin{figure}[htp]
	\centering
    \subfloat[]{\includegraphics[width=0.45\textwidth]{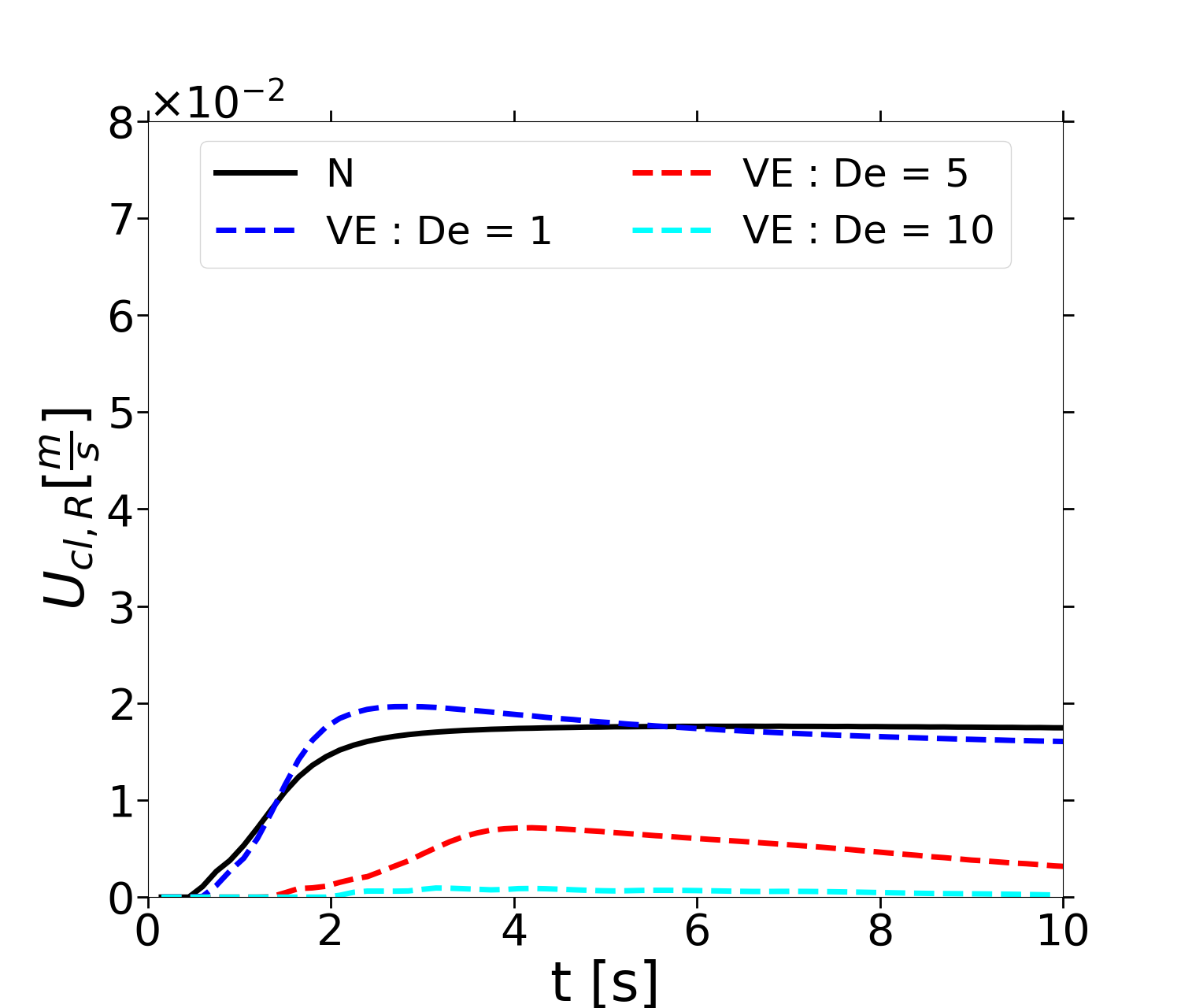}}
	\subfloat[]{\includegraphics[width=0.45\textwidth]{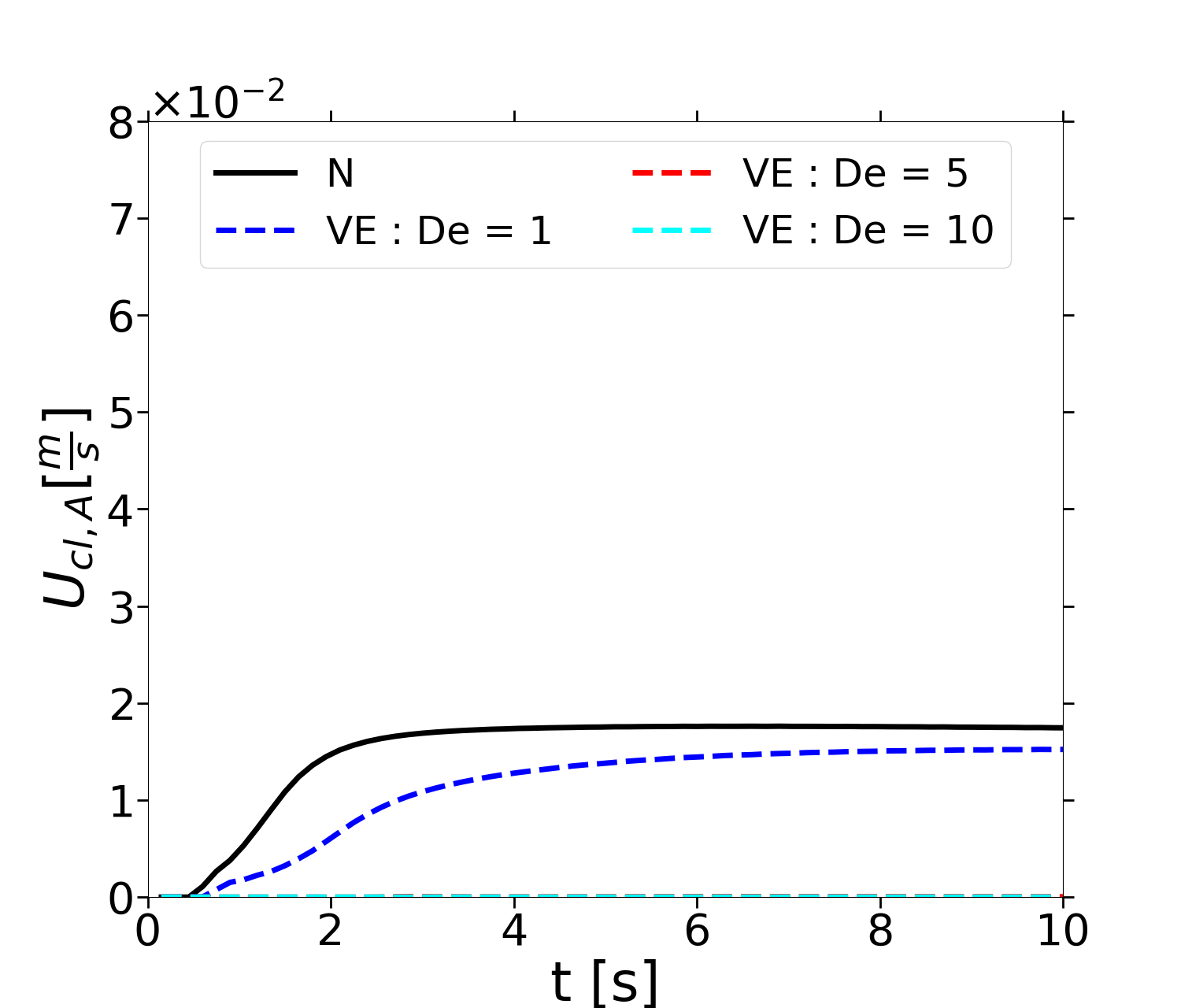}}\\
    \subfloat[]{\includegraphics[width=0.45\textwidth]{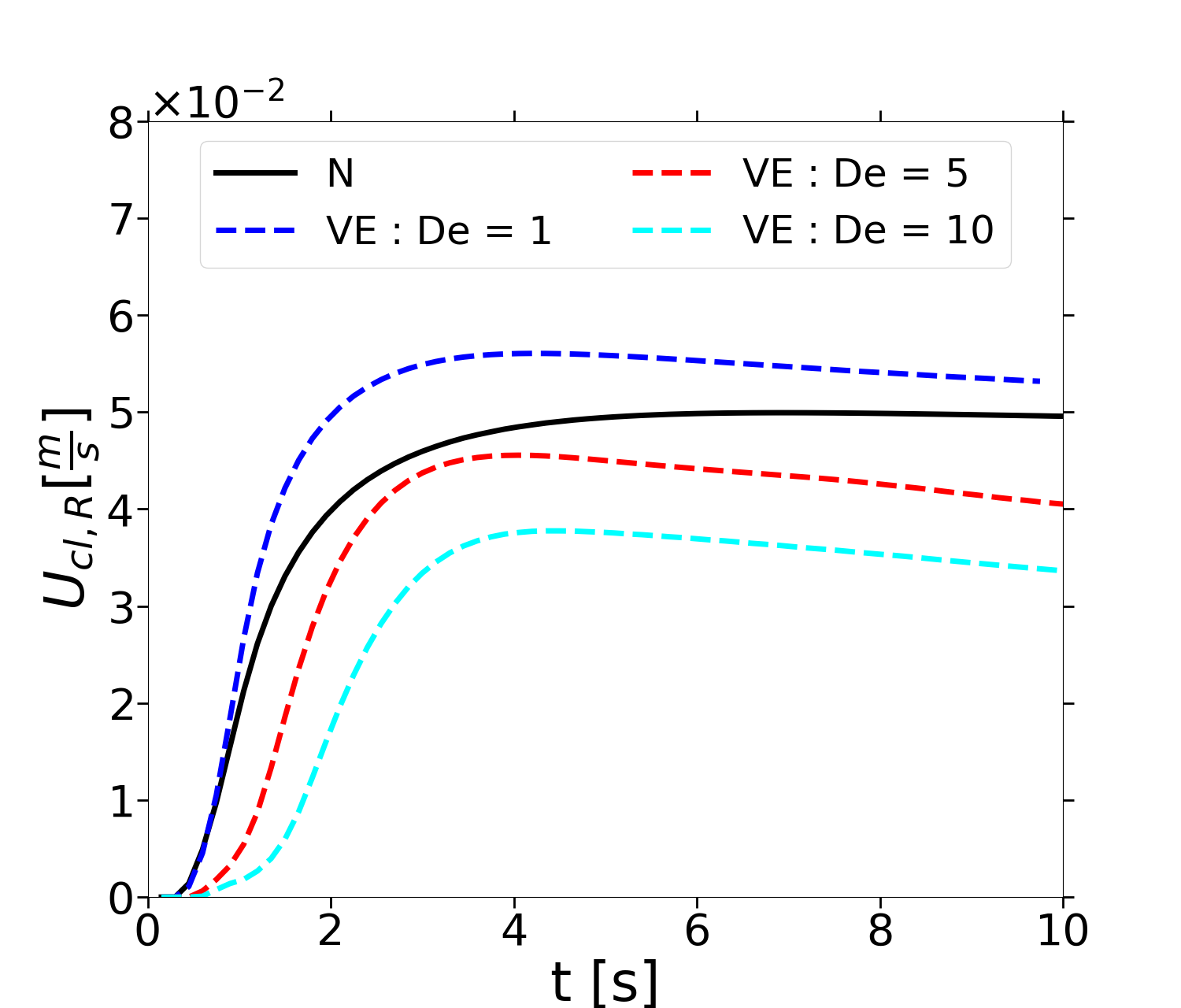}}
	\subfloat[]{\includegraphics[width=0.45\textwidth]{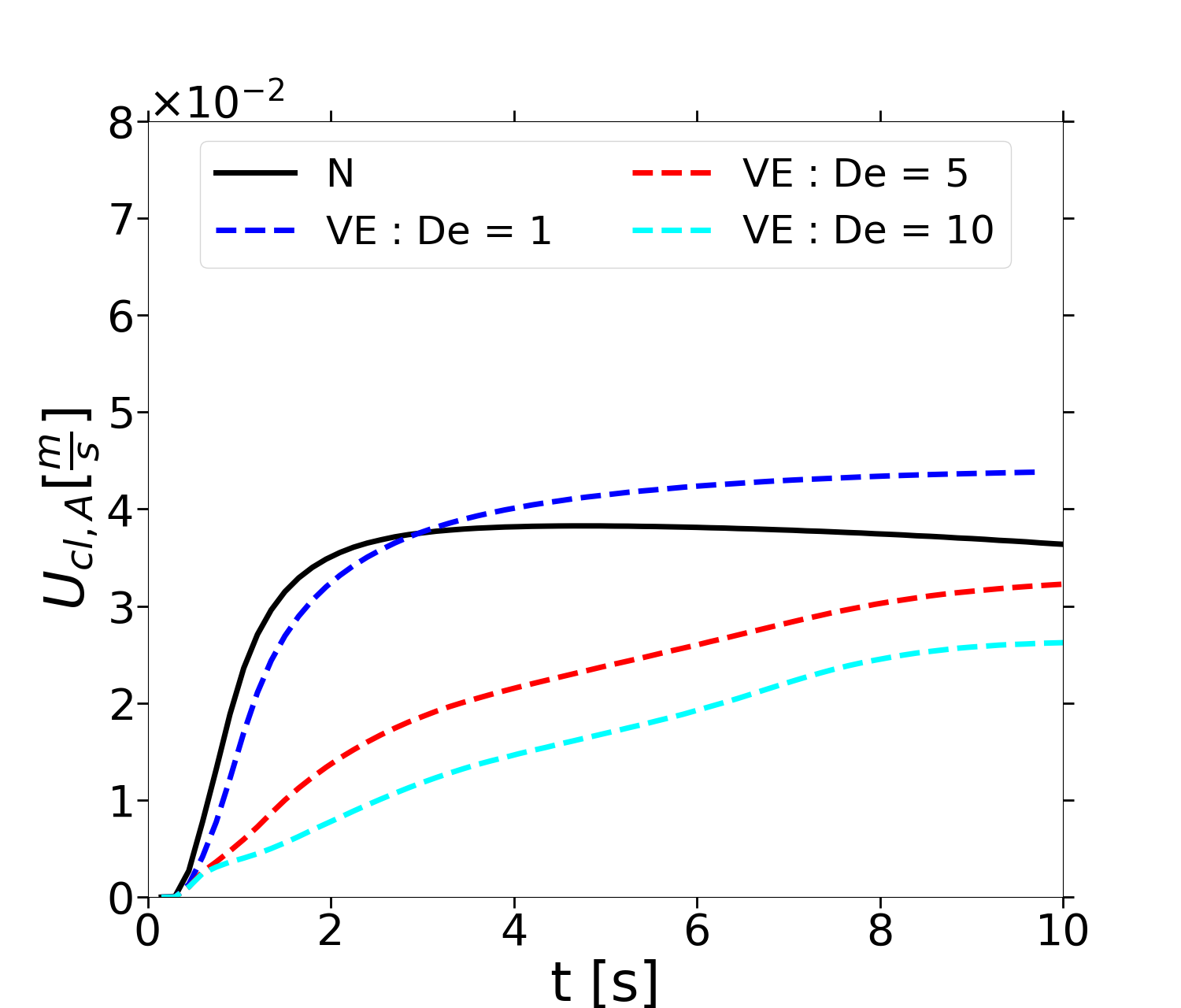}}\\
    \subfloat[]{\includegraphics[width=0.45\textwidth]{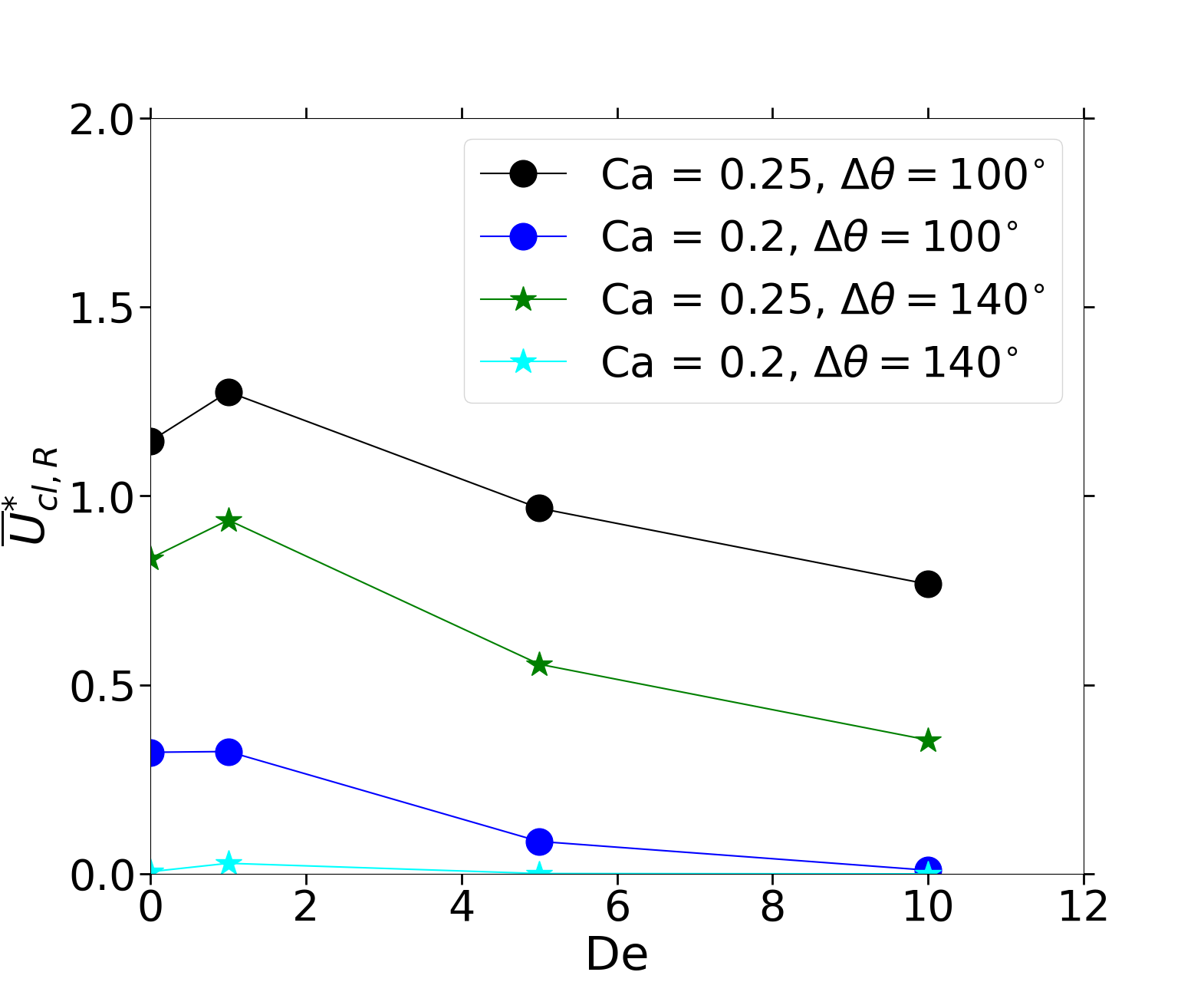}}
    \subfloat[]{\includegraphics[width=0.45\textwidth]{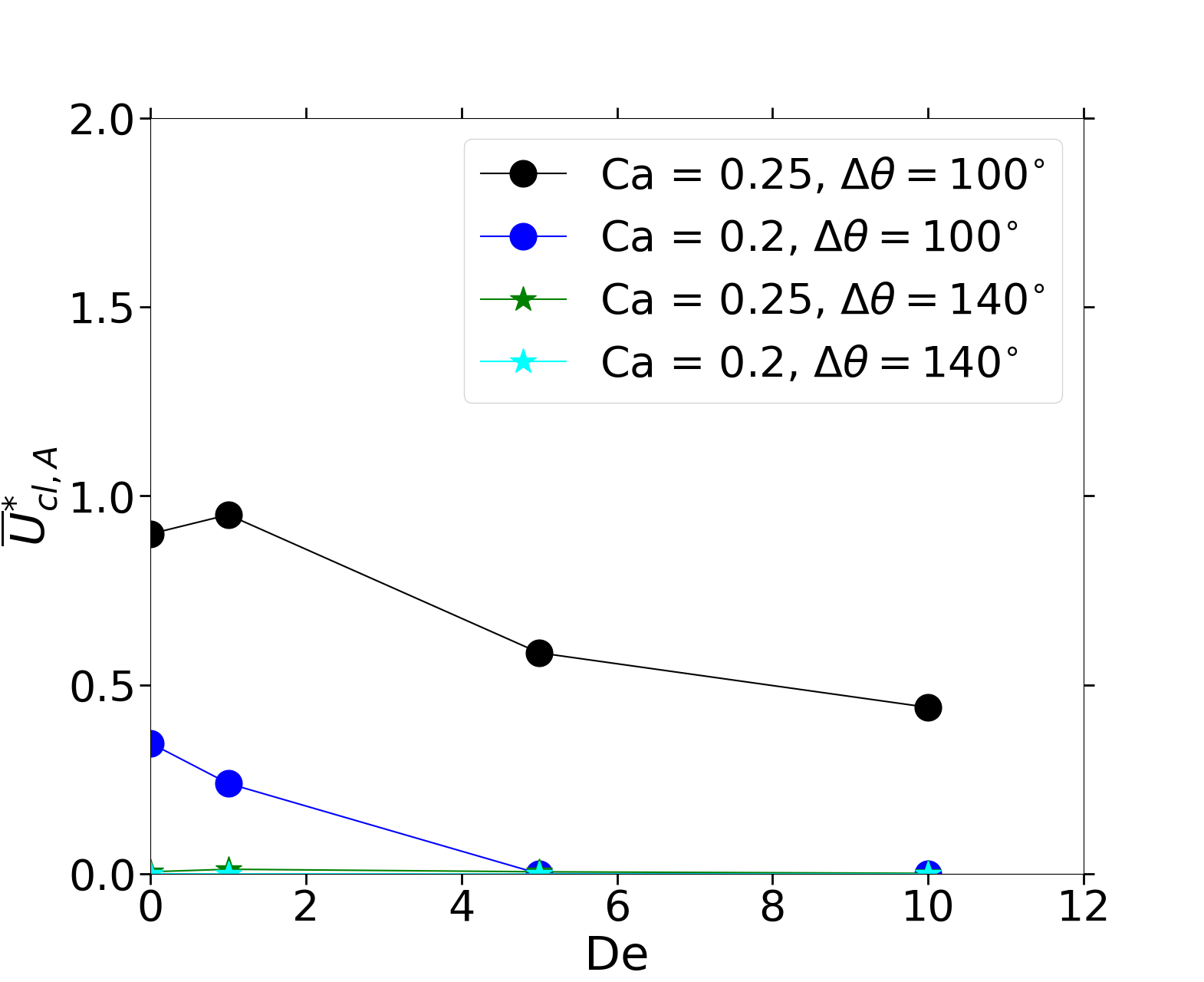}}
	\caption{\green{The effect of $De$ on the time evolution of $U_{cl,R}$ ($X^{t=0}_{cl,R}=3.436$) and $U_{cl,A}$ ($X^{t=0}_{cl,A}=4.564$) at constant capillary numbers in the V/N system. N and VE refer to Newtonian and Giesekus drops respectively. (a) $U_{cl,R}$ for $Ca=0.2$ with $\Delta\theta=100^{\circ}$ (b) $U_{cl,A}$ for $Ca=0.2$ with $\Delta\theta=100^{\circ}$ (c) $U_{cl,R}$ with $Ca=0.25$ and $\Delta\theta=100^{\circ}$ (d) $U_{cl,A}$ with $Ca=0.25$ and $\Delta\theta=100^{\circ}$ (e) $\overline{U}^{*}_{cl,R}$ vs $Ca$ (f) $\overline{U}^{*}_{cl,A}$ vs $Ca$}.}
	\label{fig:fig8}
\end{figure}

The effect of increasing the droplet's elasticity at a fixed capillary number has been investigated by changing $De$ over the surfaces with $\Delta\theta=100^{\circ}$ and $\Delta\theta=140^{\circ}$. The main effect observed is that bending of both contact lines is decreased significantly with increasing Deborah number to $De=5-10$. Thus, we have a non-monotonic behavior with respect to $De$ in the receding contact line; the elasticity of the droplet enhances the viscous bending for droplets with small elasticity, but the viscous bending is decreased for droplets with larger $De$. This non-monotonic behaviour in the receding contact angle also changes the receding contact line's velocity trend, and the velocity of the receding contact line decreases for $De=[5,10]$ in comparison to the Newtonian counterpart in all cases, see Fig. \ref{fig:fig8}. Further increasing the elasticity of the droplet depending on the hysteresis window of the surface and capillary number brings the contact lines to a halt, \green{see Figs. \ref{fig:fig8}(a) and \ref{fig:fig8}(b)} and for $De=10$, $Ca=0.2$, and $\Delta\theta=100^{\circ}$.

The advancing dynamic contact angle of the viscoelastic droplet generally attains a smaller steady-state value with further increasing the elasticity to $De=10$ (Fig. \ref{fig:fig7}), and the velocity of the advancing contact line decreases with increasing the elasticity of the droplet, \green{see Figs. \ref{fig:fig8}(b), \ref{fig:fig8}(d) and \ref{fig:fig8}(f)}. The decrease in the velocity of the advancing contact line is in line with the decreasing $\theta_{D,A}$. \green{The same trend for $U_{cl,R}$ exists in the receding contact line for $De=[5-10]$ but $\theta_{cl,R}$ gains a larger value, see Figs. \ref{fig:fig8}(a), \ref{fig:fig8}(c) and \ref{fig:fig8}(d).}
\begin{figure}[tb]
	\centering
    \subfloat[]{\includegraphics[width=0.45\textwidth]{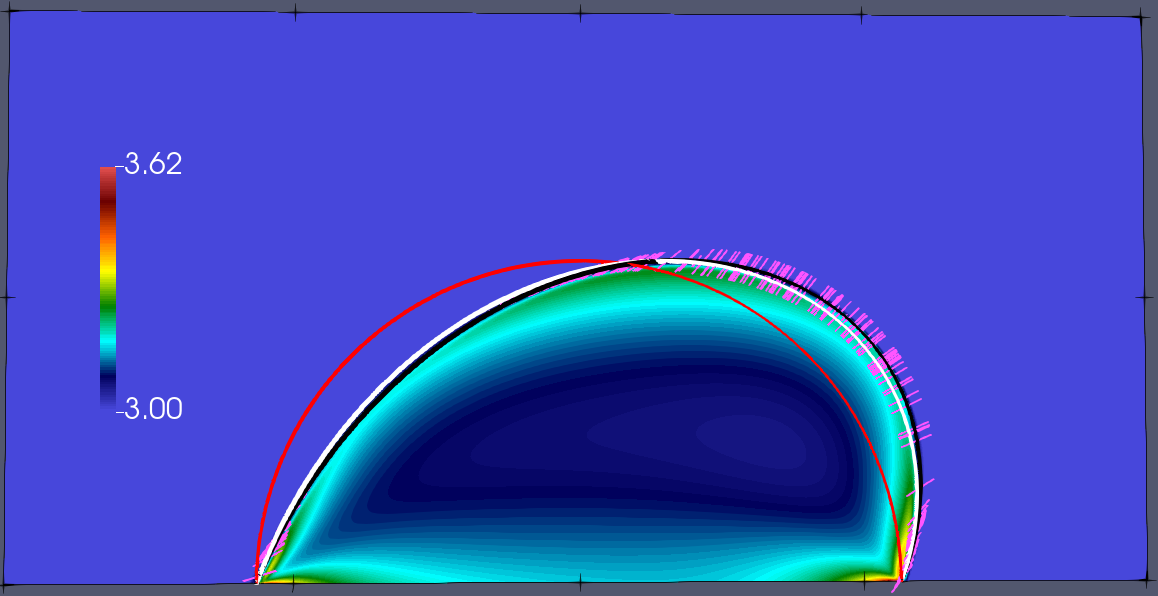}}
    \quad
    \subfloat[]{\includegraphics[width=0.45\textwidth]{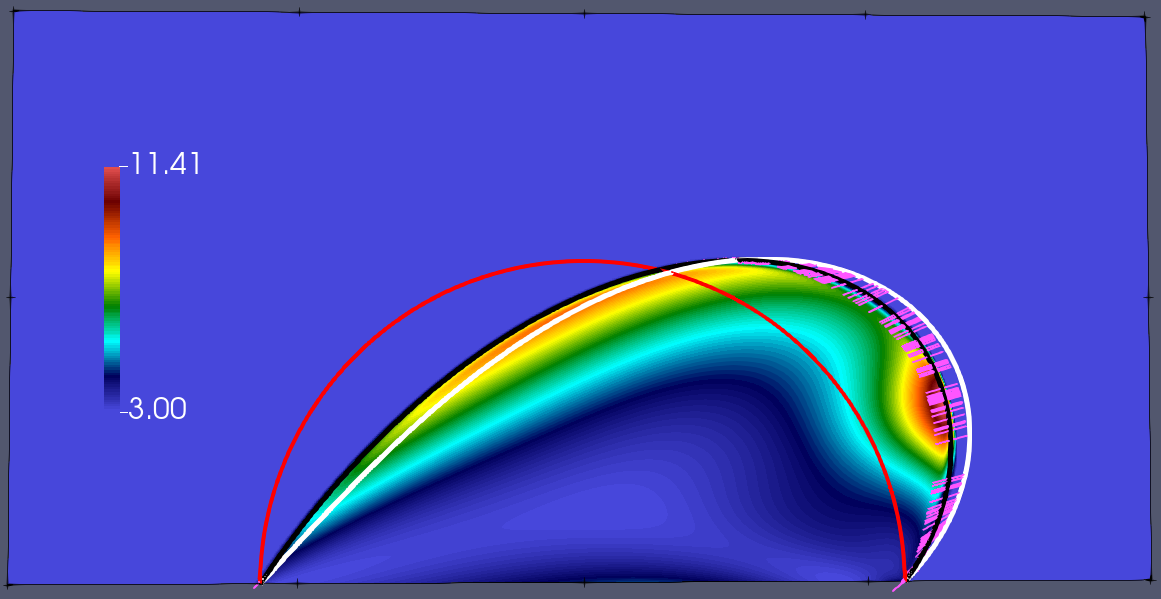}}
    \quad
    \subfloat[]{\includegraphics[width=0.45\textwidth]{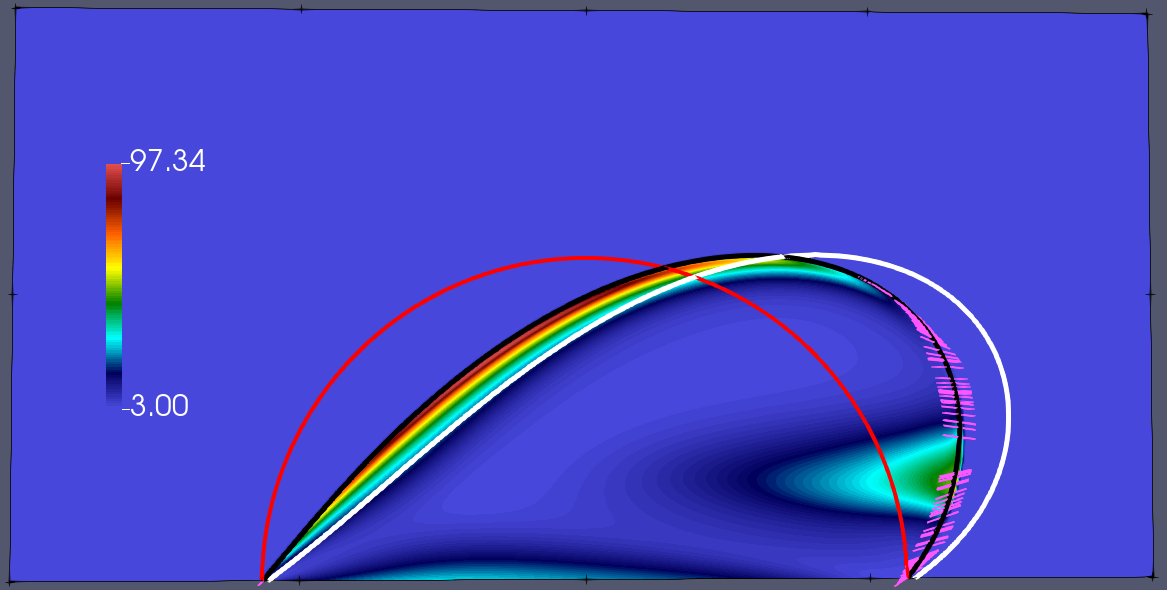}}
    \quad
    \subfloat[]{\includegraphics[width=0.45\textwidth]{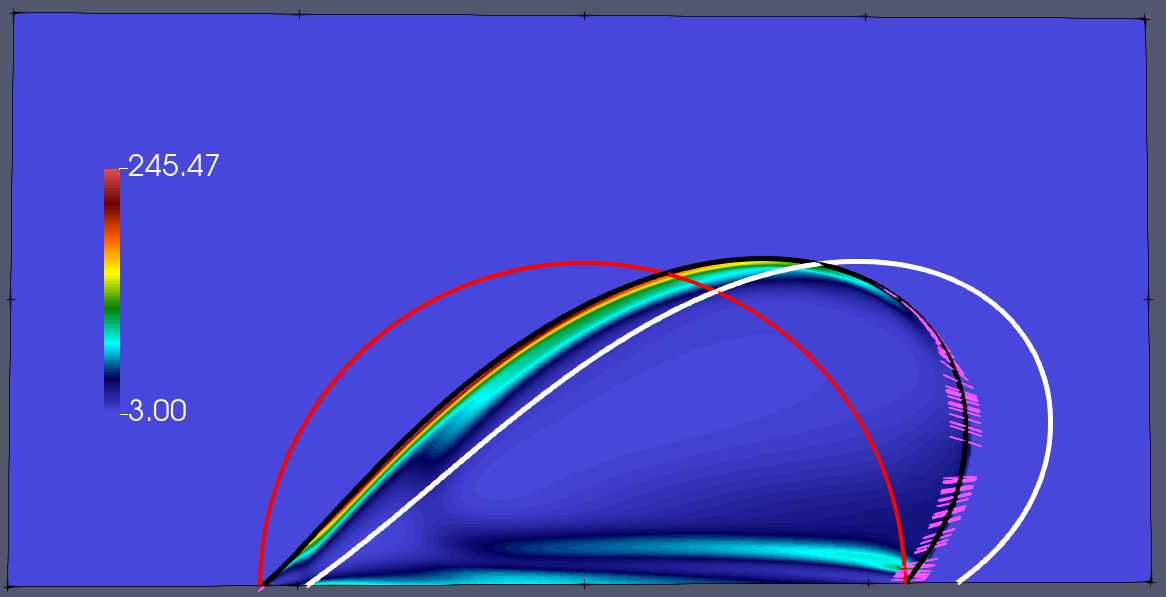}}
	\caption{ Time evolution of viscoelastic drops with $Ca=0.2$ over the surface with $\Delta\theta=100^{\circ}$. The interface $\phi=0$ of the Newtonian and Giesekus drops are depicted with a white and black, respectively. The trace of the conformation tensor has been visualized on the background, and the principal eigenvector of the conformation tensor is represented by purple line segments.
 (a) $De=10$ at $t\approx0.15\hspace{0.05cm}$s, 
 (b) $De=10$ at $t\approx0.6\hspace{0.05cm}$s, 
 (c) $De=10$ at $t\approx1.8\hspace{0.05cm}$s, 
 (d) $De=10$ at $t\approx6\hspace{0.05cm}$s.}
	\label{fig:fig9}
\end{figure}

To gain a better understanding of the elastic effects, we have visualized the trace of the conformation tensor $\mathbf{c}$ defined as $\mathbf{\boldsymbol{c}}=(\lambda_{H}/\mu_{p})\mathbf{\boldsymbol\tau_{p}}+\mathbf{\boldsymbol{I}}$, which is a measure of stretching of the polymer molecules, in figure \ref{fig:fig9} for the highest Deborah number $De=10$ at $Ca=0.2$ and $\Delta\theta=100^{\circ}$ (corresponding to drops in Figs. \ref{fig:fig7}(c) and \green{\ref{fig:fig8}(a,b)}). The results for $De=5$ are very similar to $De=10$.
The interface (defined by $\phi=0$) of the viscoelastic drop is shown with a black contour, and the corresponding Newtonian drop by a white contour, with the trace of the conformation tensor in the background, see Fig.\ref{fig:fig9}. Furthermore, we show by arrows the principal direction of the conformation tensor, representing the orientation of the dumbbells. The pull of the polymers affects the interface position most in regions where the arrows are perpendicular to the interface (\citep{KineticBird1987}, pp. 64-69). 

At early times (Figs. \ref{fig:fig9}a), the viscoelastic drop speeds up and deforms faster than the Newtonian counterpart. The polymer molecules are most stretched in the vicinity of the receding and advancing contact lines at $t\approx0.15\hspace{0.05cm}$s, but the dumbbells have acute angle with the interface there, and hence do not influence the interface. At $t\approx0.6\hspace{0.05cm}$s (Fig. \ref{fig:fig9}(b)), the receding and advancing contact lines of the Newtonian drop start moving already. However, the polymeric stresses have now built up (which takes a finite time proportional to $\lambda_{H}$). Hence, deformation and time evolution of the viscoelastic droplet have been retarded, and both contact lines of the viscoelastic droplets remain pinned for $De=5-10$. Furthermore, the dumbbells are becoming perpendicular to the interface in $\varphi\approx[140^{\circ},160^{\circ}]$ and because polymers are also very elongated here, the stresses should have a large influence on the interface.

\begin{figure}[tb]
	\centering
    \subfloat[]{\includegraphics[width=0.36\textwidth]{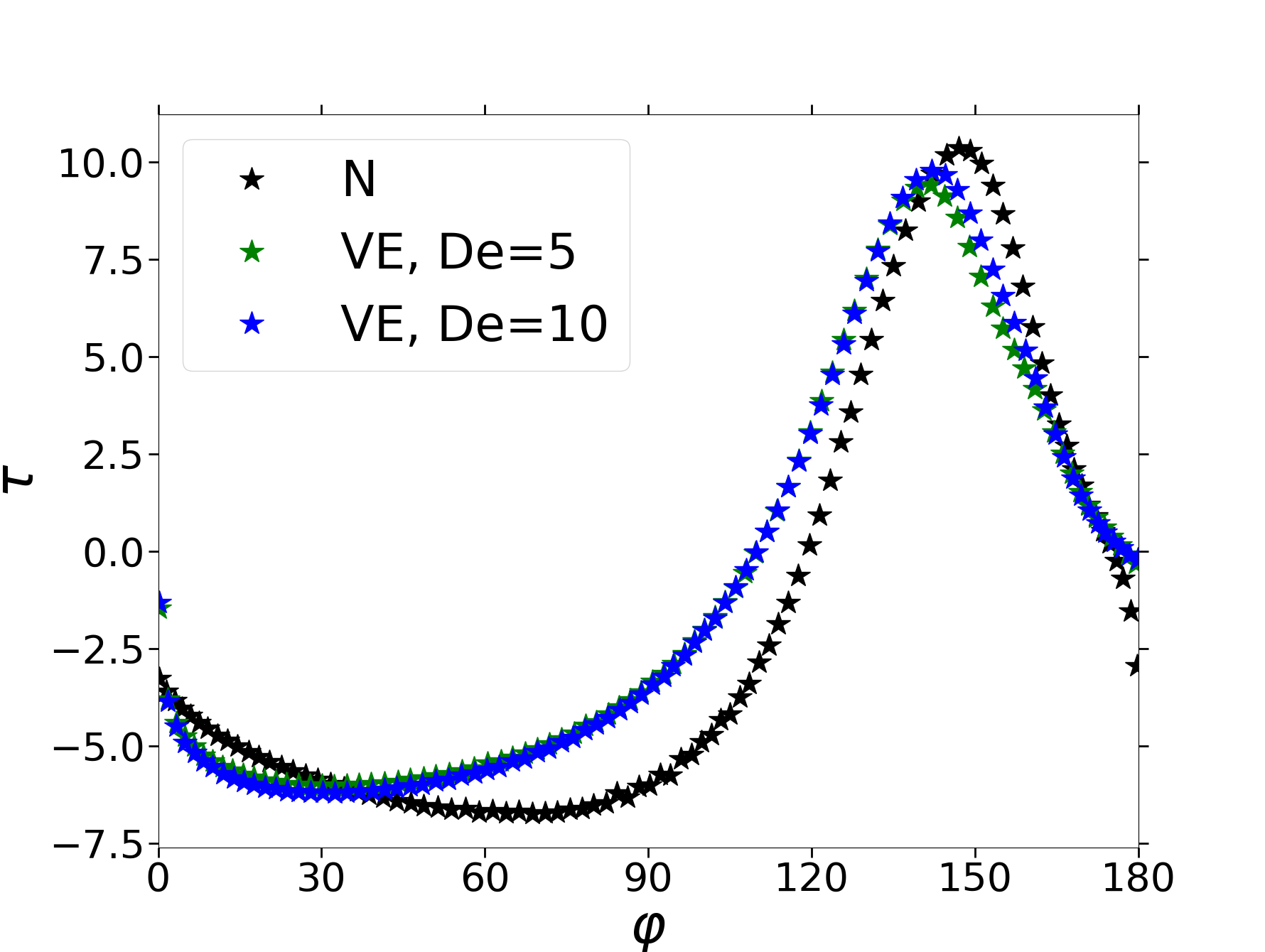}}
	\subfloat[]{\includegraphics[width=0.36\textwidth]{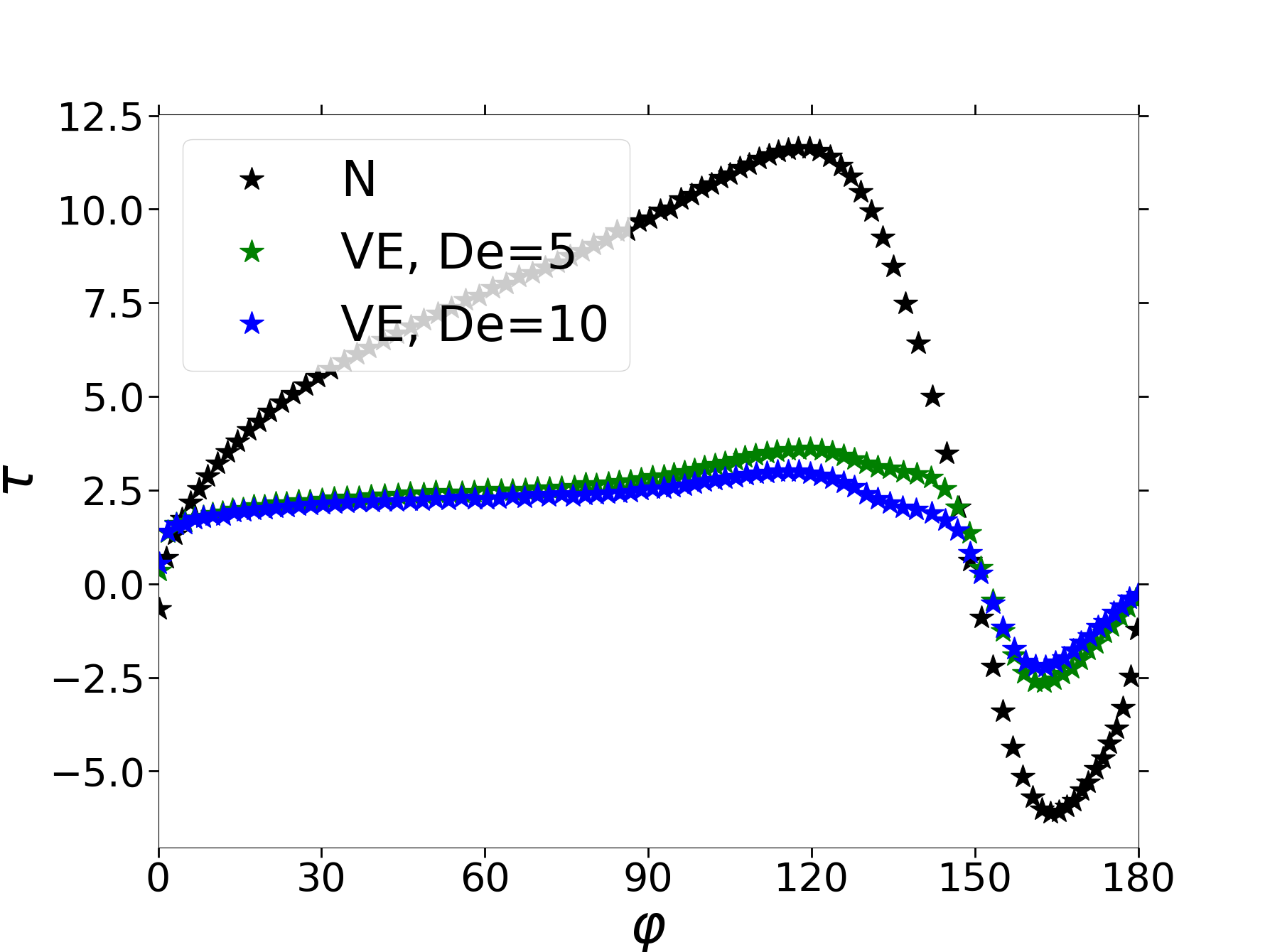}}\\
    \subfloat[]{\includegraphics[width=0.36\textwidth]{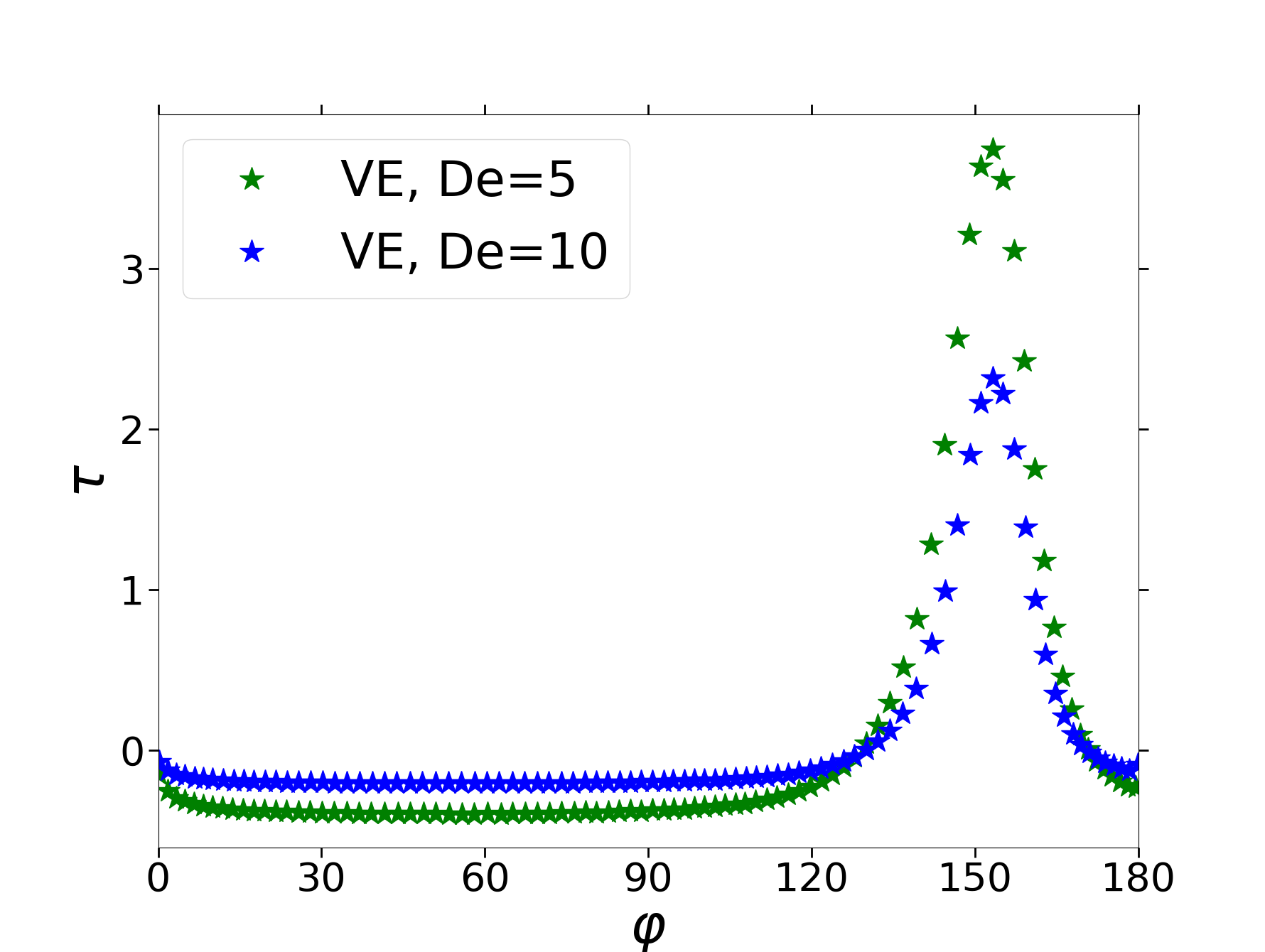}}
	\subfloat[]{\includegraphics[width=0.36\textwidth]{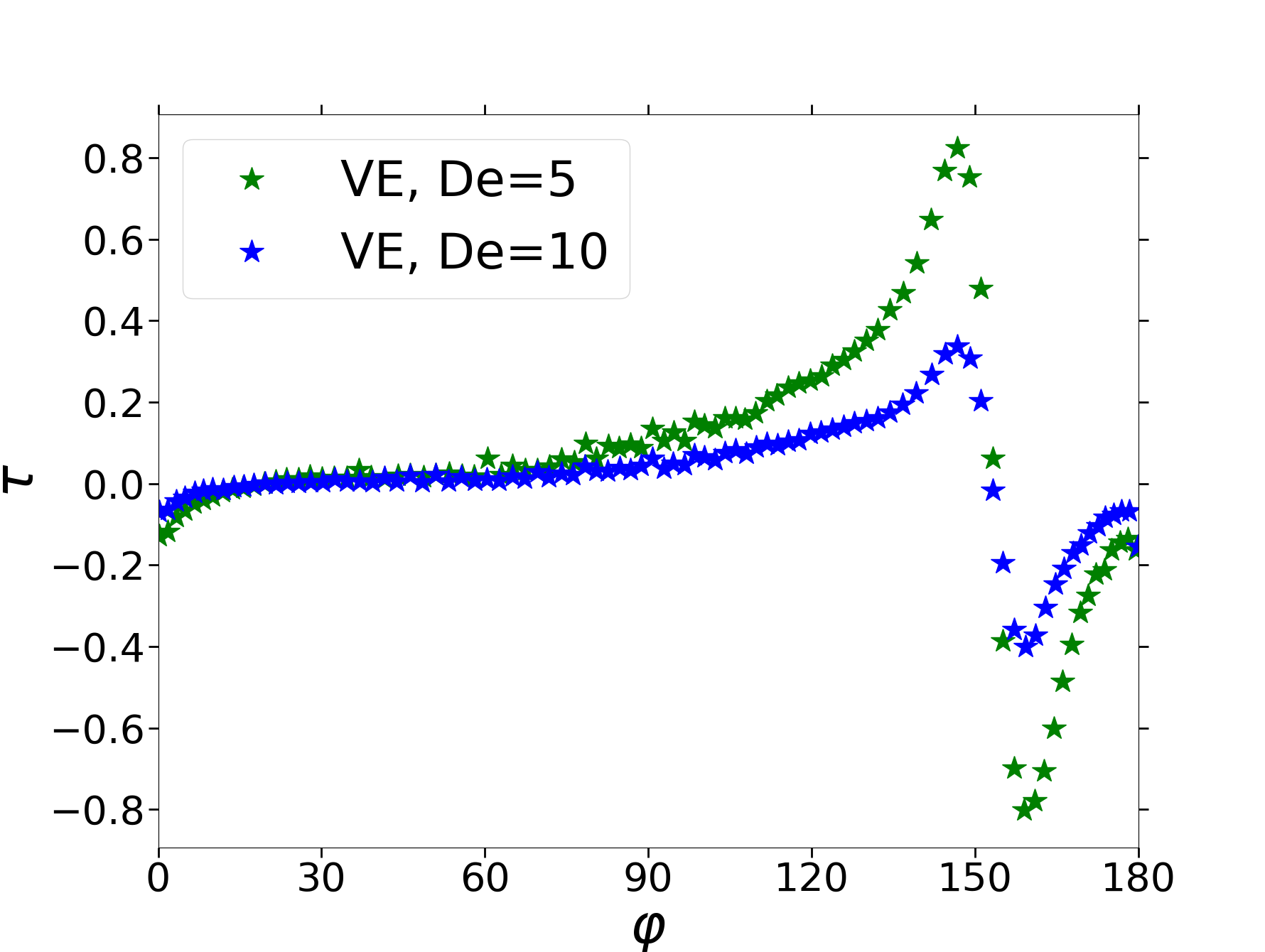}}\\
    \subfloat[]{\includegraphics[width=0.36\textwidth]{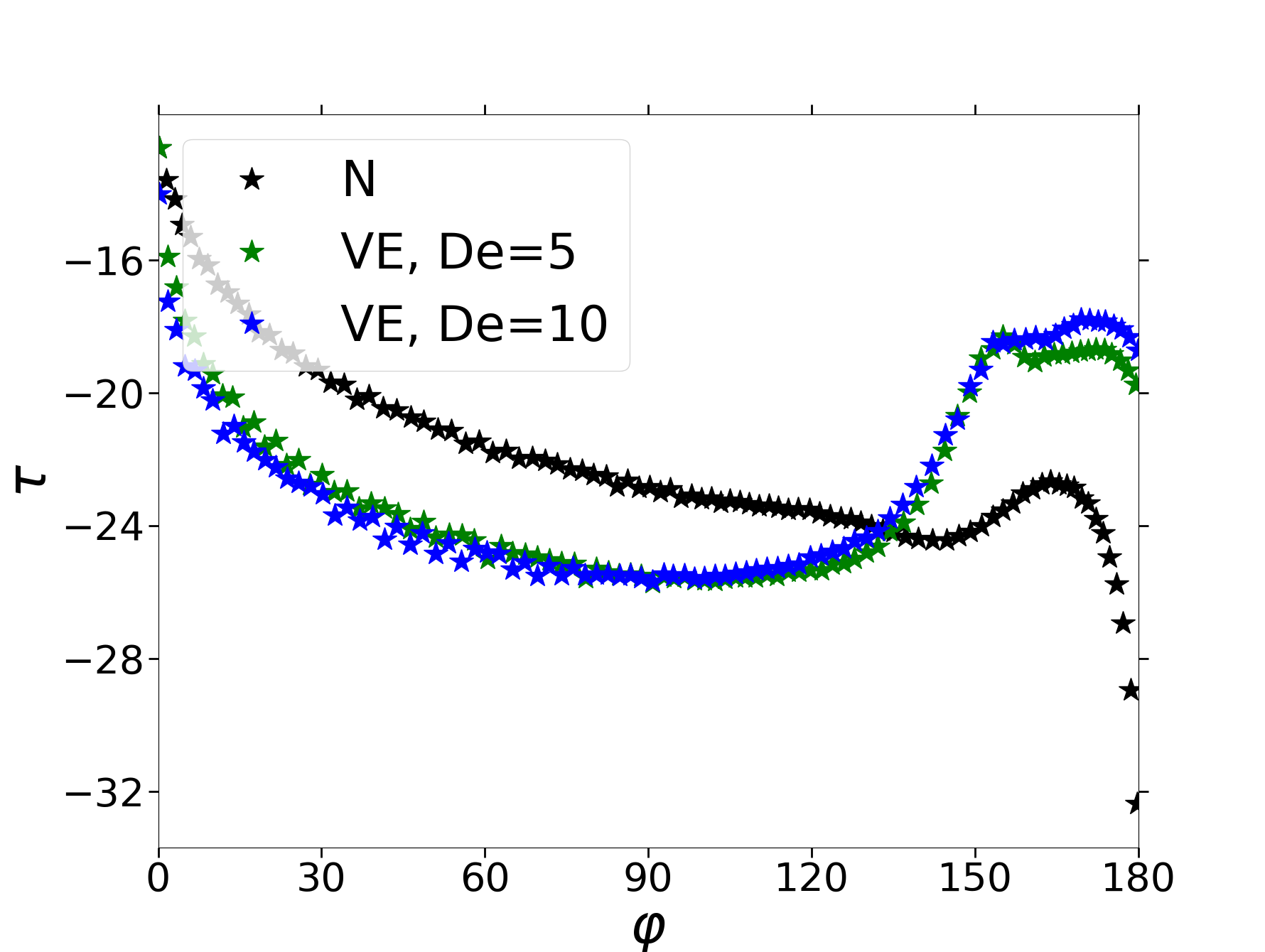}}
	\caption{Different components of the stresses along $\phi=0$ at time $t=0.6\hspace{0.05cm}$s for drops with $Ca=0.2$ over the surface with $\Delta\theta=100^{\circ}$. (a) $\tau^{n}_{vi}$ (b) $\tau^{t}_{vi}$ (c) $\tau^{n}_{p}$ (d) $\tau^{t}_{p}$ (e) $\tau^{n}_{pre}$ 
 }
	\label{fig:fig10}
\end{figure}

To compare the relative importance of polymeric and flow-induced stresses, we again show them (Eq. \ref{NS34}) along the interface $\phi=0$, separated into normal and tangential components, in Fig. \ref{fig:fig10}. Increasing viscoelasticity to $De=5-10$ suppresses the tangential component (Fig. \ref{fig:fig10}(b)) of the viscous stress along the interface ( $\left[\frac{\tau^{t}_{vi}|_{VE}}{\tau^{t}_{vi}|_{N}}\right]_{max}\approx0.3$), and this decrease is mainly responsible for the delay in the deformation of the viscoelastic droplets. The change in viscous stresses due to viscoelasticity has a larger magnitude than the previously analysed polymeric stresses (Fig. \ref{fig:fig10}(c,d)) which are only active in the segment $\varphi\approx[110^{\circ},170^{\circ}]$. Further, the polymeric stresses are largest for $De=5$ and decrease at $De=10$, while the contact line stays pinned longest for $De=10$. Hence, the retardation and pinning of the viscoelastic droplet is mainly caused by velocity modification in viscoelastic flows around the interface, rather than by the polymeric stresses themselves. 

 Moreover, another important rheological property of the viscoelastic fluid is the strain-hardening which usually occurs in extensional flow, and can be attributed to the elongational viscosity (\cite{FluidBird1987}, p.p 132) defined by $\mu_{e}\sim\mu_{e}(N_{1},\dot{\epsilon})$ where $N_{1}=\tau_{xx}-\tau_{yy}$ is the first normal stress and $\dot{\epsilon}$ is the elongational strain rate. Since the modification of the flow field is considerable in particular for $De\ge5$ when the polymeric stresses are built up, we use the flow topology parameter \citep{Lee2007} to find extensional flow regions inside the droplets:
\begin{eqnarray}
\xi=\frac{|\boldsymbol D|-|\boldsymbol \Omega|}{|\boldsymbol D|+|\boldsymbol \Omega|}
\label{NS35}
\end{eqnarray}

where $\boldsymbol D = \frac{\nabla{{\mathbf{u}}}+\nabla{{\mathbf{u}}^T}}{2}=\frac{\Dot{\boldsymbol\gamma}}{2}$ and $\boldsymbol \Omega=\frac{\nabla{{\mathbf{u}}}^{T}-\nabla{{\mathbf{u}}}}{2}$ are the deformation and vorticity tensors respectively, and the magnitude of a tensor $\boldsymbol Q$ can be computed by $|\boldsymbol Q|=\sqrt{\frac{1}{2}\sum_{i}\sum_{j}Q^{2}_{ij}}$. The flow parameter can vary between $\xi=-1$ corresponding to pure solid-like rotation to $\xi=1$ corresponding to pure extensional flow, and $\xi=0$ presenting pure shear flow.

\begin{figure}[tbp]
	\centering
    \subfloat[]{\raisebox{0.17cm}{\includegraphics[width=0.45\textwidth]{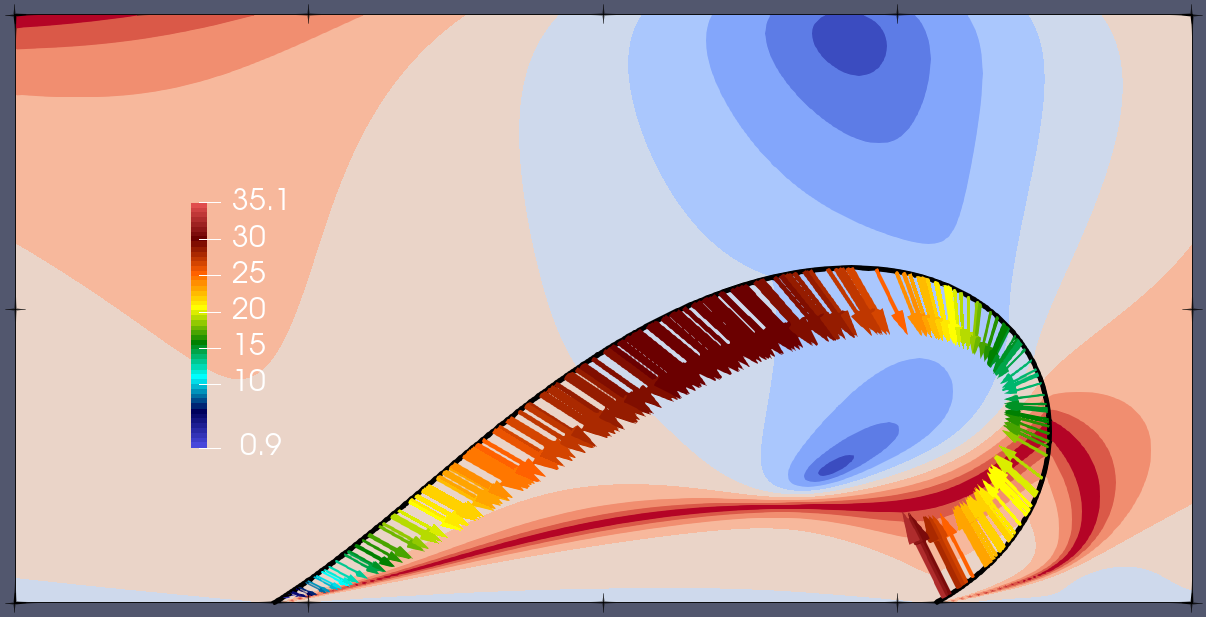}}}
    \quad
    \subfloat[]{\includegraphics[width=0.45\textwidth]{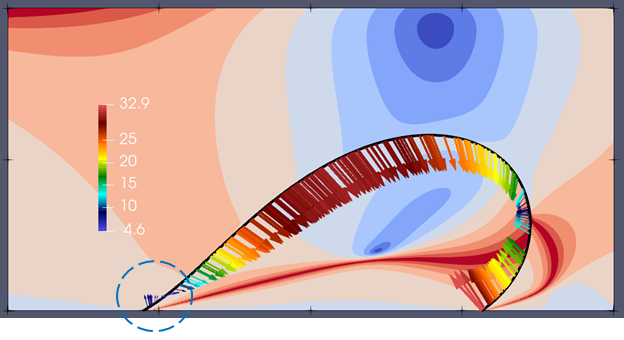}}\\
    \subfloat[]{\includegraphics[width=0.45\textwidth]{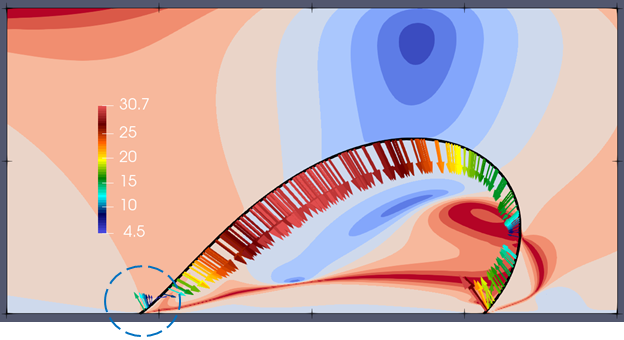}}
    \quad 
  \subfloat[]{\includegraphics[width=0.45\textwidth]{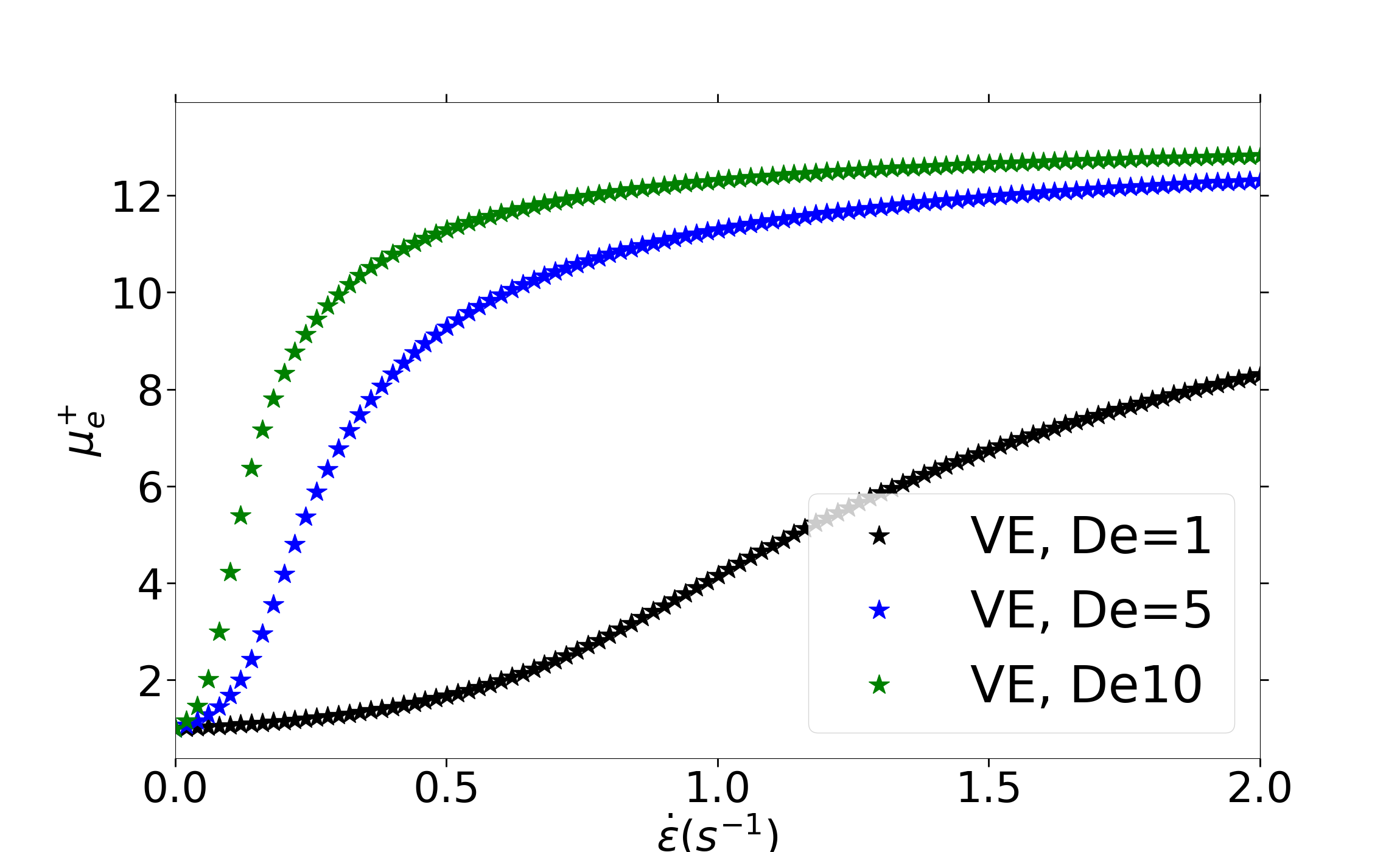}}\\
    \subfloat[]{\includegraphics[width=0.45\textwidth]{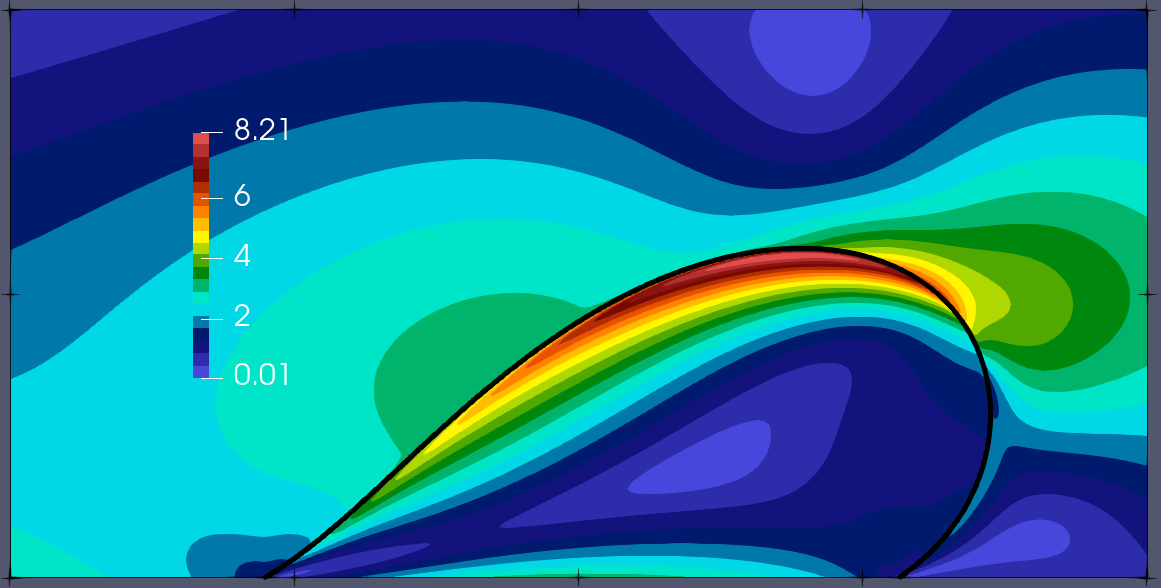}}
    \quad
    \subfloat[]{\includegraphics[width=0.45\textwidth]{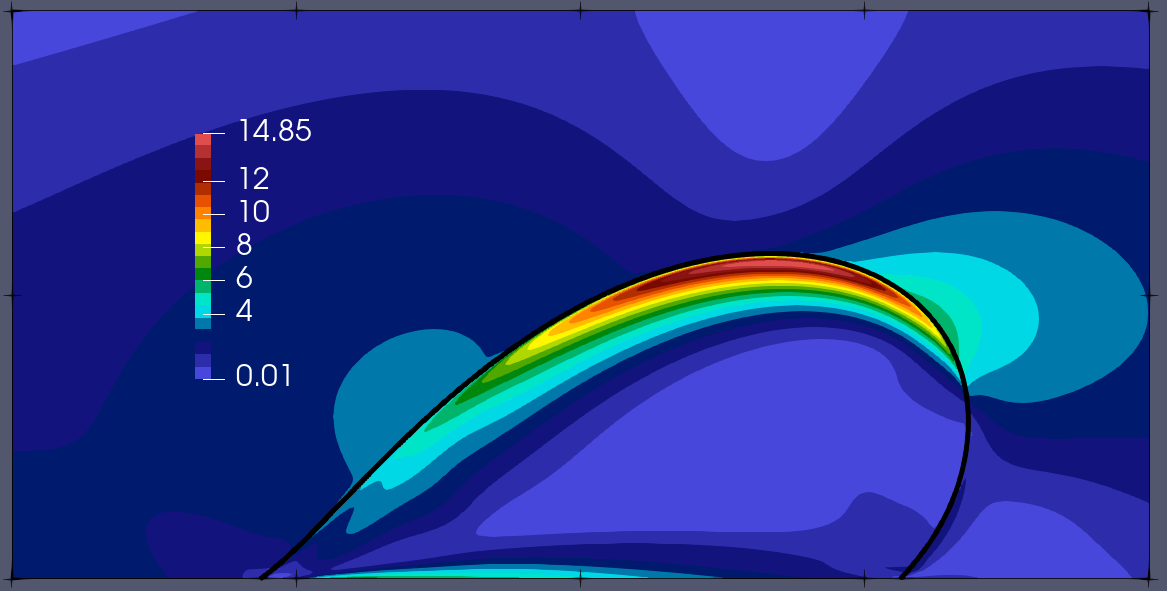}}\\
    \subfloat[]{\includegraphics[width=0.45\textwidth]{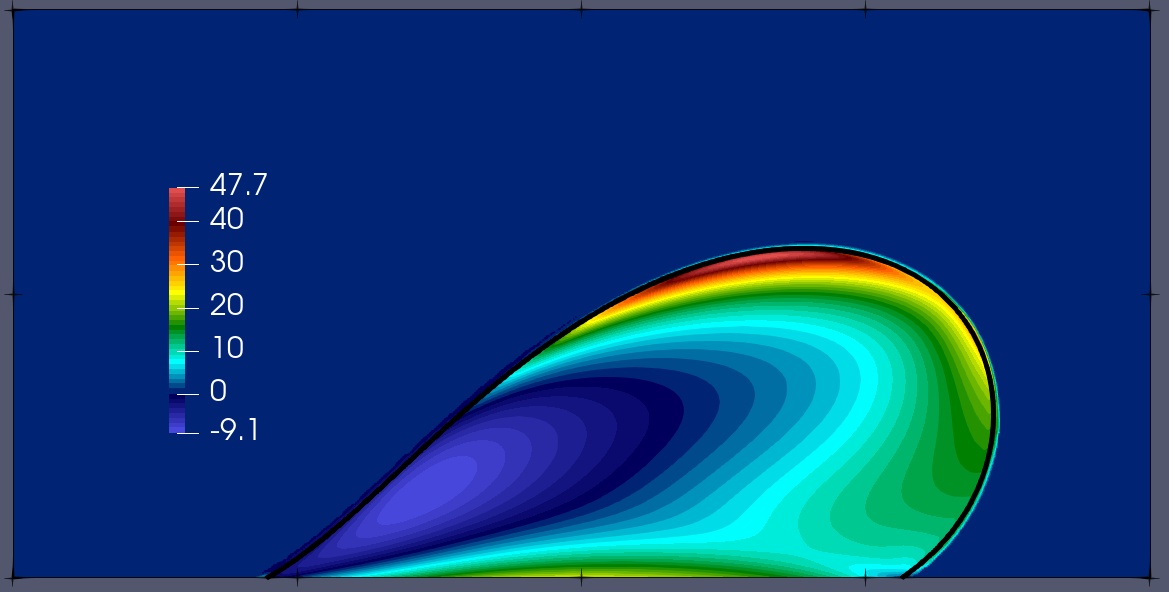}}
    \quad
    \subfloat[]{\includegraphics[width=0.45\textwidth]{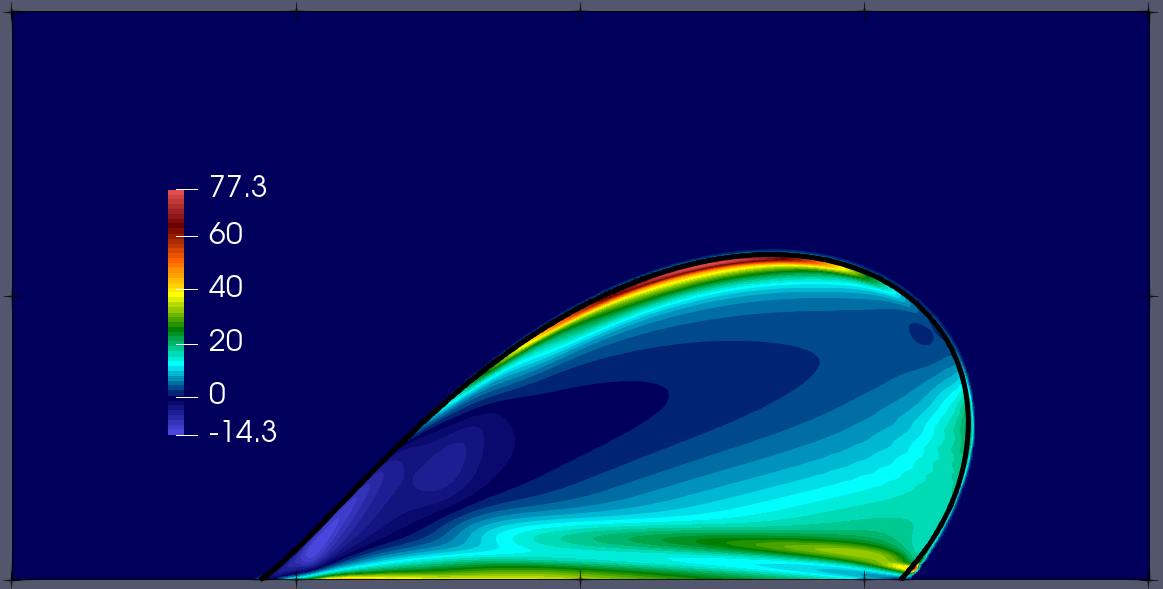}}

	\caption{Different flow quantities of the Newtonian and Giesekus drops with Ca = 0.2 over $\Delta\theta=140^{\circ}$ at $t\approx6\hspace{0.05cm}$s.(a) The flow parameter ($\xi$) on the background and the total stress vector (${\mathbf{n}}_{s}\cdot\boldsymbol\tau$ in Eq. \ref{NS34}) along the interface $\phi=0$ for the Newtonian droplet with $Ca=0.2$ and $\Delta\theta=140^{\circ}$ at $t\approx6\hspace{0.05cm}$s, (b) the same for a Giesekus drop with $Ca=0.2$, $De=1$, (c) the same   at $Ca=0.2$, $De=5$. (d) The steady-state dimensionless elongational viscosity as a function of the elongational strain rate for the Giesekus model with different $De$ corresponding to $Ca=0.2$ and $\Delta\theta=140^{\circ}$.(e) The magnitude of strain rate ($|\Dot{\boldsymbol\gamma}|$) for the droplet at $De=1$, and (f) the same at $De=5$. (g) The first normal stress difference at $De=1$, and (h) the same at $De=5$.}
	\label{fig:fig12}
\end{figure}

To further contrast the effects at small and large elasticity, we compare the flow topology of the viscoelastic droplets at $De=0$ (Newtonian case, Fig. \ref{fig:fig12} (a)), $De=1$ (Fig. \ref{fig:fig12} (b)) and $De=5$ (Fig. \ref{fig:fig12}(c)). The flow type inside the viscoelastic droplet with $De=5$ is more extensional (red color) at the front where viscoelastic stresses pull the droplet interface, in comparison to the other droplets. This is the extensional flow region, where extensional viscosity may be higher. The extensional viscosity as a function of the extensional strain rate at steady-state for the Giesekus model ($\alpha\neq0$) can be expressed as:  (\citep{FluidBird1987}, p.p 368)
\begin{eqnarray}
\mu^{+}_{e}=\frac{\mu_{e}-\mu_{s}}{3\mu_{p}}=\frac{1}{6\alpha}\Big[3+\frac{1}{\lambda_{H}\dot{\epsilon}}\Big[\sqrt{1-4(1-2\alpha)\lambda_{H}\dot{\epsilon}+4(\lambda_{H}\dot{\epsilon})^{2}}-\nonumber \\\sqrt{1+2(1-2\alpha)\lambda_{H}\dot{\epsilon}+(\lambda_{H}\dot{\epsilon})^{2}}\Big]\Big]
\label{NS36}
\end{eqnarray}
Fig. \ref{fig:fig12}(d) presents $\mu^{+}_{e}$ as a function of strain rate\footnote{The magnitude of strain rate in the extensional flow dominant regions in Figs. \ref{fig:fig12}(b) and \ref{fig:fig12}(c) is mostly less or equal to 2 ($|\Dot{\boldsymbol\gamma}|\leq2$)} $\dot{\epsilon}$ for three Deborah numbers. The magnitude of $\mu^{+}_{e}$ for $De\geq5$ increases steeply already at small elongational strain rates. The strain rate for $De=5$ shown in figure \ref{fig:fig12} (f) is highest in shear-dominated regions at the top, but also moderate in extensional flow areas at the front, and therefore we should expect a large extensional viscosity resisting deformation at the front. As observed, the shear strain rate is always higher in the viscoelastic fluid than in the Newtonian fluid, and this is a consequence of keeping the total viscosity ratio constant ($\lambda_\mu=1$); the solvent viscosity of the viscoelastic droplet is smaller. The magnitude of $N_{1}$ is largest in the top side where shear flow dominates and increases with $De$
,but those regions of the interface do not directly influence the contact line movement.

\begin{figure}[tbp]
	\centering
     \subfloat[]{\includegraphics[width=0.45\textwidth]{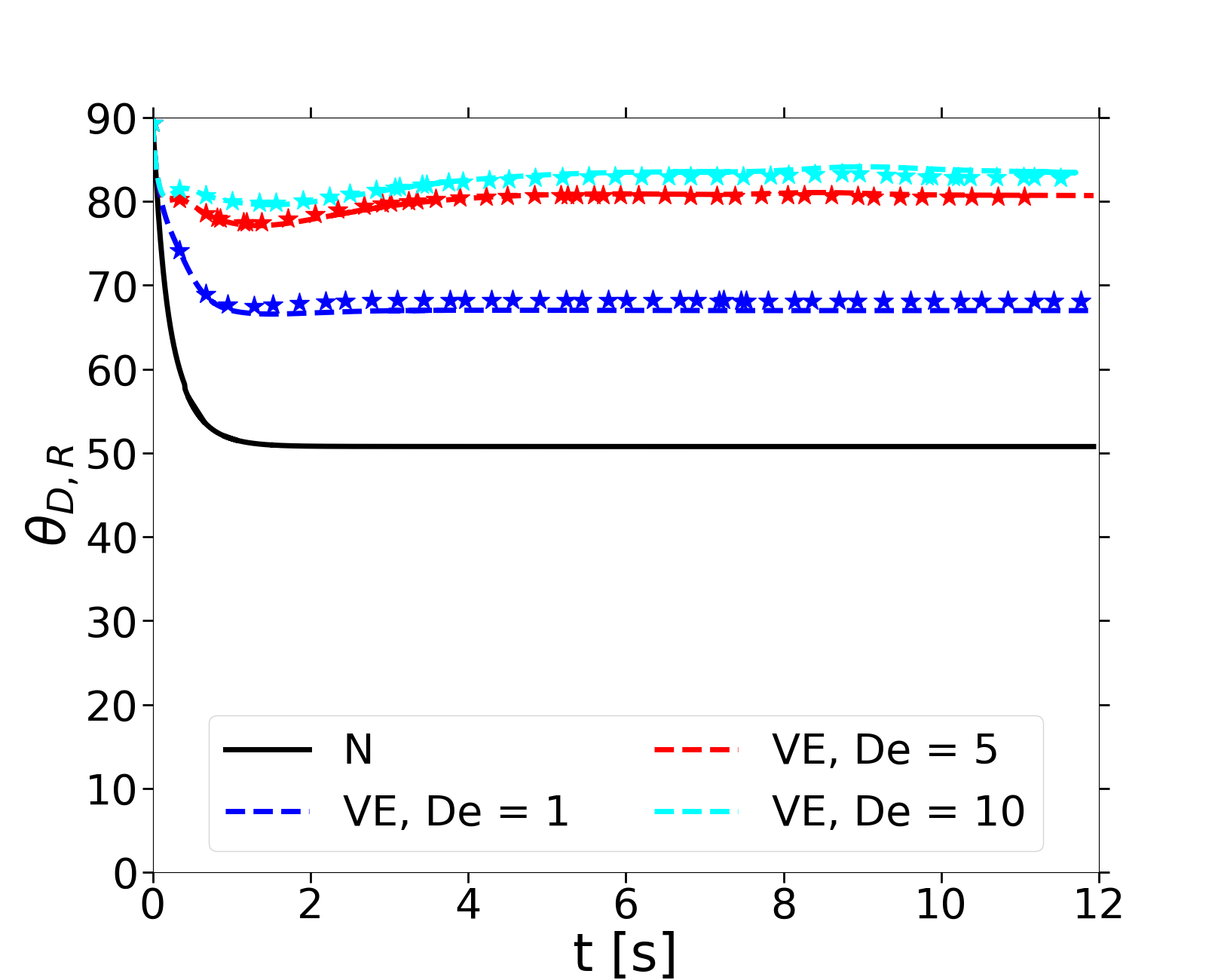}}
     \subfloat[]{\includegraphics[width=0.45\textwidth]{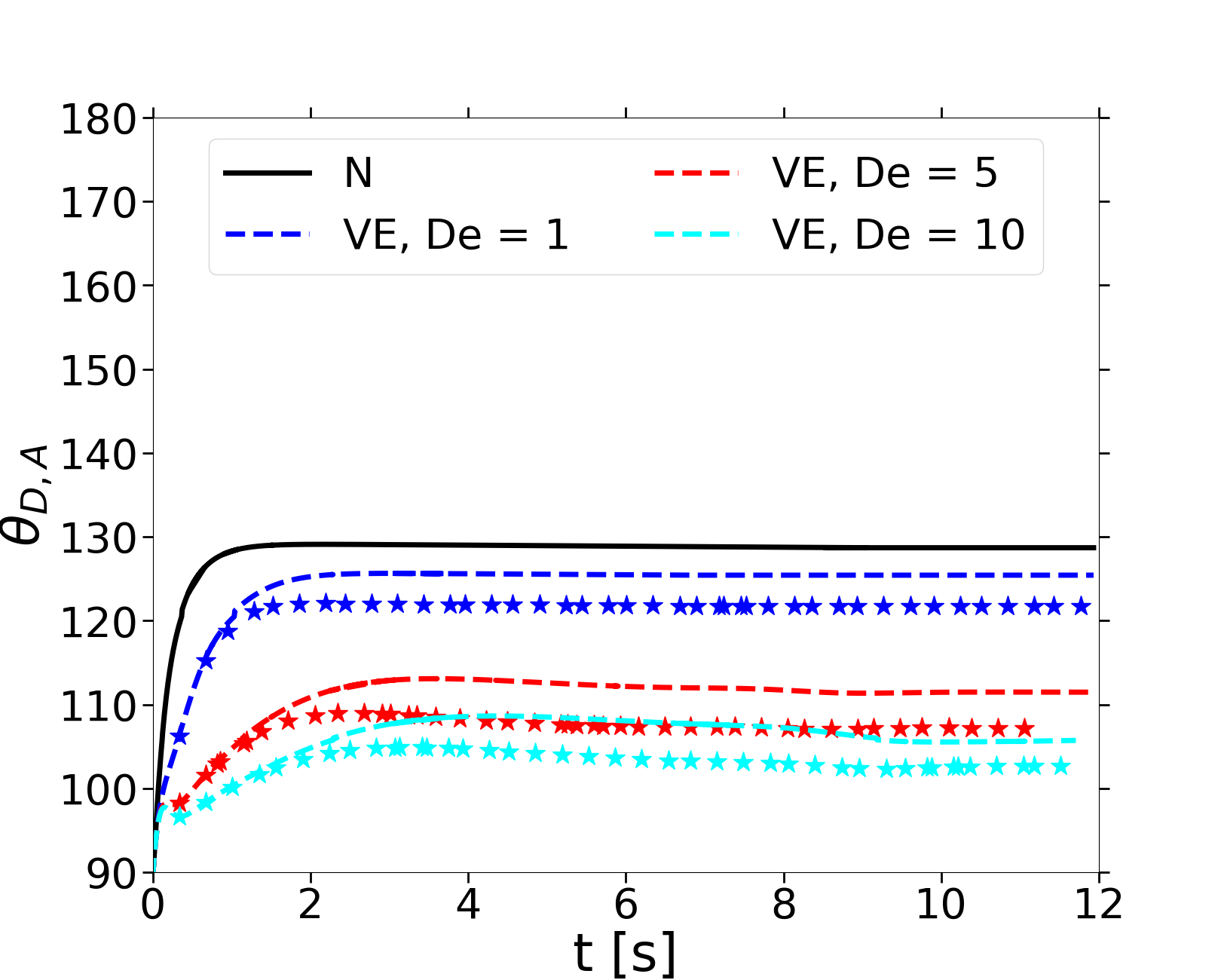}}\\
     \subfloat[]{\includegraphics[width=0.45\textwidth]{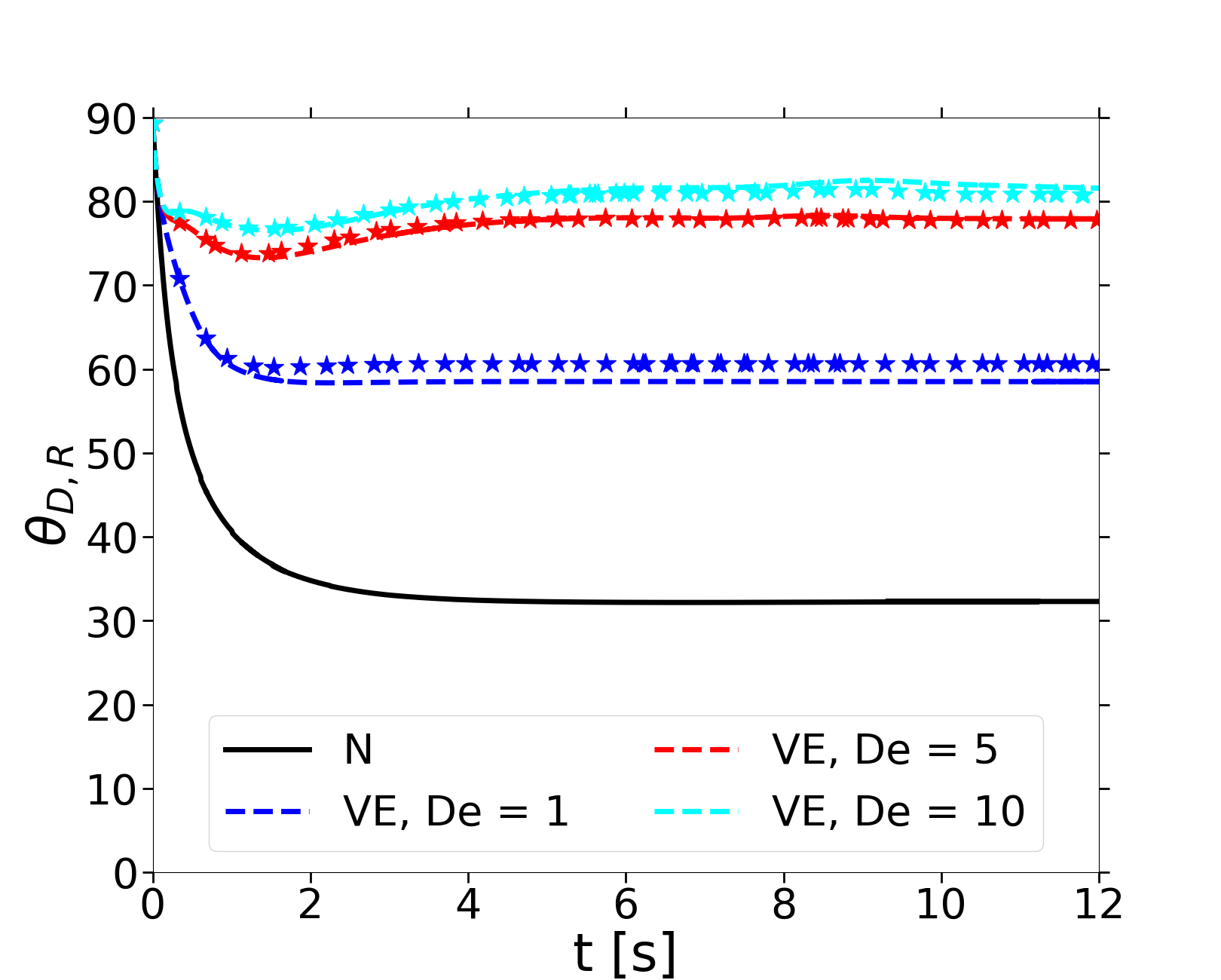}}
     \subfloat[]{\includegraphics[width=0.45\textwidth]{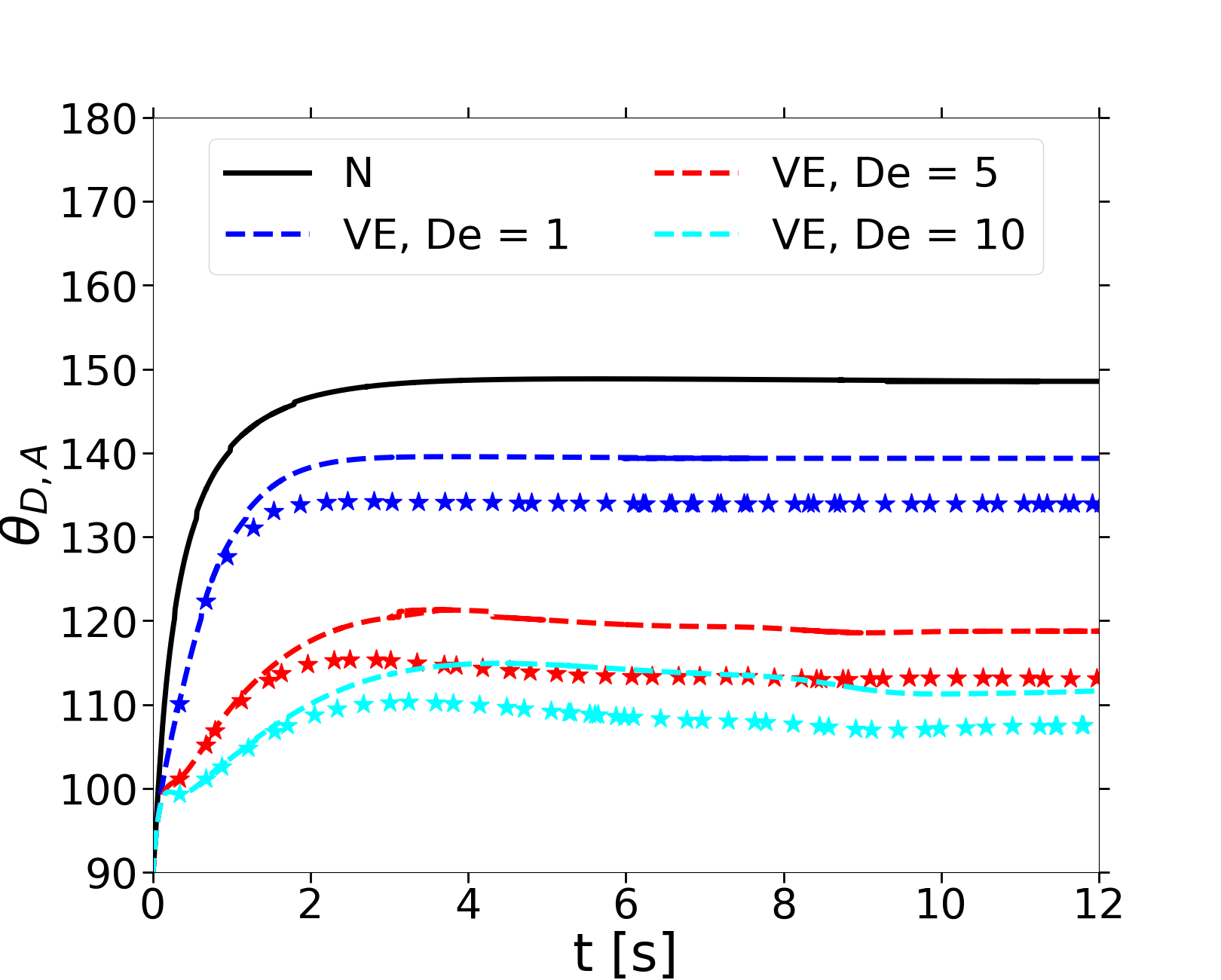}}\\
     \subfloat[]{\includegraphics[width=0.45\textwidth]{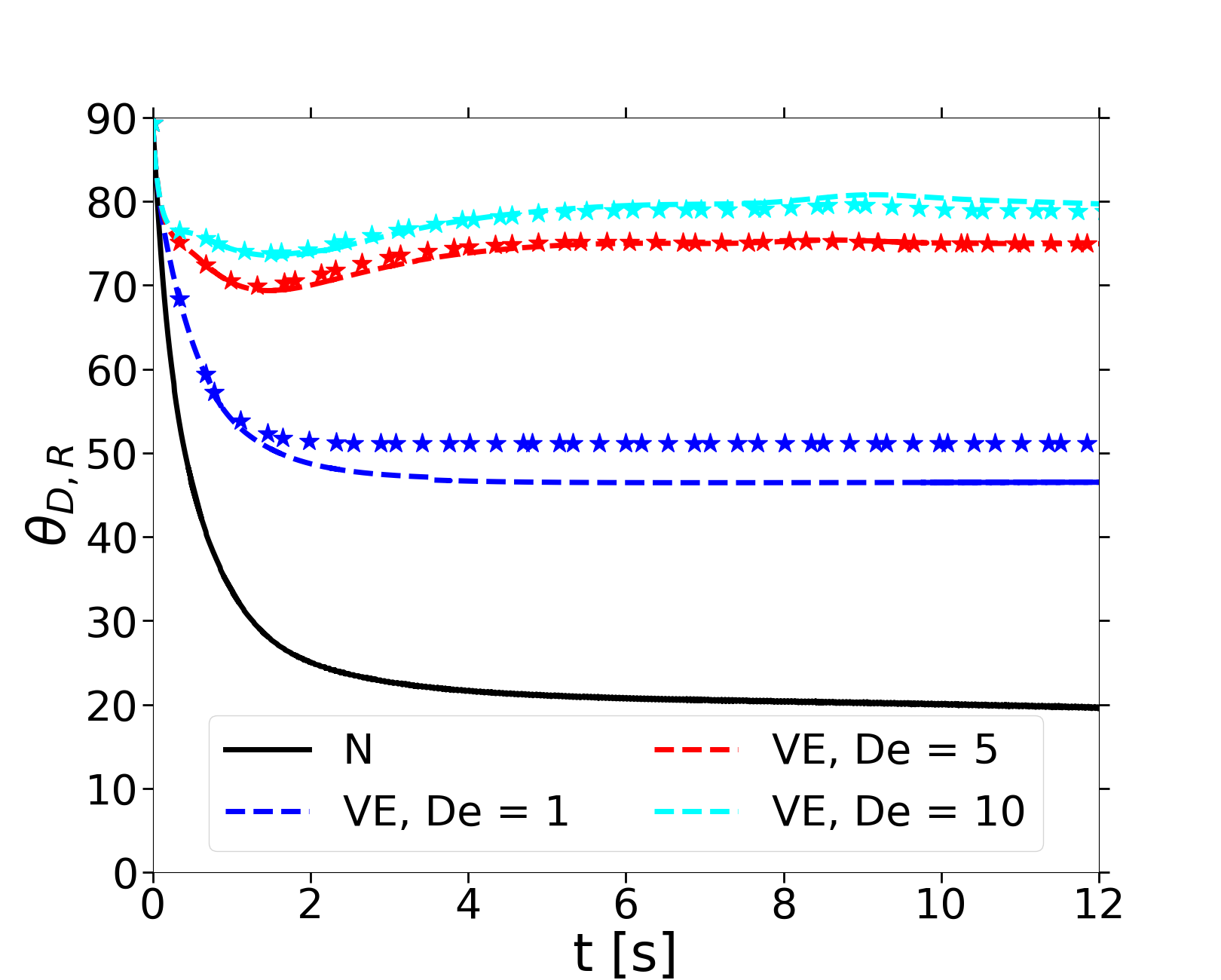}}
     \subfloat[]{\includegraphics[width=0.45\textwidth]{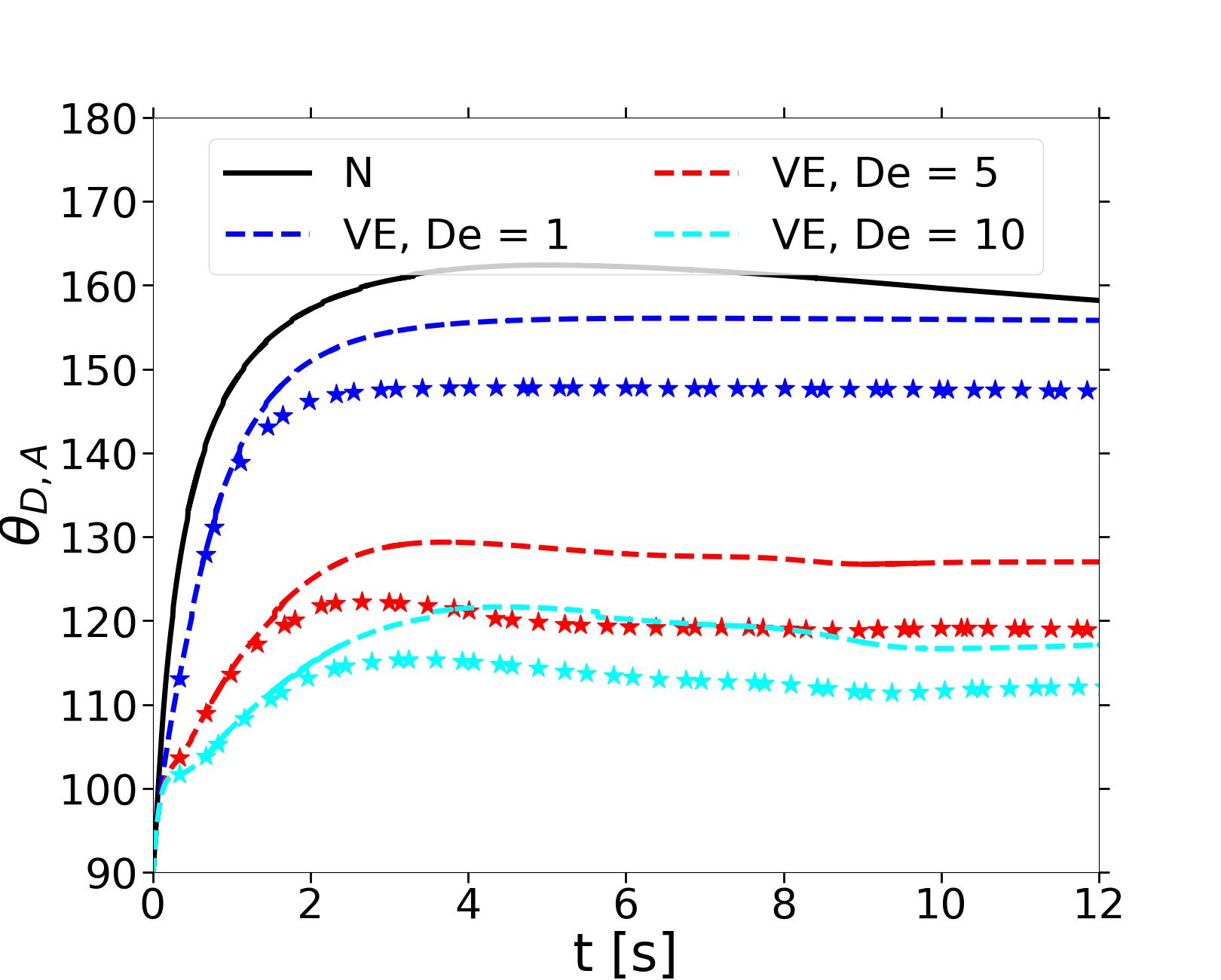}}
	\caption{The time evolution of dynamic contact angle $\theta_{D}$ over a surface with $\Delta\theta=140^{\circ}$ in the N/V system. N and VE refer to Newtonian and Giesekus matrices respectively. (a) $\theta_{D,R}$ for $Ca=0.15$ (b) $\theta_{D,A}$ for $Ca=0.15$ (c) $\theta_{D,R}$ for $Ca=0.2$ (d) $\theta_{D,A}$ for $Ca=0.2$ (e) $\theta_{D,R}$ for $Ca=0.25$ (f) $\theta_{D,A}$ for $Ca=0.25$. The symbols (\dashed) and ($\star$) correspond to $\alpha=0.05$ and $\alpha=0.1$ respectively.}
	\label{fig:fig13}
\end{figure}

The magnitude and tendency (direction) of the total stress vector in Eq. \ref{NS34} along the interface $\phi=0$ is depicted by arrows on top of figures \ref{fig:fig12} (a,b,c):(${\mathbf{n}}_{s}\cdot\boldsymbol\tau$, where ${\mathbf{n}}_{s}$ is the unit vector normal to the interface. The stresses are mostly pulling the interface inwards for both Newtonian (\ref{fig:fig12}a) and viscoelastic (\ref{fig:fig12}b,c) droplets. However, in the vicinity of the receding contact line marked with circle, the stresses are pulling the interface outward for the viscoelastic droplets. The magnitude of these outward stresses increases with $De$ (elasticity of the droplet) so that they are bringing about a larger $\theta_{D,R}$. The outward stresses around the receding contact line were not observed when capillary number was lowered to $Ca=0.15$, at least until $De=10$. 

Concluding, the non-monotonic behavior in the receding dynamic contact angle with increasing $De$ can be attributed to the combination of two effects: the outward stresses in the vicinity of the receding contact line, and the inception of the strain-hardening occurring for $De\ge5$. Also the smaller advancing dynamic contact angle at steady-state might be attributed to strain-hardening.
\begin{figure}[t]
	\centering
     \subfloat[]{\includegraphics[width=0.45\textwidth]{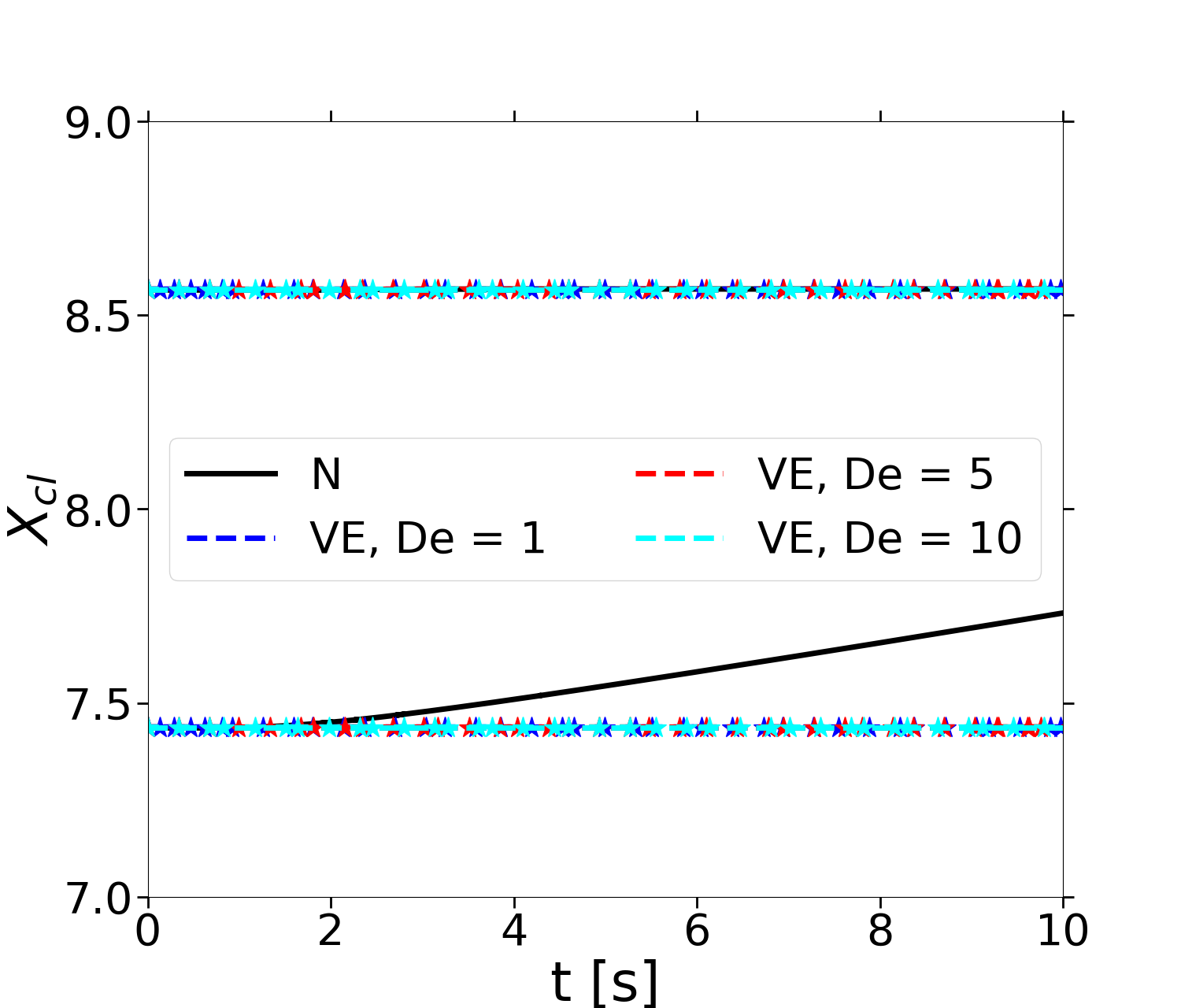}}
     \subfloat[]{\includegraphics[width=0.45\textwidth]{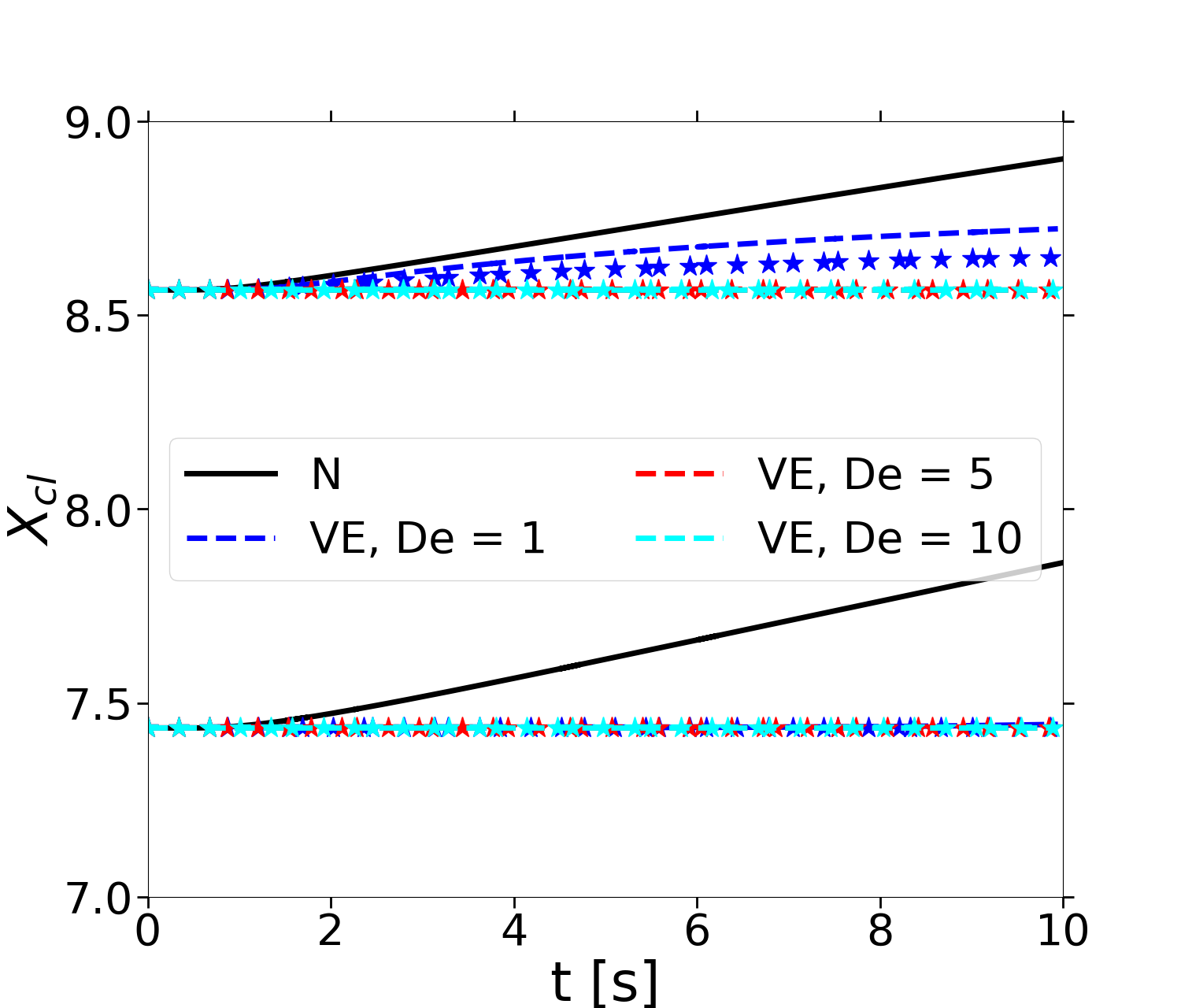}}
	\caption{The time evolution of $X_{cl,R}$ ($X^{t=0}_{cl,R}=7.436$) and $X_{cl,A}$ ($X^{t=0}_{cl,A}=9.564$) at different $De$ in the N/V system. N and VE refer to Newtonian and Giesekus matrices respectively. (a) $Ca=0.25$ over a surface with $\Delta\theta=140^{\circ}$ (b) $Ca=0.25$ over a surface with $\Delta\theta=100^{\circ}$. The symbols (\dashed) and ($\star$) correspond to $\alpha=0.05$ and $\alpha=0.1$ respectively.}
	\label{fig:fig15}
\end{figure}

\subsection{Effect of matrix viscoelasticity on a Newtonian droplet}\label{sec5_2}
In this section, we consider the effect of matrix viscoelasticity on the deformation and depinning of a Newtonian droplet (N/V system) in the presence of contact line hysteresis. The same boundary and initial conditions have been applied as previously, with the polymeric stresses are set to zero in the beginning of the simulation in the Giesekus medium. Figs. \ref{fig:fig13}(a,c,e) present the time evolution of $\theta_{D,R}$ of a Newtonian drop in the Giesekus medium over a surface with $\Delta\theta=140^{\circ}$, and it can be seen that the elasticity of the matrix has suppressed the drop's deformation in the receding side significantly for $De\ge5$, in other words, the viscous bending in the receding side has weakened drastically. The effect of changing capillary number on the steady-state value of $\theta_{D,R}$ is limited and small in the range $Ca=[0.15,0.25]$ used in our simulations when $De\ge5$. The same trend can be observed on the advancing side, where the viscoelasticity of the medium decreases the viscous bending in particular for $De=5-10$; Furthermore, the shear-thinning also plays an important role at higher De by further reducing the viscous bending, see Figs. \ref{fig:fig13}(b,d,f). 

It is worth mentioning that the decreased viscous bending results in that both contact angles are insensitive to capillary number \green{for the Newtonian drop in the viscoelastic matrix with high elasticity}. For a Newtonian droplet in a Newtonian matrix, when increasing capillary number from $Ca=0.15$ to $Ca=0.25$, both receding and advancing contact angles \green{vary} by $\propto 30^{\circ}$. For a Newtonian droplet in a viscoelastic matrix (at $De=5$ and $De=10$), the change with capillary number is less than $10^{\circ}$. Both contact lines are pinned at $Ca=0.15-0.25$ over the surface with $\Delta\theta=140^{\circ}$ in a viscoelastic matrix (Fig. \ref{fig:fig15}(a)), while the receding contact line of the Newtonian drop in the Newtonian medium moves with a significant velocity at $Ca=0.25$. If the hysteresis window is decreased to $\Delta\theta=100^{\circ}$ (Fig. \ref{fig:fig15}(b)), the advancing contact line of the least elastic \green{medium} ($De=1$) starts moving at $t\approx0.85\hspace{0.05cm}$s and the receding contact line remains pinned, while both contact lines start to move simultaneously in a Newtonian matrix \green{with appreciable velocities}.    

\begin{figure}[t]
	\centering
	\subfloat[]{\includegraphics[width=0.45\textwidth]{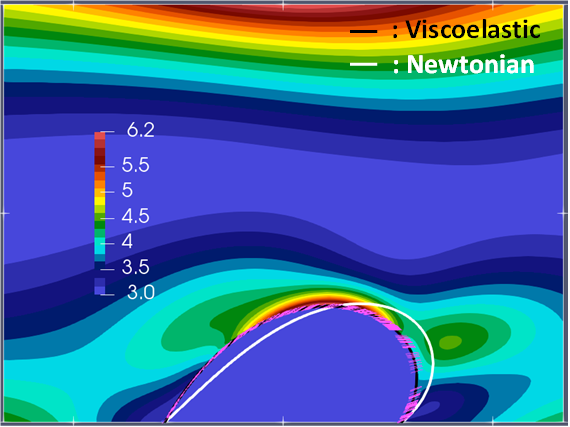}}
    \quad
    \subfloat[]{\includegraphics[width=0.45\textwidth]{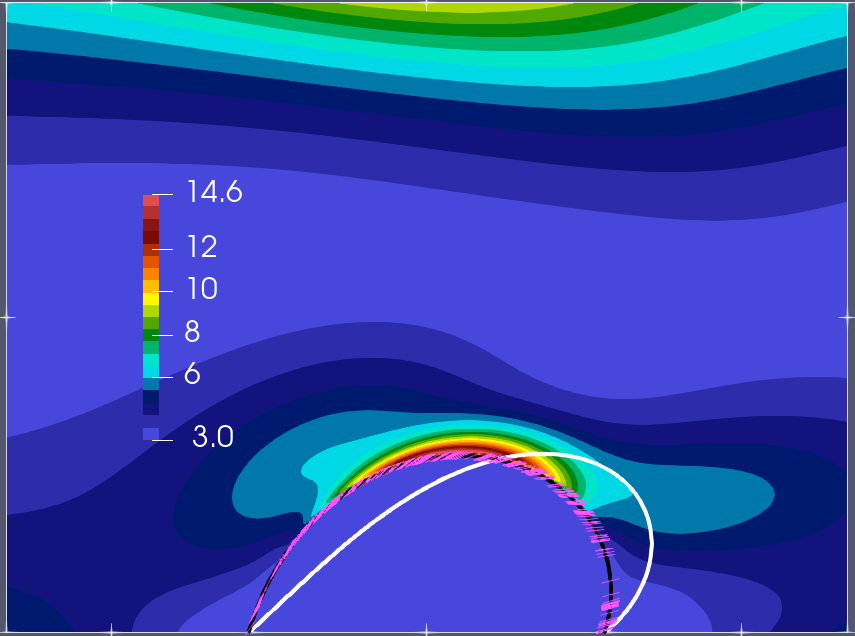}}
    \quad
    \subfloat[]{\includegraphics[width=0.45\textwidth]{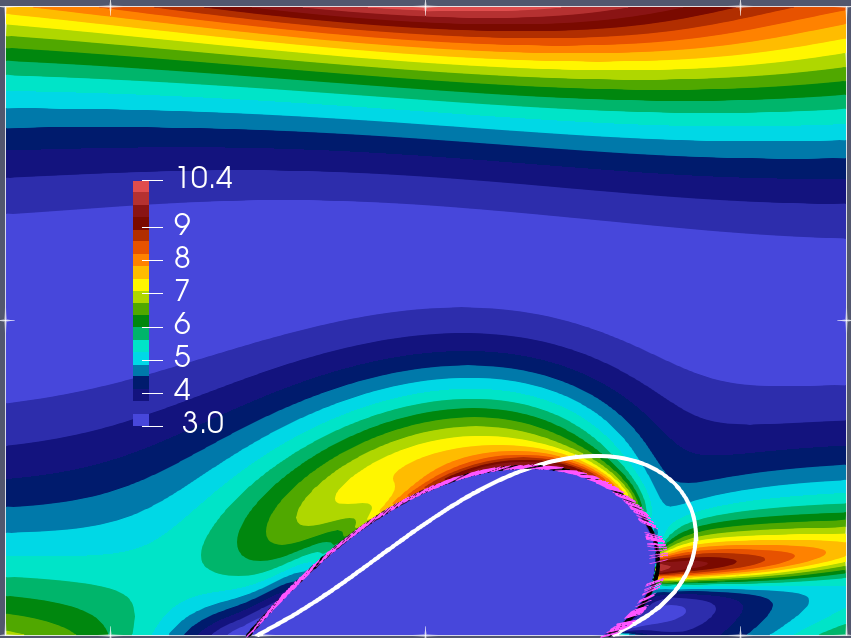}}
    \quad
    \subfloat[]{\includegraphics[width=0.45\textwidth]{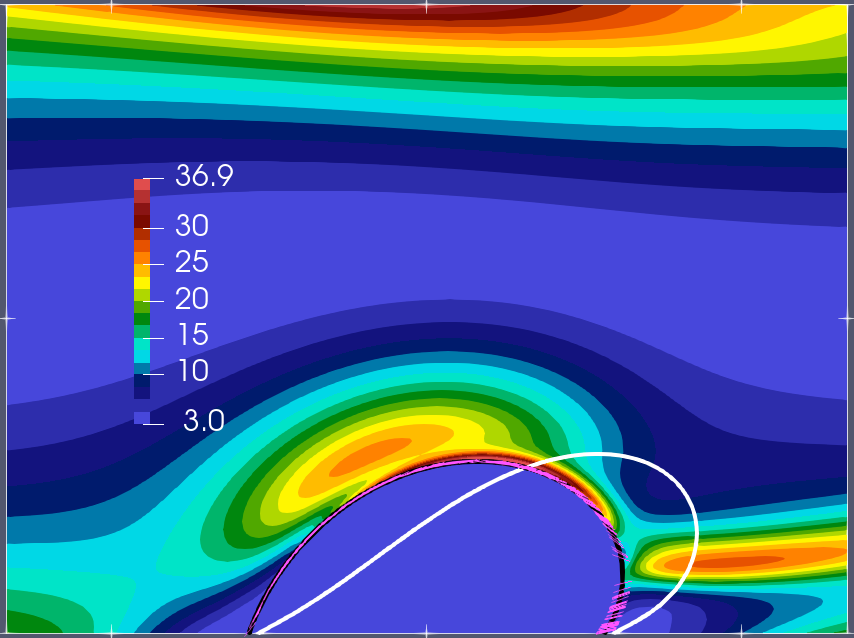}}
	\caption{ Time evolution of the Newtonian drop in viscoelastic matrix with $Ca=0.25$ over the surface with $\Delta\theta=100^{\circ}$. The interface $\phi=0$ of the Newtonian and Giesekus drops are depicted with a white and black lines, respectively. The trace of the conformation tensor has been visualized on the background, and the principal eigenvector of the conformation tensor is represented by purple line segments. (a) $De=1$ at $t\approx0.6\hspace{0.05cm}$s, (b) $De=5$ at $t\approx0.6\hspace{0.05cm}$s, (c) $De=1$ at $t\approx1.8\hspace{0.05cm}$s, (d) $De=5$ at $t\approx1.8\hspace{0.05cm}$s.}
	\label{fig:fig16}
\end{figure}
\begin{figure}[t]
	\centering
    \subfloat[]{\includegraphics[width=0.36\textwidth]{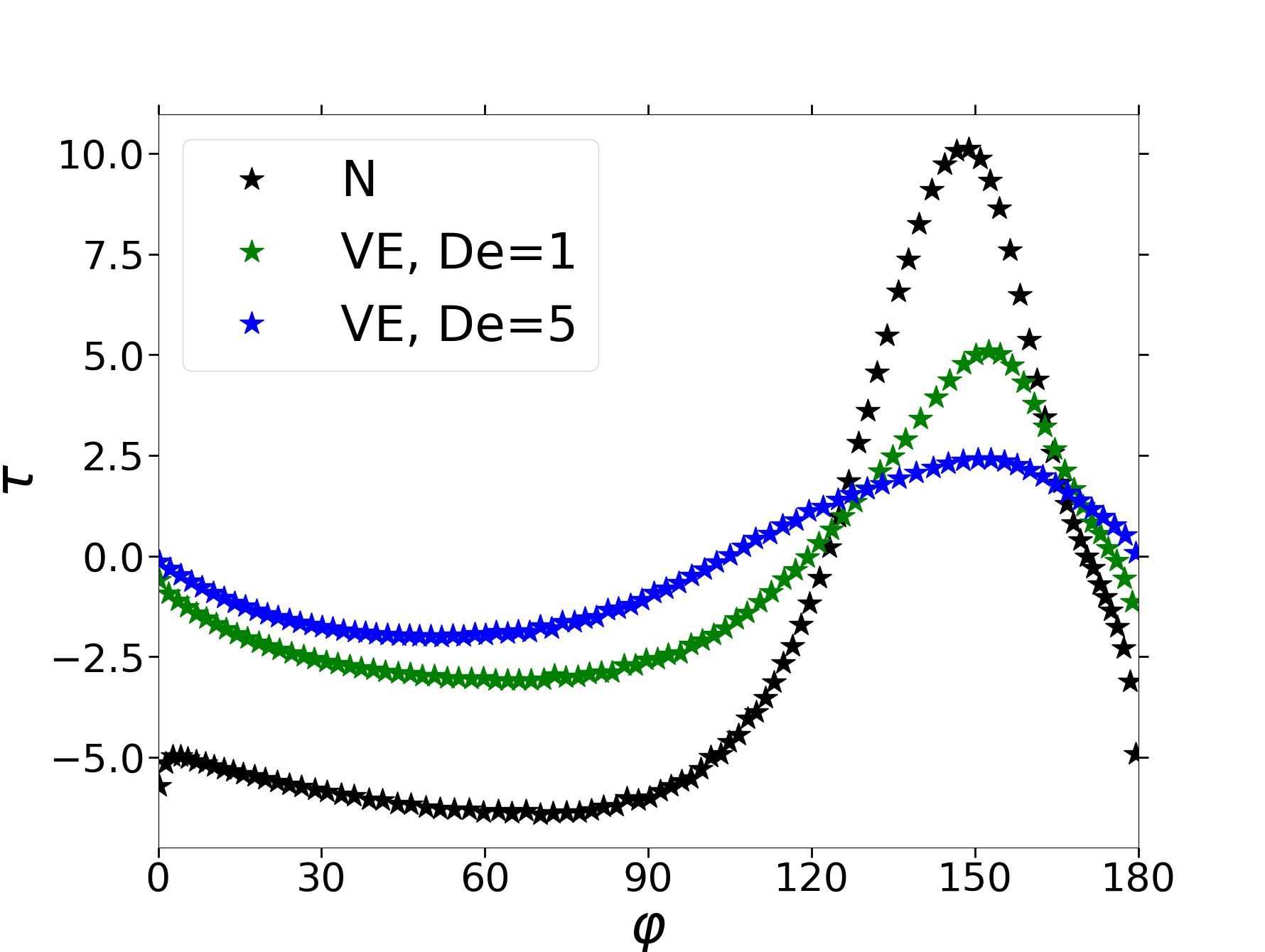}}
	\subfloat[]{\includegraphics[width=0.36\textwidth]{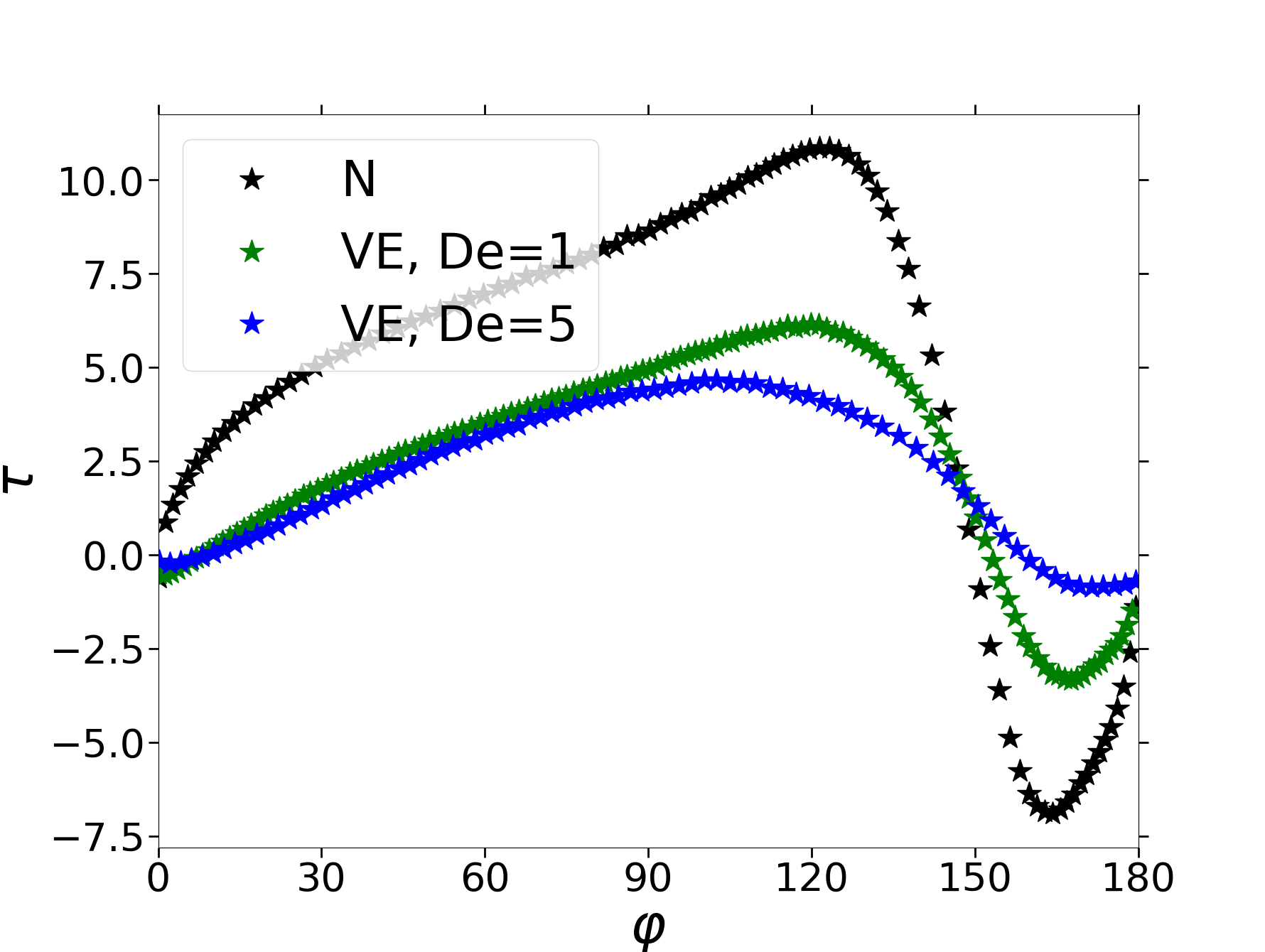}}\\
    \subfloat[]{\includegraphics[width=0.36\textwidth]{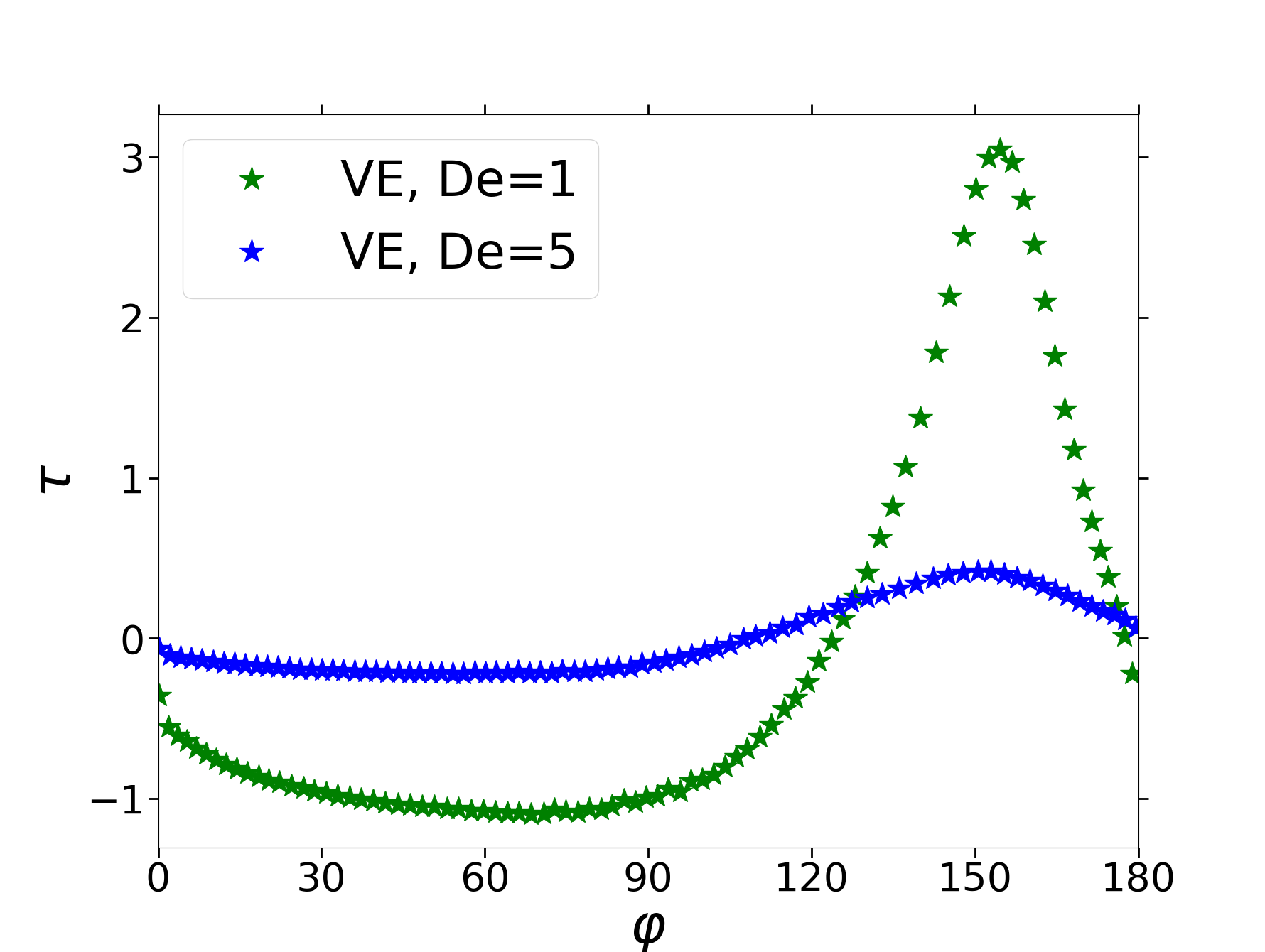}}
	\subfloat[]{\includegraphics[width=0.36\textwidth]{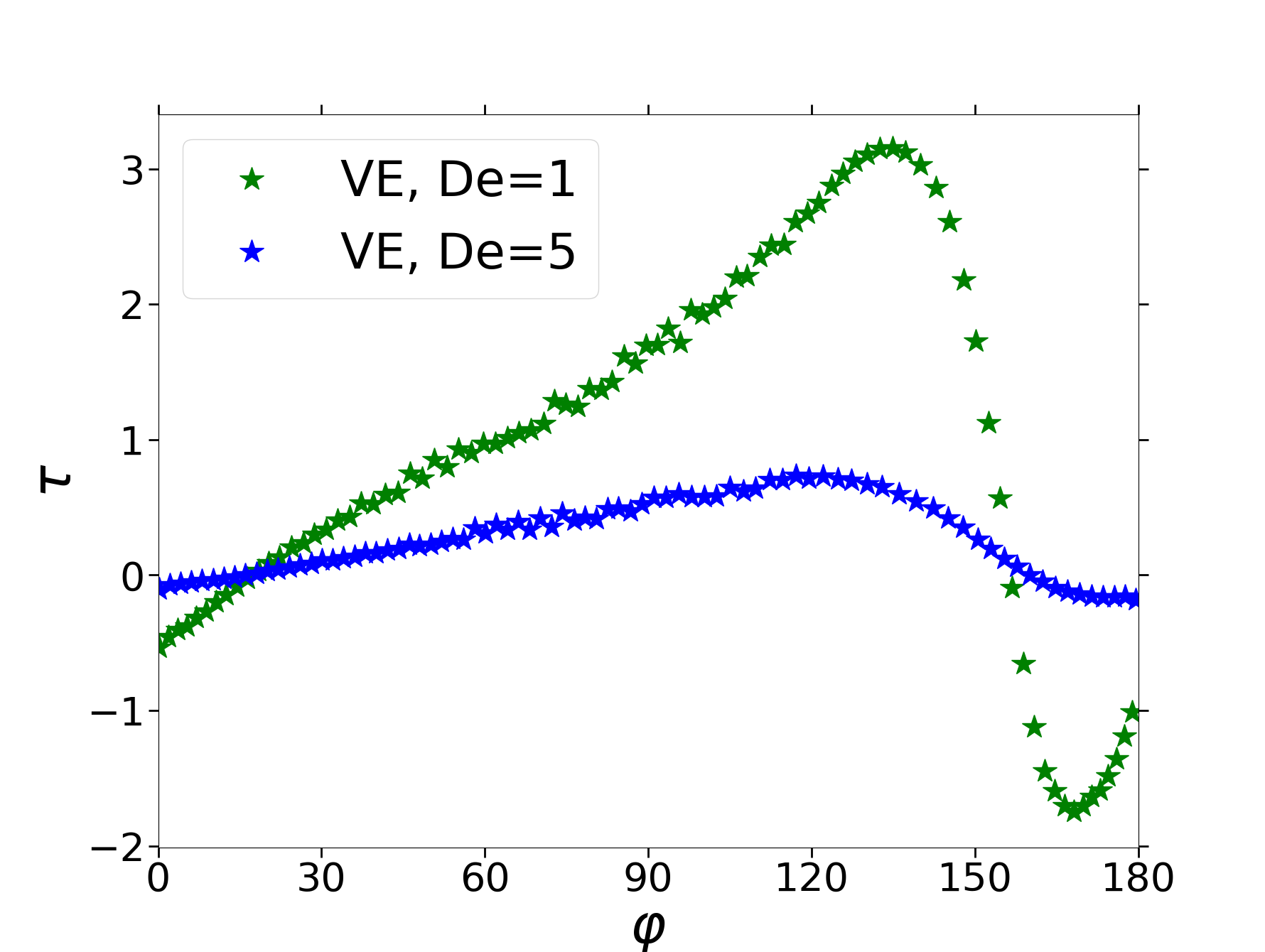}}\\
    \subfloat[]{\includegraphics[width=0.36\textwidth]{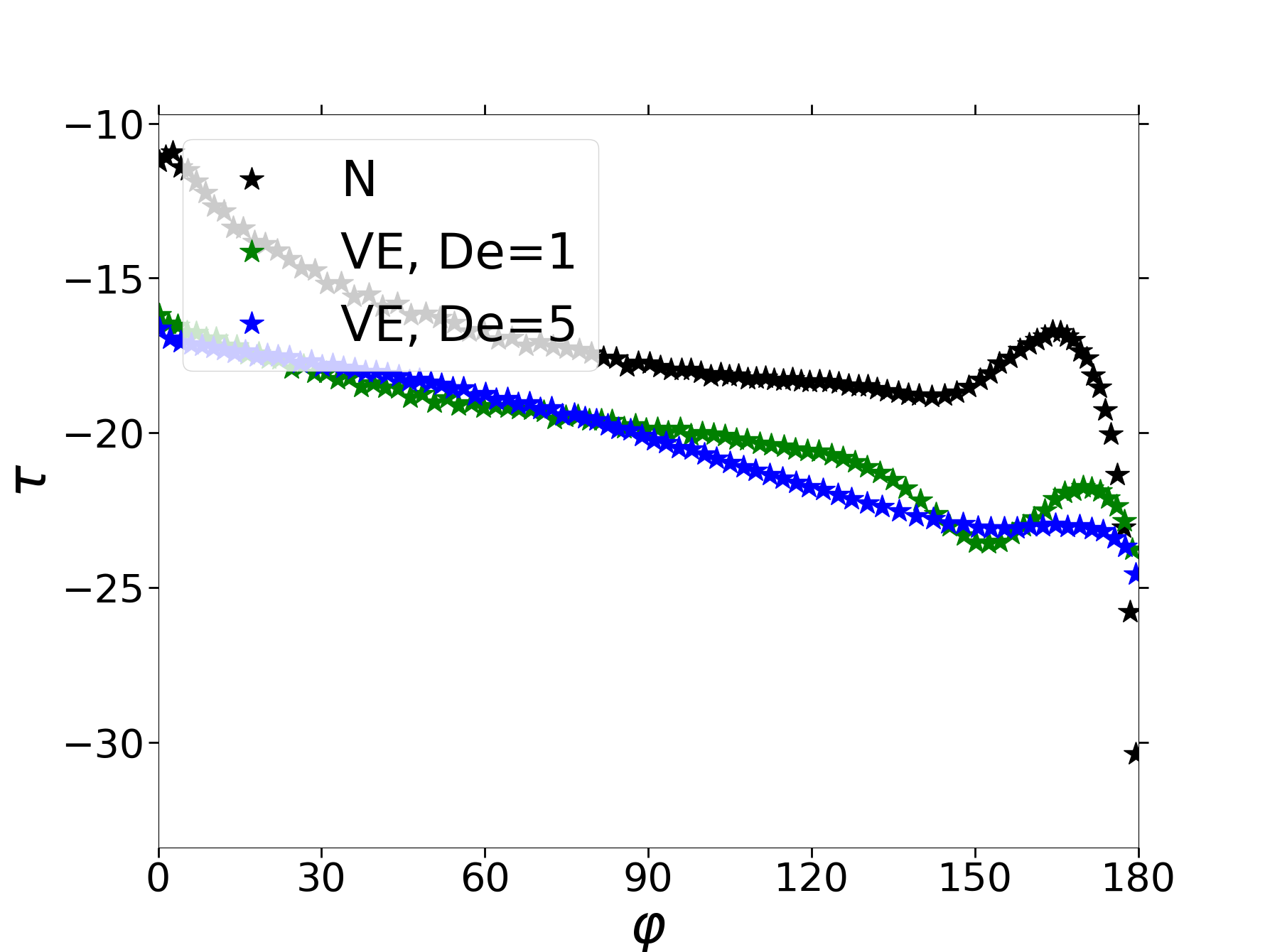}}
	\caption{Different components of the stresses along $\phi=0$ at time $t=0.6\hspace{0.05cm}$s for drops with $Ca=0.25$ over the surface with $\Delta\theta=100^{\circ}$ in N/V system. (a) $\tau^{n}_{vi}$ (b) $\tau^{t}_{vi}$ (c) $\tau^{n}_{p}$ (d) $\tau^{t}_{p}$ (e) $\tau^{n}_{pre}$.}
	\label{fig:fig17}
\end{figure}
The trace of conformation tensor along with principal stress orientation is shown at $De=1$ \ref{fig:fig16}(a,c) and $De=5$ \ref{fig:fig16}(c,d), respectively, over a surface with $\Delta\theta=100^{\circ}$. The stretching of polymer molecules is small in the vicinity of the contact lines, while large stretching occurs around the interface for $\varphi\approx[54^{\circ},138^{\circ}]$ since dumbbells are no longer parallel to the interface in this portion (as they were in the V/N case, fig. \ref{fig:fig9}), this implies that large polymeric stresses is expected over a large portion of the interface.

The normal component of the viscous stress $\tau^{n}_{vi}$ has been lessened significantly due to the elasticity of the matrix at $t=0.6\hspace{0.05cm}$s for $\varphi\approx[0,120^{\circ}]$ responsible for pulling the drop inward and for $\varphi\approx[127^{\circ},164^{\circ}]$ responsible for pulling the drop outward, see Fig. \ref{fig:fig17}(a); in the same manner, the tangential component of the viscous stress $\tau^{t}_{vi}$ has been weakened significantly for all values of $\varphi$, see Fig. \ref{fig:fig17}(b). Figs. \ref{fig:fig17}(c) and \ref{fig:fig17}(d) present that both normal and tangential components of the polymer stresses corresponding to the matrix with $De=5$ have been lessened in comparison to the medium with $De=1$. It should be noted that the same trend exists between the the polymer stresses at $t\approx1.8\hspace{0.05cm}$s for mediums with $De=1$ and $De=5$. These variations in the stresses along the interface ($\phi=0$) with increasing the elasticity of the matrix reveals that the immobilization of the drop in the matrix with $De\ge5$ is due to the lessened viscous and polymeric stresses along the interface.

\begin{figure}[t]
	\centering
    \subfloat[]{\includegraphics[width=0.45\textwidth]{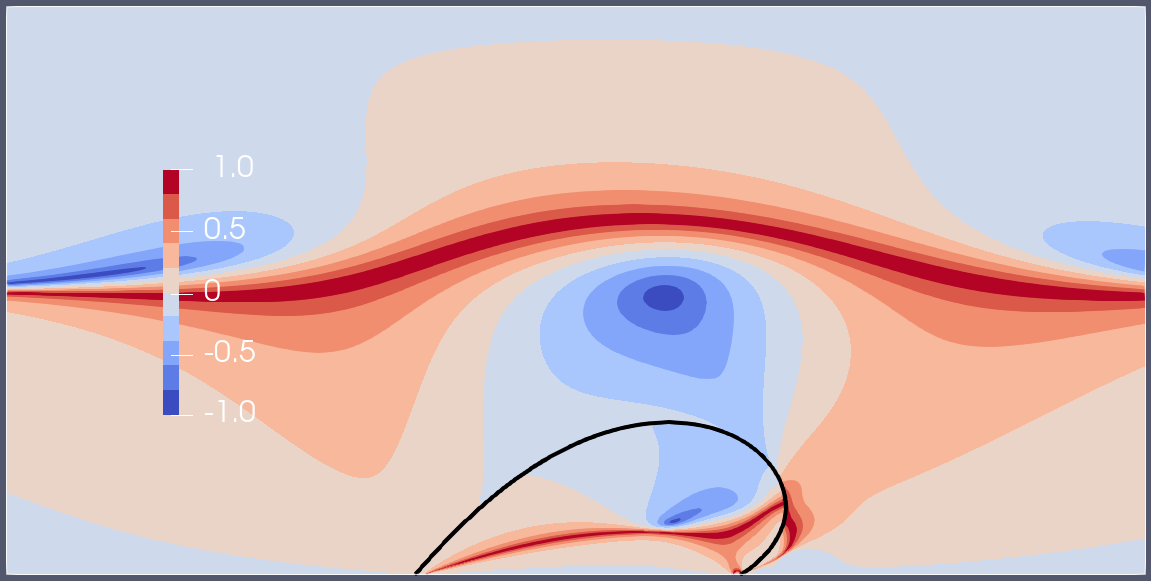}}
    \quad
    \subfloat[]{\includegraphics[width=0.45\textwidth]{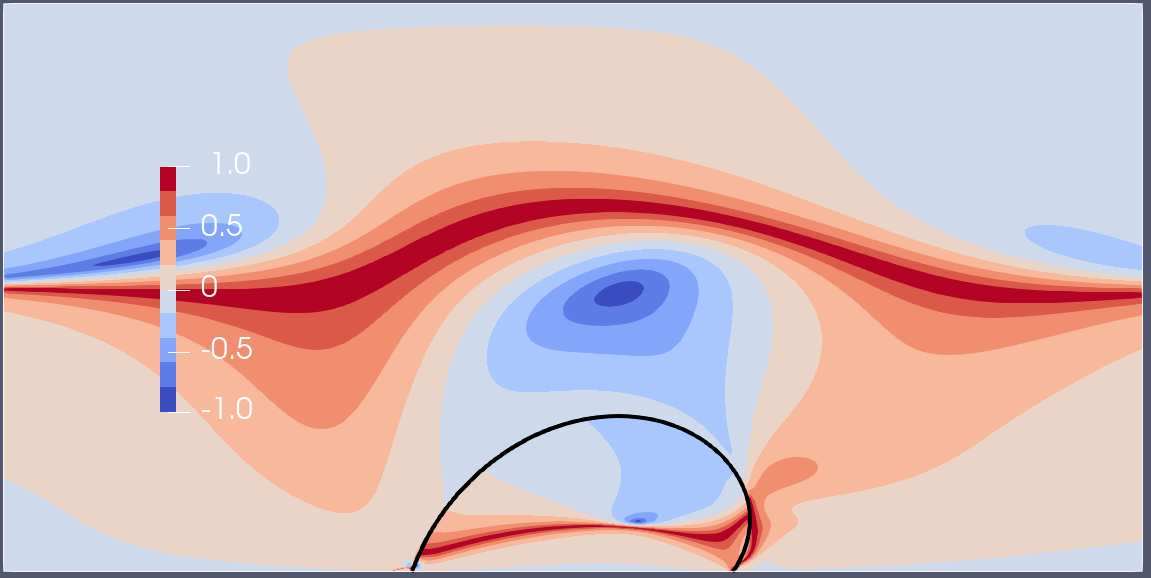}}\\
    \subfloat[]{\includegraphics[width=0.45\textwidth]{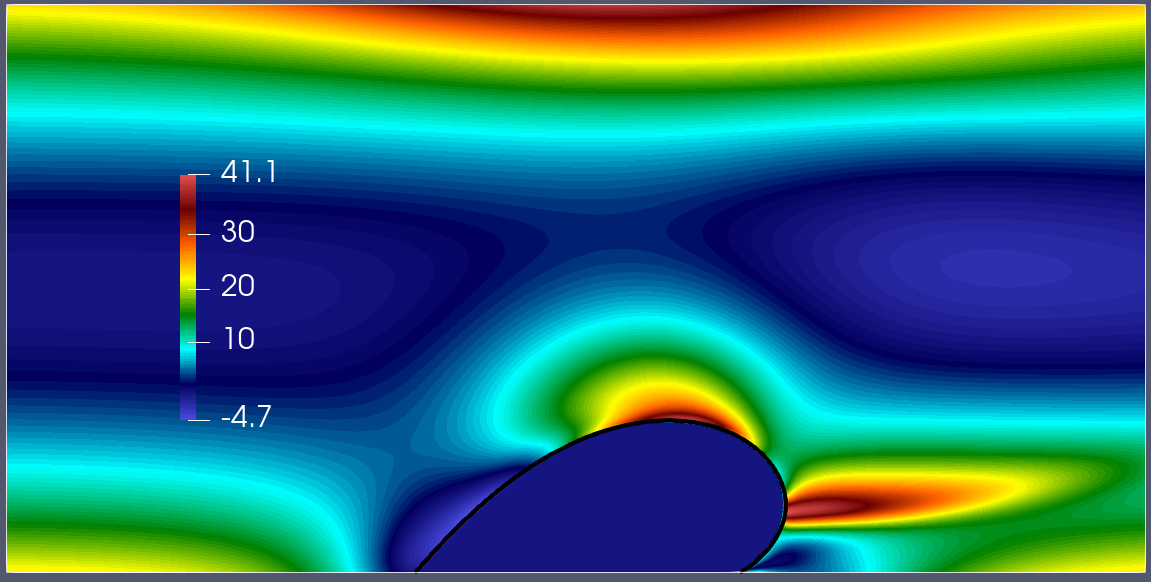}}
    \quad
    \subfloat[]{\includegraphics[width=0.45\textwidth]{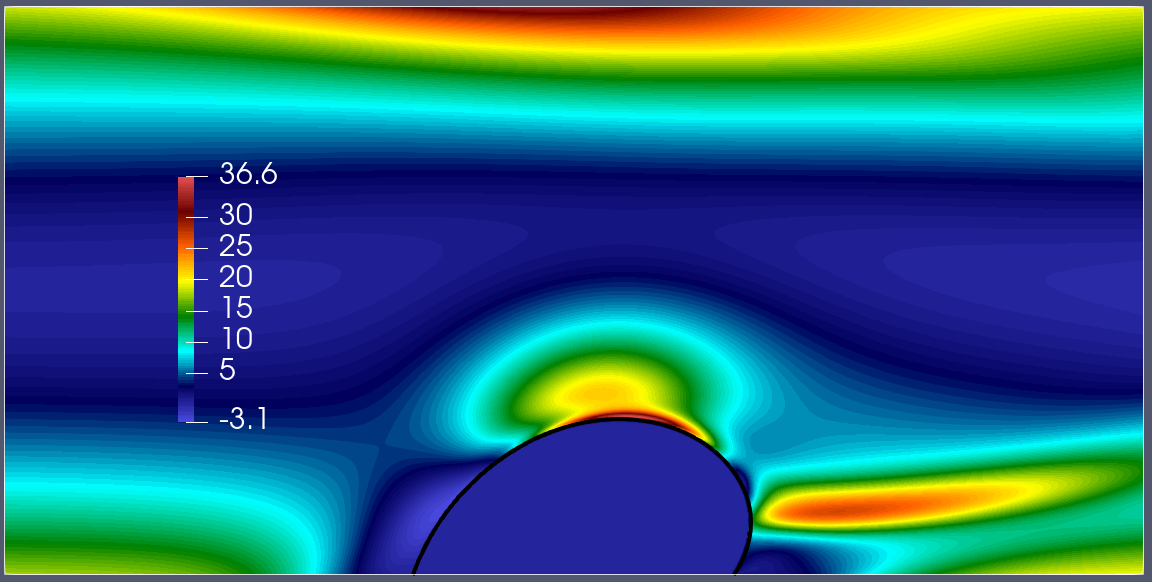}}\\
    \subfloat[]{\includegraphics[width=0.45\textwidth]{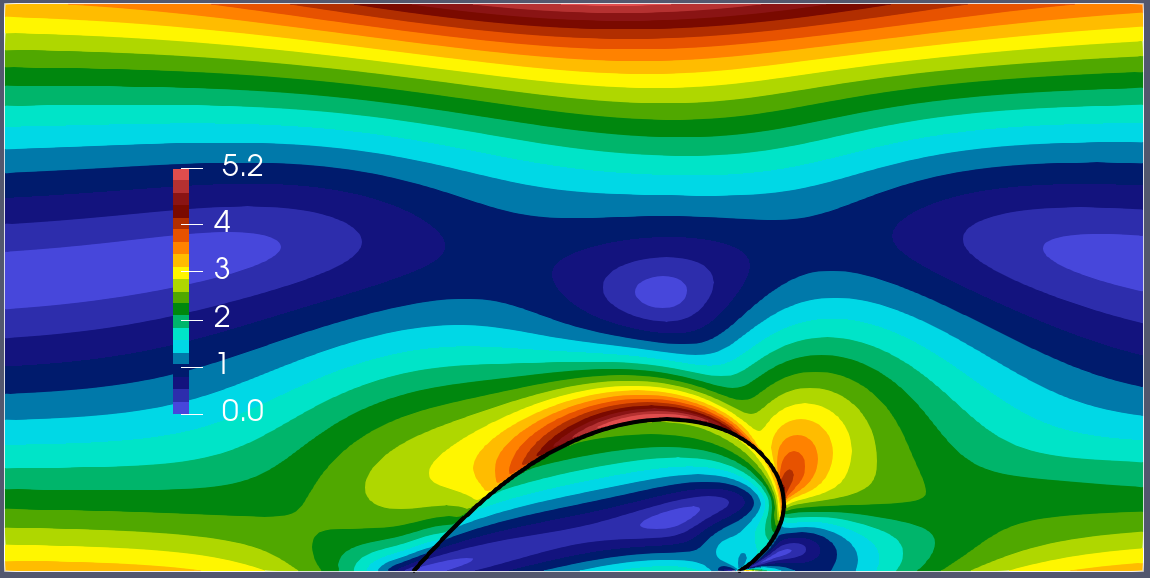}}
    \quad
    \subfloat[]{\includegraphics[width=0.45\textwidth]{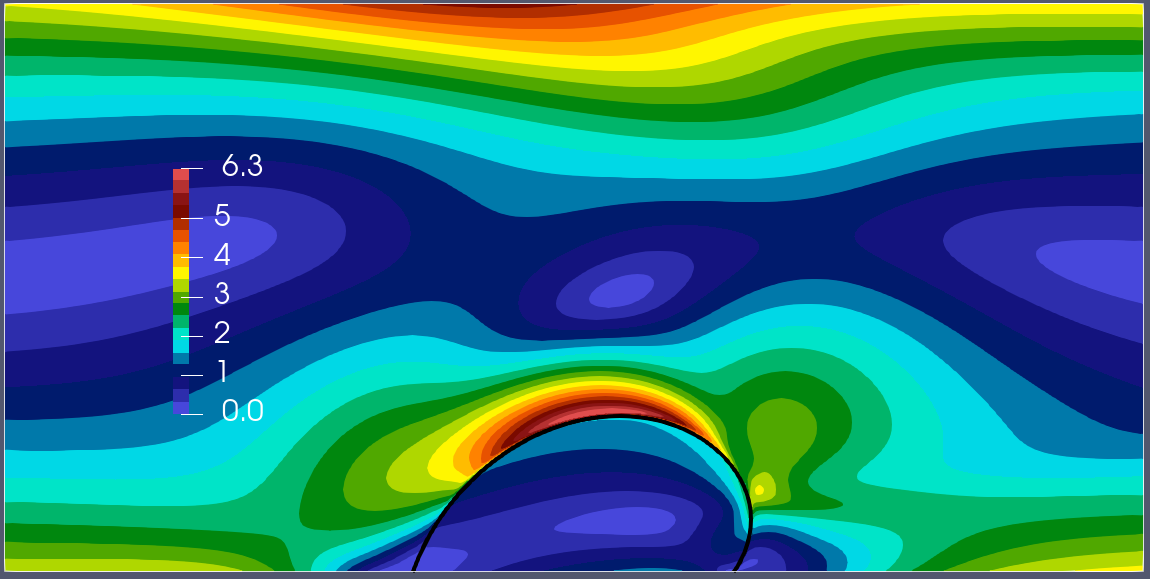}}\\
	\caption{Different flow quantities of the Newtonian drop in viscoelastic matrices with $Ca=0.25$ over the surface with $\Delta\theta=100^{\circ}$ at $t\approx1.8\hspace{0.05cm}$s. (a) flow parameter ($\xi$) with $De=1$, (b) flow parameter ($\xi$) with $De=5$, (c) first normal stress ($N_{1}$) with $De=1$, (d) first normal stress ($N_{1}$) with $De=5$, (e) the magnitude of strain rate ($|\Dot{\boldsymbol\gamma}|$) with $De=1$, (f) the magnitude of strain rate ($|\Dot{\boldsymbol\gamma}|$) with $De=5$.}
	\label{fig:fig18}
\end{figure}

The flow type becomes more extensionally dominated in the vicinity of the receding and advancing contact lines with increasing the matrix elasticity ($De\ge5$), in particular near the interface, see Figs. \ref{fig:fig18}(a) and \ref{fig:fig18}(b). Large first normal stress $N_{1}$ is accompanied with dominating shear flow in the matrix around the interface for $\varphi\approx[35^{\circ},145^{\circ}]$ so that the shear-thinning is the effective mechanism in this region of the interface, and this shear-thinning effect is reflected in decrease in $\theta_{A}$ with increasing $\alpha$, see Figs. \ref{fig:fig13}(b), \ref{fig:fig13}(d), \ref{fig:fig13}(f). The strain rate is very small and almost negligible within the matrix with $De\ge5$ in the vicinity of the interface for $\varphi\approx[0^{\circ},22^{\circ}]$ and $\varphi\approx[170^{\circ},180^{\circ}]$, see Fig. \ref{fig:fig18}(f) so that the observed $tr(\boldsymbol c)\approx3$ in this region (Fig. \ref{fig:fig16}(d)), is the consequence of the small strain rate in these regions. 

Summarizing, also viscoelasticity in the matrix phase in the N/V system suppresses the deformation of the droplet and pins the contact lines, and does so more efficiently than the viscoelasticity of the droplet in a V/N system. In both cases, the effect of viscoelasticity mainly stems from changes in the flow field and weakening the flow gradients around the droplet such as tangential viscous shear, which results in decreasing viscous bending and hence decreased movement of the contact lines.

\section{Conclusions}
In the present study, two-dimensional numerical simulations have been conducted to investigate the effect of viscoelasticity either in the drop or in the matrix on the depinning and deformation of a drop adhering to the wall in Poiseuille flow at low Reynolds number, under the influence of contact angle hysteresis. The results indicate that the viscoelastic properties have a significant impact on the drop's deformation and depping of the contact lines.

We first compare Newtonian and viscoelastic droplets at a low constant Deborah number ($De=1$) and different capillary numbers. The elasticity of the drop causes the viscoelastic drop to deform faster than the Newtonian counterpart at the early initial times, but the Newtonian drop then catches up. The viscoelastic drop continues to deform, and its receding dynamic contact angle reaches a smaller final value than the Newtonian counterpart \green{prior to depinning}; the advancing dynamic contact angle of a viscoelastic drop reaches a smaller value if the receding contact angle is pinned, otherwise it attains a larger final value. This final stage of the deformation depends on both $Ca$ and $\Delta\theta$ for a fixed $De$.

Secondly, we investigated the effect of increasing elasticity ($De=5$ and $De=10$) over the surfaces with $\Delta\theta=140^{\circ}$ and $\Delta\theta=100^{\circ}$. In the first stage of the deformation, the effect of increasing $De$ is negligible, but in the second stage, the time evolution of the dynamic contact angle at both receding and advancing sides is hindered significantly with non-monotonic behavior in the receding dynamic contact angle. This trend is attributed to the emergence of the pulling outward stresses changing to inward along the interface depending on both $Ca$ and $De$ and the inception of the strain-hardening inside the drop at both receding and advancing sides. 

Finally, we scrutinized the effect of matrix viscoelastic properties in N/V system over surfaces with $\Delta\theta=140^{\circ}$ and $\Delta\theta=100^{\circ}$. The viscoelasticity of the matrix lessens the deformation of the Newtonian drop, in other words, the viscous bending has been weakened in both receding and advancing sides due to the presence of polymer molecules in the medium, and these reductions are more pronounced for $De\ge5$. These reductions are bring the drops in the N/V system to halt in the matrix with large elasticity ($De\ge5$). Furthermore, the shear-thinning effect reduces the the viscous bending further in the advancing side by increasing $\alpha$ from 0.05 to 0.1. 
\section{Declaration of competing interest}
The authors declare that they have no known competing financial interests or personal relationships that could have appeared to influence the work reported in this paper.
\section{Acknowledgments}
This work was funded by European Research Council (ERC) through Starting grant no. 852529 MUCUS, by the Swedish Research Council through grant VR 2017-0489, and by Swedish e-Science Research Centre. We acknowledge the computing time on the supercomputer Beskow at the PDC center, KTH provided by SNIC (Swedish National Infrastructure for Computing),Sweden. 
\section{DATA AVAILABILITY}
The data that support the findings of this study are available from the corresponding author upon reasonable request.
\renewcommand{\thefigure}{A\arabic{figure}}
\renewcommand{\theequation}{A\arabic{equation}}
\setcounter{equation}{0}
\setcounter{figure}{0}
\section*{Appendix A:}\label{appenda}
The velocity of the contact line can be related to the static and dynamic contact angles over a surface without hysteresis \citep{Yue2011}, and a similar relation holds for the velocity of contact line over a surface with $\theta_{A}$ and $\theta_{R}$. The phase-field variable in our simulation is a function of $(x,y,t)$, and it is a function of two variable if we constrain it to the wall $y=0$ so that it can be wrriten as
\begin{eqnarray}
\phi=\phi(x,t) \hspace{0.5cm} on \hspace{0.5cm} \partial\Omega_{w}
\label{app_1}
\end{eqnarray}
it is initialized at $t=0$ ($\phi_{0}=\phi(x,0)$) on the wall and evolved over time by Eq. \ref{NS13}. The level curves of $\phi(x,t)=C_{\phi}$, where $-1\lesssim C_{\phi}\lesssim1$, are curves defined on the $(x,t)$ plane for different values of $C_{\phi}$. If we take the total differential of Eq. \ref{app_1} and put it equal zero $d\phi=\frac{\partial{\phi}}{\partial{x}}dx+\frac{\partial{\phi}}{\partial{t}}dt=0$, we can get a differential equation for the level curves on the $(x,t)$ plane
\begin{eqnarray}
\frac{dx}{dt}=-\frac{\frac{\partial{\phi}}{\partial{t}}}{\frac{\partial{\phi}}{\partial{x}}}
\label{app_2}
\end{eqnarray}
since we are interested on the time evolution of the contact line ($X_{cl}(t)$) on the $(x,t)$ plane and its velocity $U_{cl}=\frac{dx}{dt}|_{\phi=0}$, we choose $C_{\phi}=0$ and evaluate the $\frac{\partial{\phi}}{\partial{t}}|_{\phi=0}$ and $\frac{\partial{\phi}}{\partial{x}}|_{\phi=0}$. We compute $\frac{\partial{\phi}}{\partial{t}}|_{\phi=0}$ by using Eq. \ref{NS13}, and we assume the interface intersecting the wall is a planar interface 
\begin{eqnarray}
-\frac{\partial{\phi}}{\partial{t}}|_{\phi=0}=\frac{1}{\mu_{f}\eta}(\frac{3}{2\sqrt{2}}\sigma\eta \mathbf{n}\cdot\nabla{\phi}+f^{\prime}_{w}(\phi))|_{\phi=0}&=& \nonumber \\
\frac{3}{2\sqrt{2}}\frac{\sigma}{\mu_{f}}|\nabla{\phi}|^{A/R}_{\phi=0}(\cos{(\theta_{d})}|^{A/R}-\frac{1}{\sqrt{2}\eta|\nabla{\phi}|_{\phi=0}}\cos{(\theta_{A/R})})
\label{app_3}
\end{eqnarray}
where $\mathbf{n}\cdot\nabla{\phi}|_{\phi=0}=|\nabla{\phi}|_{\phi=0}\cos{(\theta_{d}})$ is used. We then need to compute $\frac{\partial{\phi}}{\partial{x}}|_{\phi=0}$
\begin{eqnarray}
\frac{\partial{\phi}}{\partial{x}}|_{\phi=0}=-|\nabla{\phi}|_{\phi=0}\cos{(\theta_{d}-\frac{\pi}{2})}=-|\nabla{\phi}|_{\phi=0}\sin{(\theta_{d}})
\label{app_4}
\end{eqnarray}
substituting Eqs. (\ref{app_3}-\ref{app_4}) in Eq. \ref{app_2}, we get the contact line velocity as
\begin{eqnarray}
\frac{\mu_{f}U_{cl}}{\sigma}=\frac{3}{2\sqrt{2}}\frac{1}{\sin{(\theta_{d}})}(\frac{1}{\sqrt{2}\eta|\nabla{\phi}|}\cos{(\theta_{A/R})}-\cos{(\theta_{d}}))
\label{app_5}
\end{eqnarray}
where $\theta_{R}$ is chosen when $\theta_{d}\leq\theta_{R}$ and $\theta_{A}$ if $\theta_{d}\ge\theta_{A}$ in the above equation. It should be noted that $U_{cl}<0$ when the fluid with $\phi=-1$ replaces the fluid with $\phi=1$ if not $U_{cl}>0$.

We take the derivative of Eq. \ref{app_5} with respect to $\theta_{d}$ since we just want to see the effect of varying $\theta_{d}$ on the $U_{cl}$ with having the same $\theta_{A/R}$
\begin{eqnarray}
\frac{\mu_{f}}{\sigma}\frac{\partial{U_{cl}}}{\partial{\theta_{d}}}=(\frac{\partial{}}{\partial{\theta_{d}}}\frac{1}{\sin{(\theta_{d}})})(\frac{3}{2\sqrt{2}}(\frac{1}{\sqrt{2}\eta|\nabla{\phi}|}\cos{(\theta_{A/R})}-\cos{(\theta_{d}})))+ \nonumber \\
\frac{3}{2\sqrt{2}}\frac{1}{\sin{(\theta_{d})}}(\frac{\cos{(\theta_{A/R})}}{\sqrt{2}\eta}\frac{\partial{}}{\partial{\theta_{d}}}\frac{1}{|\nabla{\phi}|}-\frac{\partial{\cos{(\theta_{d}})}}{\partial{\theta_{d}}})&=& \nonumber \\
\frac{-\cos{(\theta_{d})}}{\sin^{2}{(\theta_{d})}}(\frac{3}{2\sqrt{2}}(\frac{1}{\sqrt{2}\eta|\nabla{\phi}|}\cos{(\theta_{A/R})}-\cos{(\theta_{d}})))+ \nonumber \\
\frac{3}{2\sqrt{2}}\frac{1}{\sin{(\theta_{d})}}(\frac{\cos{(\theta_{A/R})}}{\sqrt{2}\eta}\frac{\partial{}}{\partial{\theta_{d}}}\frac{1}{|\nabla{\phi}|})+\frac{3}{2\sqrt{2}}&=& \nonumber \\
-\frac{\mu_{f}U_{cl}}{\sigma}\cot{(\theta_{d})}+\frac{3}{2\sqrt{2}}(1+\frac{\cos({\theta_{A/R})}}{\sin{(\theta_{d})}}\frac{\partial{}}{\partial{\theta_{d}}}\frac{1}{\sqrt{2}\eta|\nabla{\phi}|})&=& \nonumber \\
-\frac{\mu_{f}U_{cl}}{\sigma}\cot{(\theta_{d})}+\frac{3}{2\sqrt{2}}(1+C_{\theta_{d}})\quad \quad \quad \quad \quad \quad \quad \quad \quad \quad \quad
\label{app_6}
\end{eqnarray}
The assumption that the $\phi=$constant profiles remain parallel (planar interface) after infinitesimal variation in $\theta_{D}$ and reaching equilibrium can simplify the Eq. \ref{app_6}. The gradient of phase-field variable $|\nabla{\phi}|$ for a planar interface in equilibrium is given by
\begin{eqnarray}
|\nabla{\phi}|_{\phi=0}=\frac{1-\phi^{2}}{\sqrt{2}\eta}|_{\phi=0}=\frac{1}{\sqrt{2}\eta}
\label{app_7}
\end{eqnarray}
with this assumption $\frac{\partial{}}{\partial{\theta_{d}}}\frac{1}{\sqrt{2}\eta|\nabla{\phi}|}$ should be small and negligible in comparison to other terms, and the Eq. \ref{app_6} can be reduced to 
\begin{eqnarray}
\frac{\mu_{f}}{\sigma}\frac{\partial{U_{cl}}}{\partial{\theta_{d}}}=-\frac{\mu_{f}U_{cl}}{\sigma}\cot{(\theta_{d})}+\frac{3}{2\sqrt{2}}
\label{app_9}
\end{eqnarray}
Eq. \ref{app_9} anticipates a large variation in the velocity with a change in $\theta_{d}$ due to the dependency on $\cot{}$ function over superhydrophobic surfaces since these surfaces usually own large contact angles.
\bibliography{mybibfile}

\end{document}